\documentstyle[12pt,psfig]{article}
\newcommand{\s}{\rm}

\newcommand{\ra}{\rightarrow}
\newcommand{\mn}{\mu \nu}

\newcommand{\be}{\begin{equation}}
\newcommand{\ee}{\end{equation}}

\newcommand{\bea}{\begin{eqnarray}}
\newcommand{\eea}{\end{eqnarray}}
\newcommand{\bef}{\begin{figure}}
\newcommand{\eef}{\end{figure}}

\newcommand{\lda}{\lambda}

\newcommand{\D}{\Delta}
\newcommand{\eps}{\epsilon}
\newcommand{\ab}{\alpha\beta}
\newcommand{\sls}{\!\!\!/}
\newcommand{\wnn}{\omega NN}
\newcommand{\snn}{\sigma NN}
\newcommand{\lgl}{\langle}

\begin{document}
%\begin{flushright}
%{\bf KUNS-1604}\\
%{\bf hep-ph/9909267}
%\end{flushright}
\begin{center}
{\Large{Thermal Photons and Lepton Pairs \\
 from  Quark Gluon Plasma
 and Hot Hadronic Matter} \\
}
\vskip .2in
Jan-e Alam$^{a,}$\footnote{On leave from Variable Energy Cyclotron Centre, 
1/AF Bidhan Nagar, Calcutta 700 064, India; Present
address: Department of Physics, University of
Tokyo, Tokyo 113 0033, Japan.}, Sourav Sarkar$^b$, Pradip Roy$^c$, 
 T. Hatsuda$^{a,}$\footnote{Present address: Department of Physics,
University of Tokyo, Tokyo 113 0033, Japan.}
 and Bikash Sinha$^{b,c}$\\
\vskip .1in

{\it a) Physics Department, Kyoto University, Kitashirakawa,
     Kyoto 606-8502, Japan}\\

{\it b) Variable Energy Cyclotron Centre,
     1/AF Bidhan Nagar, Calcutta 700 064
     India}\\

{\it c) Saha Institute of Nuclear Physics,
           1/AF Bidhan Nagar, Calcutta 700 064
           India}\\
\end{center}

\begin{center}
{\bf Abstract}\\
\end{center}

The formulation of the real and virtual photon production rate 
from strongly interacting matter is presented in the framework
of finite temperature field theory.  The changes in the hadronic
spectral function induced by temperature are discussed within the
ambit of the Walecka type model, gauged linear and non-linear
sigma models, hidden local symmetry approach  and QCD sum rule approach.
Possibility of observing the direct thermal photon and lepton pair
from quark gluon plasma has been contrasted with those from hot 
hadronic matter with and without medium effects for various
mass variation scenarios. 
At SPS energies, in-medium effects of different magnitude 
on the hadronic properties 
for the Walecka model, Brown-Rho scaling and Nambu scaling
scenarios  are conspicuously visible through the low invariant mass 
distribution of dilepton and transverse momentum spectra of photon.
However, at RHIC energies the thermal photon (dilepton)
spectra originating from Quark Gluon Plasma overshines those from
hadronic matter for large transverse momentum (invariant mass) irrespective
of the models used for evaluating the finite temperature effects
on the hadronic properties. It is thus expected that both at RHIC and LHC 
energies the formation of Quark Gluon Plasma in the initial stages may
indeed turn out to be a realistic scenario.


\newpage
\tableofcontents
\newpage
\section*{List of Symbols}

In the following we list some of the symbols which have been used.
\vskip 0.2 in
\begin{tabular}{ll}
$A_{\mn}$ & transverse projection tensor\\
$B_{\mn}$ & longitudinal projection tensor\\
${\bar D}_{\mn}^{0}$ & free vacuum propagator for spin 1 particles\\
${\bf D_{\mn}^0}$ & matrix propagator for spin 1 particles at finite temperature\\
$D_{\mn}^{ij}$ & $ij$th element of ${\bf D_{\mn}^0}$\\ 
$D_{\mn}^{R}$ & retarded propagator for spin 1 particles\\
${\cal F}$ & forward scattering amplitude\\
${F_{\mn}}$ & field tensor for electromagnetic field\\
${F^{\mn}_{l,r}}$ & non-abelian field tensor for left 
(right) handed field $V_{l,r}$\\
$f_{BE}$ & Bose distribution\\
$f_{FD}$ & Fermi distribution\\
$g_{\mn}$ & the metric tensor (1,-1,-1,-1)\\
$G^0$ & time-ordered free thermal propagator for nucleons\\
$G^H_F$ & Hartree nucleon propagator; vacuum part\\ 
$G^H_D$ & Hartree nucleon propagator; medium part\\ 
$H_{\mn}$ & photon tensor\\
$J_\mu^h(J_\nu^l)$ & hadronic (leptonic) electromagnetic current\\
$L_{\mn}$ & leptonic tensor\\
${\cal M}$ & invariant amplitude\\
${M_B}$ & Borel mass\\
${M_N}$ & nucleon mass\\
${m_V}$ & mass of vector meson $V$\\
$M$ & the invariant mass\\
$P_{\mn}^{R}$ & retarded improper self energy for spin 1 particles\\
$V$ & vector mesons\\
${\cal V}$ & three volume\\
$W_{\mn}^{R}$ & retarded electromagnetic current correlation function\\
$Z(\beta)$ & partition function at temperature $T=1/\beta$\\
$\alpha$ & electromagnetic coupling constant, $e=\sqrt{4\pi\alpha}$\\
$\alpha_s$ & strong coupling constant, $g_s=\sqrt{4\pi\alpha_s}$\\
$\epsilon^{\alpha\beta\mn}$ & totally antisymmetric tensor,
with $\epsilon^{0123}=1$\\
$\epsilon$ & thermodynamic energy density\\
$\Gamma_V$ & width of the vector meson $V$\\
$\Omega$ & four-volume\\
$\Pi_{\mn}^{R}$ & retarded proper self energy for spin 1 particles\\
$\rho_{\mn}$ & spectral function for spin 1 particles\\
$\varrho_{\mn}$ & non-abelian field tensor for the $\rho$ meson\\
\end{tabular}
\vskip .2in
The subscripts $L$ and $T$ will be used to denote the longitudinal and transverse
components respectively of a 2-ranked tensor in a medium.
\newpage
\addtolength{\baselineskip}{0.5\baselineskip}
\section{Introduction}
The QCD renormalization group calculation 
predicts that strongly interacting systems at very high temperature
and/or density are composed of weakly interacting quarks and gluons
~\cite{collins,km,evs} due to 
asymptotic freedom and the Debye screening of colour charge. 
On the other hand at low temperature and
density the quarks and gluons are confined within the hadrons.
Therefore, a phase transition is  expected to take place
at an intermediate value of temperature and/or density~\cite{hsatz,ldm}.
This transition is actually observed in lattice QCD
numerical simulations \cite{ukawa} at high temperature. 
A system of thermalized
strongly interacting matter where the properties of the system
are governed by the quark and gluon degrees of freedom is
called Quark Gluon Plasma (QGP).
One expects that ultra-relativistic heavy ion 
collisions (URHIC) at CERN/SPS, BNL/RHIC and CERN/LHC
might create conditions conducive 
for the formation and study 
of QGP~\cite{mueller,hwa,wong} (for recent development
see Ref.~\cite{qm99}). 

Various model calculations have been
performed to look for observable signatures of this state of matter.
However, among various signatures of QGP, photons and
dileptons are known to be advantageous as these signals 
probe the entire
volume  of  the  plasma,  with little interaction and thus, are
better markers of the space-time history of the evolving fireball.
This  is  primarily  so  because electromagnetic interaction is strong
enough to lead to detectable signal, yet it is weak enough to allow the
produced particles (real photons and dileptons) to  escape the system
without further interaction, carrying information 
of the fundamental constituents and their momentum distribution 
in the thermal bath.
The real and virtual photon (dilepton) emission rate from QGP is 
determined by the fundamental theory of strong interaction, 
 QCD. The dominant processes for the photon production from
 QGP are the annihilation ($q\bar{q}\rightarrow g\gamma$)
and Compton processes ($q(\bar{q})\rightarrow q(\bar{q})\gamma$).
The emission rate resulting from these reactions have been evaluated
~\cite{kapusta,baier,aurenche} in the framework of Hard Thermal
Loops (HTL) resummation in QCD~\cite{braaten,frenkel}. 

The disadvantage with photons is the substantial background from 
various processes (thermal and non-thermal)~\cite{janepr,cassing,pvr,qm93}. 
Among these, 
the contribution from hard QCD processes is well understood in the 
framework of perturbative QCD and the yield from hadronic decays 
e.g. $\pi^0/\eta\,\ra\,\gamma\,\gamma$ can be estimated by invariant mass 
analysis. On the other hand, photons from the thermalized 
hadronic gas pose a more
difficult task to disentangle. 
Therefore, to detect the electromagnetic signals from QGP 
it is very important to estimate
photons (both real and virtual) from hot and dense hadronic 
gas along with the possible modification
of the hadronic properties below the critical temperature of the 
phase transition.
However, the progress in our understanding of hot and 
dense hadronic matter has been retarded due to the  
 nonperturbative QCD dynamics
 in the low energy regime. Because of this severe 
constraint considerable amount of work has been done on model 
building (see e.g. Refs.~\cite{meissner},~\cite{nowak},
~\cite{brhcnp},~\cite{book}) 
in order to study the low energy hadronic states. 
 Nevertheless, in URHIC,  hadronic matter is expected to be formed
after a phase transition from QGP.  Even if such a
phase transition does not occur, realization of hadronic matter at
high temperature ($\sim$ 150 -- 200 MeV) and/or baryon density (a few
times normal nuclear matter density) is inevitable. As a result
the study of hadronic interactions at high temperature and density
assumes great significance. 
 Also, there are several other aspects
where medium effects may play an important role. For example spontaneously
broken chiral symmetry of the normal hadronic world is expected to be restored 
at high temperature and density
and this will be 
reflected in the thermal modification of the hadronic spectral 
function~\cite{pisarski1,pisarski2,hk1,plb185,hatsuda1}.
These modifications can be studied by analyzing photon, dilepton as well
as hadronic spectra. 

Various investigations have addressed the issue of temperature and density
dependence of hadronic spectra  within different models over the
past several years.  In particular, in-medium  QCD sum rules 
are useful  to make constraints on the hadronic spectral functions 
at finite temperature and density \cite{thatsuda}.
Brown and Rho (BR)~\cite{rho} argued that requiring chiral symmetry 
(in particular the QCD
trace  anomaly)  yields  an  approximate  scaling relation between
various effective hadronic masses,
$m_N^\ast/m_N\sim m_{\sigma}^\ast/m_{\sigma}\sim
m_{\rho,\omega}^\ast/m_{\rho,\omega}
\sim f_\pi^\ast/f_\pi$,
which  implies,  that  all
hadronic masses (except pseudoscalar mesons) decrease  with temperature. 
In the Walecka model approach~\cite{vol16,chin},
the vector meson mass gets shifted due to the 
 decrease of the nucleon mass which appears through thermal loops in the
 vector meson self energy \cite{sxk}.
 The reduction in $\rho$ meson mass has also been observed
in the  gauged linear sigma model~\cite{rdp} at low temperature, however, 
near the chiral transition point it shows an upward trend.
The nonlinear sigma model~\cite{abhijit,cs}, claimed to be the closest
low energy description of QCD shows the opposite
trend, {\it {i.e.}} $m_\rho^\ast$ increasing with temperature. A 
similar qualitative behaviour of the $\rho$ mass has been observed in the  
hidden local symmetry approach~\cite{harada}. 
The relation between the self energy and the forward
scattering amplitude has also been utilized to study the 
change of hadronic properties in the medium~\cite{st},
where the effects of non-zero temperature is rather small~\cite{eik}.
Thus,  there  exists  a lot of controversy in the literature about
the  finite  temperature   properties of hadrons. 
In view of this in the present
article we shall consider various scenarios for the shift in the hadronic 
spectral function at finite temperature and evaluate its effects on the 
experimentally measurable quantities (as these issues
should be settled by experiment), the photon and dilepton
spectra originating from a thermalized system formed after URHIC.

At this point we would like to clarify 
the scope of the present work. Our aim is 
to contrast the photon (both real and virtual) emission rate 
from the following two nuclear collision scenarios:
\begin{center}
{\bf (i) A\,+A\,$\ra$QGP$\ra$Mixed Phase$\ra$Hadronic Phase\\
or\\
(ii) A\,+\,A\,$\ra$Hadronic Phase},
\end{center}
by taking into account the finite temperature effects on the hadronic 
masses and decay widths, calculated within the ambit of various  
models  for hadronic interactions. 
The properties of matter formed after the nuclear collisions
at SPS would most likely be determined by soft interactions, 
where the application of perturbative QCD is not comfortably
justified and it is rather difficult to make any definite
conclusion about the formation of QGP here (for a recent
review see ~\cite{gyulassy}).
Therefore, at SPS we consider both the scenarios (i) and (ii).
However, at RHIC the centre of mass energy of the collision 
increases by an order of magnitude where in addition to the soft
processes the semi-hard processes contribute significantly
to the initial energy density. We have considered scenario (i)
exclusively for RHIC energies. This is because, as shown later,
the initial temperature estimated is too high for
the hadronic degrees of freedom to survive.
At even higher energies at LHC, the predictions
of perturbative QCD become more reliable and  the theoretical
estimate for the initial energy density is rather high~\cite{eskola}
where the formation of QGP is beyond any reasonable doubt!

The models considered to evaluate 
the medium effects are the Walecka model, the gauged linear and
non-linear sigma models, the hidden local symmetry 
approach, BR scaling and Nambu
scaling scenarios ~\cite{brpr}.
The QCD sum rules have been used to constrain
the hadronic spectral function at non-zero temperature.
Other models {\it e.g}, those 
proposed by Rapp {\it et al}~\cite{rcw} and by Klingl
{\it et al}~\cite{klingl}, where the effects of non-zero
baryon density (baryonic chemical potential) is dominant over 
finite temperature effects are not discussed in the present article. 
This is because
the aim of the present work is to study the hot, baryon free (central
rapidity) region of the ultra-relativistic heavy ion collisions.
As will be shown in the next section, both the photon and
dilepton emission rates are proportional to the retarded  
electromagnetic current correlation function. These correlators or the 
spectral functions can be constructed in vacuum from 
the experimental data obtained in $e^+e^-\rightarrow
hadrons$ (or from hadronic decays of 
the $\tau$ lepton) for various isovector and isoscalar
channels~\cite{shuryak,shifman}. The in-medium 
spectral function of the vector meson is obtained,
within the ambit of various models,
by modifying the pole and the continuum structure, 
resulting from its interaction with the constituents
of the thermal bath.

We organize the paper as follows. In section 2 we review the formalism
of the emission of real and virtual photons from a thermalized system of
strongly interacting particles. In section 3 we introduce
the HTL resummation technique and discuss the specific
reactions which are used to calculate the photon and dilepton
spectra. Section 4 is devoted to the discussion
of the properties of hadrons at finite temperature. The gauged linear 
sigma model, the gauged non-linear sigma model and the hidden local symmetry 
approach have been described in section 5.
The QCD sum rule approach has been discussed in section 6.
We discuss space time evolution dynamics in section 7. 
In section 8 we present the results of our calculations and
finally in section 9, we give a summary and outlook.

%%%%%%%%%%%%%%%%%% formalism.tex%%%%%%%%%%%%%%%%%%%%%%%%%%%%%%
\setcounter{equation}{0}
\def\theequation{2.\arabic{equation}}

\section{Electromagnetic Probes - Formulation}

The importance of the electromagnetic probes for the study of thermodynamic
state  of the evolving matter was first proposed by Feinberg in 
 1976~\cite{feinberg}. While for most purposes one can 
calculate  the  emission rates in a classical framework, Feinberg
showed  that  the  emission  rates  can   be   related   to   the
electromagnetic   current-current  correlation  function  in  a
thermalized  system  in  a  quantum  picture   and,   more
importantly,  in  a nonperturbative manner. 
Generally the production of a particle which interacts weakly 
with the constituents of the thermal bath (the constituents 
may interact strongly among themselves)
can always be expressed
in terms of the discontinuities or imaginary parts of the self
energies of that particle~\cite{ruuskanen,bellac}. 
In this section, therefore, we look at the
connection between the  electromagnetic  emission  rates  (real
and virtual photons)
and  the photon spectral function ( which is 
connected with the discontinuities in the interacting propagators) 
in a thermal system~\cite{weldon90}, 
which in turn is connected to the hadronic electromagnetic current-current
correlation function~\cite{mclerran} through Maxwell equations. 
It will be shown that 
the photon emission rate can be obtained from the dilepton
emission rate by appropriate modifications. 

We begin our discussion with the dilepton production rate.
Following Weldon~\cite{weldon90} 
let us define $A^\mu$ as the exact Heisenberg photon field which 
is the source of the leptonic current $J^l_\mu$. 
To lowest order in the electromagnetic coupling, the scattering matrix
element, $S_{HI}$,  for the transition  
$\mid I\rangle\,\rightarrow\,\mid H;l^+l^-\rangle$ is given by
\be
S_{HI}=e\,\int\,\langle H;l^+l^-\mid J^l_\mu(x)\,A^\mu(x) \mid I\rangle\,
d^4x\,e^{iq\cdot x},
\label{shi1}
\ee
where $\mid I\rangle$ is the initial state corresponding to the two
incoming nuclei, $\mid H;l^+l^-\rangle$ is the final state 
which corresponds to a lepton pair plus anything, 
the parameter $e$ is the
renormalized charge and $q=(q_0,\vec{q})$ is the 
four momentum of the lepton pair.
Since we assume that the lepton pair does not interact
with the emitting system, the matrix element can be factorized 
in the following way
\begin{equation}
\langle H;l^+l^-\mid J^l_\mu(x)\,A^\mu(x) \mid I\rangle=
\langle H\mid A^\mu(x)\mid I\rangle\langle\,l^+l^-\mid J^l_\mu(x)\mid 0\rangle.
\end{equation}
where $\mid 0\rangle$ is the vacuum state.
Putting the explicit form of the current $J^l_\mu$ in terms of the
Dirac spinors ($\bar{u}(p_1)$ and $v(p_2)$) we obtain
\begin{equation}
S_{HI}=e\,\frac{\bar{u}(p_1)\gamma_\mu\,v(p_2)}{{\cal V}\sqrt{2E_12E_2}}\int
d^4x\,e^{iq\cdot x}\langle H\mid\,A^\mu(x)\mid I\rangle.
\end{equation}
where $\gamma_\mu$ denote the Dirac matrices, $E_i=\sqrt{p_i^2+m^2}$, 
with $i=1\,\&\,2$ are the energy of the leptons 
and ${\cal V}$ is the volume of the system.

The dilepton multiplicity $N$ from a thermal system is obtained 
by summing over the final states and averaging over the initial 
states with a weight factor $Z(\beta)^{-1}\,e^{-\beta\,E_I}$; 
\begin{equation}
N=\frac{1}{Z(\beta)}\sum_I\,\sum_H\,\mid S_{HI}\,\mid^2\,e^{-\beta\,E_I}
\frac{{\cal V}\,d^3p_1}{(2\pi)^3}\,\frac{{\cal V}\,d^3p_2}{(2\pi)^3},
\end{equation}
where $E_I$ is the total energy in the initial state, $Z(\beta)$ is
the partition function and $\beta=1/T$ is the inverse temperature. 
After some algebra $N$ can be written in a compact form as follows:
\begin{equation}
N=e^2\,L^{\mu\nu}\,H_{\mu\nu} \frac{d^3p_1}{(2\pi)^3E_1}
\,\frac{d^3p_2}{(2\pi)^3E_2},
\end{equation}
where $L_{\mu\nu}$ is the leptonic tensor defined by
\begin{eqnarray}
L^{\mu\nu}&\equiv&\frac{1}{4}\,
\sum_{spins}\,\bar{u}(p_1)\gamma^\mu\,v(p_2)\bar{v}(p_2)
\gamma^\nu\,u(p_1)\nonumber\\
&=&\,p_1^\mu\,p_2^\nu+p_2^\mu\,p_1^\nu-\frac{q^2}{2}\,g^{\mn},
\end{eqnarray}
and $H_{\mu\nu}$ is the photon tensor
\be
H_{\mn}\equiv\frac{1}{Z(\beta)}\,e^{-\beta\,q_0}
\int d^4x\,d^4y\,e^{iq\cdot (x-y)}\sum_H
\,\langle H\mid A_\mu(x)\,A_\nu(y)\mid H\rangle\,e^{-\beta\,E_H}.
\ee
To obtain the above equation we have used 
the resolution of identity $1= \sum_I \mid I\rangle\langle I\mid$ and
the energy conservation equation $E_I=E_H+q_0$, where
$q_0$ is the energy of the lepton pair and $E_H$ is the energy
of the rest of the system.
Using translational invariance of the matrix element we can write
\begin{equation}
H_{\mu\nu}=\Omega\,e^{-\beta\,q_0}D^{>}_{\mu\nu}(q),
\end{equation}
where $\Omega$ is the four volume of the system. $D^>_{\mu\nu}$
is the component $iD^{21}_{\mn}$ of the exact
photon propagator $\bf D_{\mn}$ in the real time formalism 
of thermal field theory which has a $(2\times2)$ 
matrix structure (see appendix).
The time ordered propagator is the $(1,1)$ component of $\bf D_{\mn}$. In 
coordinate space it is defined as
\bea
iD^{11}_{\mn}(x)&\equiv&\frac{1}{Z(\beta)}\sum_H
\,\langle H\mid T\{A_\mu(x)\,A_\nu(0)\}\mid H\rangle\,e^{-\beta\,E_H}\nonumber\\
&\equiv&\theta(x_0)D_{\mn}^>(x)+\theta(-x_0)D_{\mn}^<(x).
\eea
where $D_{\mn}^<$ is the component $iD_{\mn}^{12}$ (see appendix), $x_0$
is the time component of the four vector, 
$x_\mu(=x_0,\vec{x})$ and $\theta(x_0)$ is the step function.
Using the integral representation of the $\theta$-functions in the 
above expression and taking the Fourier transform we get~\cite{dolan}
\be 
D^{11}_{\mn}(q_0,\vec q)=\int_{-\infty}^{\infty}\,\frac{d\omega}{2\pi}\,
\left[\frac{D^>_{\mn}(\omega,\vec q)}{q^0-\omega+i\epsilon}
-\frac{D^<_{\mn}(\omega,\vec q)}{q^0-\omega-i\epsilon}\right].
\label{top}
\ee
Using the Kubo Martin Schwinger (KMS) relation in momentum space,
\be
D^>_{\mn}(\omega,\vec q)=e^{\beta\,\omega}D^<_{\mn}(\omega,\vec q),
\ee
we have
\be
D^>_{\mn}(q_0,\vec q)=-\frac{2}{1+e^{-\beta q_0}}{\s{Im}}D^{11}_{\mn}(q_0,\vec q).
\ee
The rate of dilepton production per unit volume ($N/\Omega$) is then obtained as 
\be
\frac{dN}{d^4x}=-\frac{2e^2}{e^{\beta q_0}+1}\,
L^{\mn}{\s Im}D^{11}_{\mn}(q_0,\vec q)
\frac{d^3p_1}{(2\pi)^3E_1}\,\frac{d^3p_2}{(2\pi)^3E_2}.
\label{dnd4x1}
\ee

Now, the spectral function of the (virtual) photon in the thermal bath is 
conventionally defined as
\be
\rho_{\mn}(q_0,\vec q)\equiv\frac{1}{2\pi Z(\beta)}\int d^4x\,e^{iq\cdot x}\sum_H
\,\langle H\mid[A_\mu(x),A_\nu(0)]\mid H\rangle e^{-\beta\,E_H},
\label{spectral1}
\ee
so that, we have~\cite{bellac,dolan}
\be
D^{11}_{\mn}(q_0,\vec q)=\int_{-\infty}^\infty 
d\omega\frac{\rho_{\mn}(\omega,\vec q)}
{q_0-\omega+i\eps}-2i\pi f_{BE}(q_0)\rho_{\mn}(q_0,q),
\ee
where $f_{BE}(q_0)=[e^{\beta q_0}-1]^{-1}$. This leads to
\be
{\s Im}D^{11}_{\mn}(q_0,\vec q)=-\pi[1+2f_{BE}(q_0)]\rho_{\mn}(q_0,\vec q).
\ee
In terms of the photon spectral function the dilepton emission rate
is obtained as
\be
\frac{dN}{d^4x}=2\pi e^2 L^{\mn}\rho_{\mn}(q_0,\vec q)
\frac{d^3p_1}{(2\pi)^3E_1}\,\frac{d^3p_2}{(2\pi)^3E_2}f_{BE}(q_0).
\label{dnd4x2}
\ee
This relation which expresses the dilepton emission rate 
in terms of the spectral function of the photon in the medium 
is an important result.

As is well known, it is not the time-ordered propagator that has the
required analytic properties in a heat bath, but rather the retarded one.
We thus introduce the retarded propagator which will enable us to express
the dilepton rate in terms of the retarded photon self energy.
The retarded photon propagator is defined as
\be
iD^R_{\mn}(q_0,\vec q)\equiv\frac{1}{Z(\beta)}\int d^4x\,e^{iq\cdot x}\theta(x_0)\sum_H
\,\langle H\mid[A_\mu(x),A_\nu(0)]\mid H\rangle e^{-\beta\,E_H},
\label{retp}
\ee
which leads to the relation
\be
\rho_{\mn}=-\frac{1}{\pi}{\s Im}D^R_{\mn}.
\label{specdr}
\ee
The above equation implies that in order to evaluate the spectral function
at $T\neq 0$ we need to know the imaginary part of the retarded propagator.
It is interesting to note that the above expression for spectral function
reduces to its vacuum value as $\beta\ra\,\infty$
since the only state which
enters in the spectral function is the vacuum~\cite{lsbrown}.

Consequently we have
\be
\frac{dN}{d^4x}=-2e^2L^{\mn}{\s Im}D^R_{\mn}(q_0,\vec q)
\frac{d^3p_1}{(2\pi)^3E_1}\,\frac{d^3p_2}{(2\pi)^3E_2}f_{BE}(q_0).
\label{dnd4x3}
\ee
This result can also be derived directly from Eq.~(\ref{dnd4x1}) 
using the relation~\cite{adas},
\be
{\s Im}\,D^{11}_{\mn}=(1+2f_{BE}){\s Im}\,D^R_{\mn}.
\label{topret}
\ee
Now the exact retarded photon propagator can be expressed in terms of the proper self 
energy through the Dyson-Schwinger equation:
\be
D^R_{\mn}=-\frac{A_{\mn}}{q^2+\Pi^R_T}-
\frac{B_{\mn}}{q^2+\Pi^R_L}+ (\zeta-1)\frac{q_\mu\,q_\nu}{q^4},
\label{drself}
\ee
where, $-i\Pi_{\mn}^R$ is the sum of all 1PI (one particle irreducible)
retarded photon self energy insertions which can be decomposed as
\be
\Pi^R_{\mn}=A_{\mn}\Pi_T^R+B_{\mn}\Pi_L^R.
\label{trlong}
\ee
Here $A^{\mn}$ and $B^{\mn}$ are transverse and longitudinal 
projection tensors respectively (see appendix) and $\Pi_T^R$ and
$\Pi_L^R$ are the transverse and longitudinal components of the retarded
photon self energy.
The presence of the parameter $\zeta$ indicates the gauge 
dependence of the propagator. Although the gauge dependence cancels out
in the calculation of physical quantities,
one should, however, be careful when extracting
physical quantities from the propagator directly, especially in the
non-abelian gauge theory.

Inserting the imaginary part of the retarded photon propagator from 
Eq.~(\ref{drself}) in Eq.~(\ref{dnd4x3}) we get
\be
\frac{dN}{d^4x}=2\pi\,e^2\,L_{\mn}(A^{\mn}\rho_T\,+\,B^{\mn}\,\rho_L)
\frac{d^3p_1}{(2\pi)^3E_1}\,\frac{d^3p_2}{(2\pi)^3E_2}\,
f_{BE}(q_0),
\label{dnd4x4}
\ee
with
\be
\rho_{T,L}\equiv-\frac{1}{\pi}\frac{{\s Im}\,\Pi^R_{T,L}}
{(q^2+{\s Re}\,\Pi^R_{T,L})^2+({\s Im}\,\Pi^R_{T,L})^2}. 
\ee 
Comparing with Eq.~(\ref{dnd4x2}) we have
\be
\rho^{\mn}=A^{\mn}\rho_T+B^{\mn}\rho_L.
\ee
It has been argued by Weldon~\cite{weldonprl} that the electromagnetic
plasma resonance occurring through the spectral function derived
above could be a signal of the deconfinement phase transition provided
the plasma life time is long enough for the establishment of the
resonance.
Using the relation
\bea
\int\prod_{i=1,2}\,\frac{d^3p_i}{(2\pi)^3E_i}
\delta^4(p_1+p_2-q)\,L^{\mn}(p_1,p_2)&=&\frac{1}{(2\pi)^6}\frac{2\pi}{3}
(q^\mu\,q^\nu-q^2\,g^{\mn})\nonumber\\
&\times&(1+\frac{2m^2}{q^2})\sqrt{1-\frac{4m^2}{q^2}},
\label{red1}
\eea
and
\bea
q^\mu A_{\mn}&=&0\nonumber\\
q^\mu B_{\mn}&=&0\nonumber\\
g^{\mn}A_{\mn}&=&2\nonumber\\
g^{\mn}B_{\mn}&=&1,
\label{red2}
\eea
the dilepton rate is finally obtained as
\be
\frac{dR}{d^4q}=-\frac{\alpha}{12\pi^3}q^2(1+\frac{2m^2}{q^2})
\sqrt{1-\frac{4m^2}{q^2}}(2\rho_T+\rho_L)f_{BE}(q_0),
\label{drd4q1}
\ee
where $m$ is the lepton mass and $\alpha$ denotes the fine structure constant. 
This is the {\it exact} expression for the dilepton emission rate 
from a thermal medium of 
interacting particles.
In most of the cases the dilepton production rate from a thermal system
is calculated with the approximation
$\Pi^R_T=\Pi^R_L \equiv \Pi^R$. 
Since $\Pi^R_{L,T}$ and $\Pi^R_{L,T}$ are both proportional
to $\alpha$, they are small
for all practical purposes (this corresponds to the free 
propagation of the virtual photon in the thermal bath). 
Therefore $\rho_{L,T}(\equiv\rho)$ is given by
\be
\rho=-\frac{1}{\pi}\frac{{\s Im}\,\Pi^R}{q^4}
=-\frac{1}{\pi}\frac{{\s Im}\Pi^{R\mu}_\mu}{3q^4}.
\label{approx}
\ee
Using Eqs.~(\ref{drd4q1}) and ~(\ref{approx})
we get
\be
\frac{dR}{d^4q}=\frac{\alpha}{12\pi^4\,q^2}(1+\frac{2m^2}{q^2})
\sqrt{1-\frac{4m^2}{q^2}}{\s Im}\Pi^{R\mu}_\mu\,f_{BE}(q_0).
\label{drd4q2}
\ee
This is the familiar result most widely used for the dilepton emission
rate~\cite{bellac}. 
It must be emphasized that this relation is valid only
to $O(e^2)$ since 
it does not account for the possible reinteractions of the virtual
photon on its way out of the bath and multiple emission of photon is ignored here.
The possibility of emission of more than one
photon has also been neglected here. However, the expression is true to all 
orders in strong interaction.

The emission rate of dilepton can also be obtained 
in terms of the electromagnetic
current-current correlation function~\cite{mclerran}.
Denoting the hadronic part of the electromagnetic  current operator
by  $J^h_\mu$,  the leptonic part by $J^l_\nu$ and the free photon
propagator by ${\bar D}_0^{\mn}$, we have the matrix element  for  this
transition :
\be
S_{HI}=e\langle\,H;l^+\,l^-\,\mid\int d^4x d^4y
J^l_{\mu}(x){\bar D}_0^{\mn}(x-y)\,J^h_{\nu}(y)\mid I\rangle.
\ee
This obviously follows from Eq.~(\ref{shi1}) by realizing that the solution
of the equation of motion of the interacting photon field is
\be
A^\mu(x)=\int d^4y {\bar D}_0^{\mn}(x-y)J^h_\nu(y).
\ee
As in the earlier case the leptonic part of the current 
can be easily factored out. 
Writing  the  Fourier transform of photon propagator and squaring
the matrix elements,
one obtains the rate of dilepton
production
\be
dR=e^2\,L^{\mn}\,W^>_{\mn}\,\frac{e^{-\beta q_0}}{q^4}\,
\frac{d^3\,p_1}{(2\pi)^3\,E_1}
\frac{d^3\,p_2}{(2\pi)^3\,E_2},
\label{drcor}
\ee
where $W^>_{\mu\nu}(q)$ is the Fourier transform of the thermal 
expectation value
of the electromagnetic current-current correlation function: 
\be
W^>_{\mu\nu}(q)\equiv\int d^4x e^{iq\cdot x}\sum_H\,
\langle H\,\mid J^h_{\mu}(x)J^h_{\nu}(0)\mid\,H\,\rangle\,
\frac{e^{-\beta\,E_H}}{Z(\beta)}.
\label{correl}
\ee
The subtleties of the thermal averaging have been elucidated earlier.
It is thus readily seen (Eq.~(\ref{drcor})) that the dilepton 
(and photon) data
yield considerable information about the thermal state 
of the hadronic system; at least the full tensor structure
of $W^{\mu\nu}$ can in principle be determined.
Now, the current-current correlation function $W^>_{\mn}$ is related to the 
retarded correlator by
\be
W^>_{\mn}=2e^{\beta q_0}f_{BE}(q_0){\s Im}W^R_{\mn}
\label{wmng}
\ee
where
\be
W^R_{\mu\nu}(q)\equiv i\int d^4x e^{iq\cdot x}\theta(x_0)\sum_H\,
\langle H\,\mid [J^h_{\mu}(x),J^h_{\nu}(0)]\mid\,H\,\rangle\,
\frac{e^{-\beta\,E_H}}{Z(\beta)},
\label{retcor}
\ee
We then get
\be
dR=2e^2\,L^{\mn}\,{\s Im}W^R_{\mn}\,\frac{1}{q^4}\,
\frac{d^3\,p_1}{(2\pi)^3\,E_1}
\frac{d^3\,p_2}{(2\pi)^3\,E_2}f_{BE}(q_0),
\label{drretcor}
\ee
We now define the (improper) photon self energy through the relation
\be
D^{R,\,\alpha\beta}=
D_0^{R,\,\alpha\beta}+D_0^{R,\alpha\mu}\,P_{\mn}^R\,D_0^{R,\nu\beta}
\ee
where $-iP^R_{\mn}$ is the sum of {\it all} self energy diagrams. 
The advantage is that $P^R_{\mn}$ can be defined in coordinate space 
as~\cite{beres}
\be
iP^R_{\mu\nu}(x)\equiv\theta(x_0)\sum_H\,
\langle H\,\mid [J^h_{\mu}(x),J^h_{\nu}(0)]\mid\,H\,\rangle\,
\frac{e^{-\beta\,E_H}}{Z(\beta)}.
\label{impself}
\ee
Taking the Fourier transform and 
comparing with Eq.~(\ref{retcor}) we see that  $P^R_{\mn}(q)=-W^R_{\mn}(q)$. Using the relations \ref{trlong}, \ref{red1} and
\ref{red2} we end up with~\cite{gale}
\be
\frac{dR}{d^4q}=\frac{\alpha}{12\pi^4\,q^2}(1+\frac{2m^2}{q^2})
\sqrt{1-\frac{4m^2}{q^2}}{\s Im}P^{R\mu}_\mu\,f_{BE}(q_0),
\label{drd4qimp}
\ee
where
\be
P^{R\mu}_\mu=g^{\mn}P^R_{\mn}=2P^R_T+P^R_L
\ee
It is important to realize 
that the analysis is essentially nonperturbative
up to this point.
To $O(e^2)$ we note that $P$ reduces to the proper self energy $\Pi$
($=P\,D_0\,D^{-1}$)
and consequently Eq.~(\ref{drd4qimp}) reduces to Eq.~(\ref{drd4q2}).
This approximation is the same as implied in Eq.~(\ref{approx}).

The connection between 
the electromagnetic current correlation function and the spectral function 
can be expressed in a straight forward way
by substituting $J^h_\mu$ and $J^h_\nu$ 
using Maxwell equation ($\partial_\alpha\partial^\alpha\,A_\mu
-\zeta^{-1}(\zeta-1)\partial_\mu\,(\partial_\alpha\,A^\alpha)=J^h_\mu$) in Eq.~(\ref{correl})
to obtain
\bea
W^>_{\mn}&=&\left(q^2g_{\mu\alpha}-\frac{\zeta-1}{\zeta}q_\mu q_\alpha\right)
D_>^{\alpha\beta}
\left( q^2g_{\beta\nu}-\frac{\zeta-1}{\zeta}q_\beta q_\nu\right)\nonumber\\
&=&2\pi\left(q^2g_{\mu\alpha}-\frac{\zeta-1}{\zeta}q_\mu q_\alpha\right)
\rho^{\alpha\beta}\left(q^2g_{\beta\nu}-\frac{\zeta-1}{\zeta}q_\beta q_\nu\right)
(1+f_{BE}).
\eea
Substituting the above equation in 
Eq.~(\ref{drcor}) 
we can recover Eq.~(\ref{dnd4x2}).
This establishes the connection between the two approaches of 
Refs.\cite{weldon90,mclerran}.

The electromagnetic decay of unstable particles ({\it e.g.} $\rho$,
$\omega$ and $\phi$) within a thermal system could provide valuable
information about the nature of the system. In a thermal medium the production
of an off-shell vector meson ($V$)  of four momentum 
$q$ (where $q^2=M^2$) and its subsequent
decay into a lepton pair leads to the dilepton emission rate as~\cite{weldon93}
\be
dR=\frac{2M}{(2\pi)^3}\rho^V_{\mn}\,P^{\mn}\,f_{BE}(q_0)
\Gamma_{V\,\ra\,l^+l^-}\,d^4q,
\ee
where $\rho^V_{\mn}$ is the spectral function of the off-shell vector 
meson given by Eq.~(\ref{spectral1}) with the photon field replaced 
by the interpolating fields for vector mesons, the exact form of which 
is not important in the present case,
$P_{\mn}(=\sum\,\epsilon_\mu\,\epsilon_\nu^\ast)=-g_{\mn}+q_\mu q_\nu/q^2$ 
is the projection operator for the vector meson $V$ and 
$\Gamma_{V\,\ra\,l^+l^-}$ is the partial decay width for the process
$V\,\ra\,l^+l^-$ in vacuum.
The spectral function is expressed in terms of the retarded vector meson
propagator (as has been done before in Eq.~(\ref{specdr}) in case of photon).
In the limit $\Pi_T^R=\Pi_L^R=\Pi^R$, the spectral function is given by
\be
\rho^V_{\mn}=\frac{1}{\pi}\frac{{\s Im}\Pi^R}{(q^2-m_V^2+{\s Re}\Pi^R)^2+
({\s Im}\Pi^R)^2}\,P_{\mn}.
\ee
Using the relation 
$P^{\mn}\,P_{\mn}=(2J+1)$, we get the dilepton emission rate due to the
decay of an unstable vector meson of spin $J$ as 
\be
\frac{dR}{d^4q}=2\frac{(2J+1)}{(2\pi)^3}\,f_{BE}\,M\Gamma_{V\,\ra\,l^+l^-}\,
\left[\frac{1}{\pi}\frac{{\s Im}\Pi^R}{(q^2-m_V^2+{\s Re}\Pi^R)^2+({\s Im}\Pi^R)^2}
\right],
\label{wel1}
\ee
where ${\rm Im}\,\Pi^R$ is the imaginary part of the self energy of 
the particle $V$ which
should be calculated within the framework of thermal field theory~\cite{weldon83}.
For a particle which does not decay in the collision volume 
(the total width $\Gamma_{\s tot}={\s Im}\Pi^R/M$ is small) the spectral
function in the above equation (term within the square bracket) becomes
$\delta(q^2-m_V^2)$, as it should be for a stable particle. 
In a  medium the width $\Gamma_{\s tot}$ 
for $V$ should be calculated with all the processes involving the creation
and annihilation of $V$, {\it i.e.} $\Gamma_{\s tot}=\Gamma_{V\ra\,{\rm all}}-
\Gamma_{{\rm all}\ra\,V}$~\cite{weldon93}.

\begin{figure}
\centerline{\psfig{figure=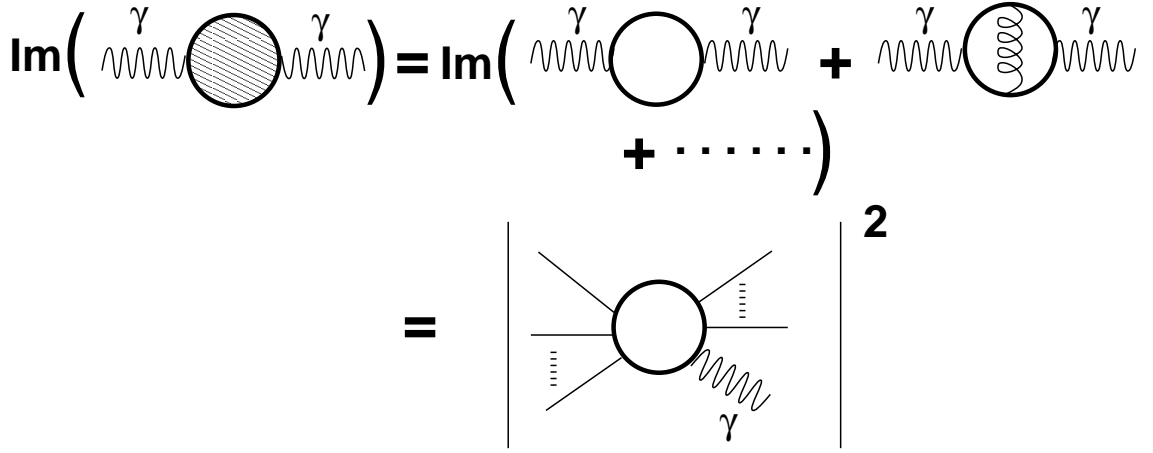,width=15cm,height=6cm,angle=-90}}
\caption{Optical Theorem in Quantum Field Theory}
\label{opt}
\end{figure}

To obtain the real photon emission rate per unit volume 
($dR$) from a system in thermal 
equilibrium we note that the dilepton emission rate differs from the
photon emission rate in the following way. The factor 
$e^2\,L_{\mn}/q^4$ which is the product of the electromagnetic vertex 
$\gamma^\ast\,\ra\,l^+\,l^-$, the leptonic current involving Dirac spinors
and the square of the photon propagator should be replaced by the factor 
$\sum\,\epsilon_\mu\,\epsilon_\nu^\ast (=-g_{\mn})$ for the real (on-shell)
photon. Finally the phase space factor 
$d^3p_1/[(2\pi)^3\,E_1]\,d^3p_2/[(2\pi)^3E_2]$
should be replaced by $d^3q/[(2\pi)^3q_0]$ to obtain
\be
dR=-\frac{e^{-\beta q_0}}{2(2\pi)^3}\,g^{\mn}\,W^>_{\mn}\,\frac{d^3q}{q_0}.
\ee
As in the case of dileptons this expression can be reduced to
\be
q_0\frac{dR}{d^3q}=\frac{g^{\mn}}{(2\pi)^3}\,{\s Im}\Pi^R_{\mn}f_{BE}(q_0). 
\ee
This result can also be obtained directly from Eq.~(\ref{drd4q2}).
The emission rate given above is correct up to order $e^2$ in 
electromagnetic interaction but exact, in principle, to all order
in strong interaction. However, for all practical purposes 
one is able to evaluate up to a finite order of loop expansion.
Now it is clear from the above results 
that to evaluate photon and dilepton emission rate from a 
thermal system we need to evaluate the 
imaginary part of the photon self energy.
The Cutkosky rules at finite temperature
or the thermal cutting rules~\cite{adas,kobes,gelis}
give a systematic procedure to calculate the imaginary part of a 
Feynman diagram. The Cutkosky rule expresses the 
imaginary part of the $n$-loop amplitude in terms of physical 
amplitude of lower order ($n-1$ loop or lower). This is shown schematically
in Fig.~(\ref{opt}).  When the imaginary part of the self energy is calculated
up to and including $L$ order loops where $L$ satisfies $x\,+\,y\,<\,L\,+\,1$,
then one obtains the photon emission rate for the reaction $x$ particles
$\ra$ $y$ particles $+\,\gamma$ and the above formalism becomes 
equivalent to the relativistic kinetic theory formalism~\cite{gk}. 

%%%%%%%%%%%%% Figure 1 %%%%%%%%%%%%%%%%%%%%%%%%%%%%%
%%%%%%%%%%%%% End of Fig.1 %%%%%%%%%%%%%%%%%%%%%%%%%%%%%
For a reaction $1\,+\,2\,\ra\,3\,+\gamma$ the photon (of energy $E$)
emission rate is given by~\cite{sourav}
\begin{eqnarray}
E\frac{dR}{d^3p}&=&\frac{{\cal{N}}}{16(2\pi)^7E}\,\int_{(m_1+m_2)^2}^{\infty}
\,ds\,\int_{t_{{\rm min}}}^{t_{{\rm max}}}\,dt\,|{\cal M}|^2\,
\int\,dE_1\nonumber\\
&&\times\int\,dE_2\frac{f(E_1)\,f(E_2)\left[1+f(E_3)\right]}{\sqrt{aE_2^2+
2bE_2+c}},
\label{rktp}
\end{eqnarray}
where
\begin{eqnarray}
a&=&-(s+t-m_2^2-m_3^2)^2\nonumber\\
b&=&E_1(s+t-m_2^2-m_3^2)(m_2^2-t)+E[(s+t-m_2^2-m_3^2)(s-m_1^2-m_2^2)\nonumber\\
&&-2m_1^2(m_2^2-t)]\nonumber\\
c&=&-E_1^2(m_2^2-t)^2-2E_1E[2m_2^2(s+t-m_2^2-m_3^2)-(m_2^2-t)(s-m_1^2-m_2^2)]
\nonumber\\
&&-E^2[(s-m_1^2-m_2^2)^2-4m_1^2m_2^2]-(s+t-m_2^2-m_3^2)(m_2^2-t)\nonumber\\
&&\times(s-m_1^2-m_2^2)
+m_2^2(s+t-m_2^2-m_3^2)^2+m_1^2(m_2^2-t)^2\nonumber\\
E_{1{{\rm min}}}&=&\frac{(s+t-m_2^2-m_3^2)}{4E}+
\frac{Em_1^2}{s+t-m_2^2-m_3^2}
\nonumber\\
E_{2{{\rm min}}}&=&\frac{Em_2^2}{m_2^2-t}+\frac{m_2^2-t}{4E}\nonumber\\
E_{2{{\rm max}}}&=&-\frac{b}{a}+\frac{\sqrt{b^2-ac}}{a}.\nonumber
\end{eqnarray}
${\cal N}$ is the overall degeneracy of the particles 1 and 2,
${\cal M}$ is the invariant amplitude of the reaction (summed over
final states and averaged over initial states), $f$ denotes
the thermal distribution functions and $s$, $t$, $u$ are the 
usual Mandelstam variables.

In a similar way the dilepton emission rate for a reaction
$a\,\bar a\,\ra\,l^+\,l^-$ can be obtained as
\bea
\frac{dR}{d^4q}&=&\int {d^3p_a\over 2E_a(2\pi)^3}f(p_a)
\int {d^3p_{\bar a}\over 2E_{\bar a}(2\pi)^3}f(p_{\bar a})
\int {d^3p_1\over 2E_1(2\pi)^3}
\int {d^3p_2\over 2E_2(2\pi)^3}\nonumber\\
&&\mid {\cal M}\mid_{a\bar a\rightarrow l^+l^-}^2
(2\pi)^4\delta^{(4)}(p_a+p_{\bar a}-p_1-p_2)
\delta^{(4)}(q-p_a-p_{\bar a}).
\eea
where  $f(p_a)$ is the appropriate 
occupation probability for bosons or fermions.
The Pauli blocking of the lepton pair in the final state has been
neglected in the above equation. 

%%%%%%%%%%%%%%%%%%%%%%%%%% htl.tex%%%%%%%%%%%%%%%%%%%%%%%%%%%%%
\setcounter{equation}{0}
\def\theequation{3.\arabic{equation}}

\section{Photon and Dilepton Emission Rate}
In this section we briefly discuss the 
HTL resummation technique and the specific reactions
considered in the present work to evaluate the electromagnetic 
probes from QGP as well as hadronic matter.

\subsection{Photon emission from QGP in the HTL approximation}
In the system formed after URHIC, numerous interactions take place among
the charged constituents, either quarks or hadrons, which lead
to the production of photons and dileptons. In this section we will
discuss the emission rate of hard ($E>T$) photons from QGP.

Naively, one expects that the properties of QGP at high temperature
($T>>T_c$) can be studied by applying perturbation theory due to the small
value of the strong coupling constant, $\alpha_s(T)$. However, QCD 
perturbation theory at high temperature is plagued by infra-red problems 
and gauge dependence of the physical quantities, e.g. the
gluon damping rate (see Refs.~\cite{thoma},\cite{rkobes} and~\cite{npa525}). 
The gauge dependence of the gluon damping rate was cured by
Braaten and Pisarski~\cite{braaten} by an effective expansion in terms of hard
thermal loops - i.e. including all the relevant loop effects in a given order 
of the coupling constant in a systematic way (for a beautiful review on gluon 
damping see Ref.~\cite{npa525}). 
But the problem of infra-red 
divergences in QCD is not solved completely by the HTL framework.
The quantities which are quadratically divergent in naive perturbation
theory such as the damping rate of fast moving fermions in QGP becomes 
logarithmically divergent in effective perturbation theory. On the 
other hand  quantities which are logarithmically divergent in the
naive perturbation theory turns out to be finite if one applies
HTL resummation method. The hard photon ($E>T$) emission rate which falls
in the second category, is the relevant quantity for the present discussions.

The thermal photon emission rate from QGP is governed 
by the following Lagrangian density:
\be
{\cal L}_{QGP}={\cal L}_{QCD}+{\cal L}_{\gamma q},
\ee
where
\bea
{\cal L}_{QCD}&=&-\frac{1}{4}\sum_{a=1}^{8}G_{\mn}^aG^{a\mn}+
\sum_{f=1}^{N_f}\bar \psi_f(i\partial\sls-
g_s\gamma^\mu G^a_\mu\frac{\lda^a}{2})\psi_f,\nonumber\\
{\cal L}_{\gamma q}&=& 
-\frac{1}{4}F_{\mn}F^{\mn}-
\sum_{f=1}^{N_f}e_f\bar \psi_f\gamma^\mu A_\mu\psi_f.
\eea
In the above, $G_{\mn}^a$ is the non-abelian field tensor for
the gluon field $G_{\mu}^a$ of color $a$, $\psi_f$ is the
Dirac field for the quark flavor $f$, $g_s$ is the color
charge, $e_f$ is the (fractional) electric charge of quark flavor $f$, 
$\lambda^a$'s are the Gell-Mann matrices, $F_{\mn}$ is the electromagnetic 
field tensor and $A^\mu$ is the photon field.
As mentioned in the introduction the dominant processes for 
photon production from QGP are the annihilation 
($q\bar{q}\rightarrow g\gamma$)
and the Compton processes ($q(\bar{q})\rightarrow q(\bar{q})\gamma$).
However, the production rate from these processes diverges due to
the exchange of massless particles. As mentioned earlier
this is a well-known problem in thermal perturbative expansion of non-abelian
gauge theory which suffers from infra-red divergences. 
One type of the divergences could be cured by taking into
account the `electric type' screening through 
the HTL approximation~\cite{braaten}. 
The non-abelian gauge theory also contains `magnetic type' divergences,
which can be eliminated if there is a screening
of the magnetic field~\cite{rmp,nair,blaizot}.  This is in sharp contrast to
Quantum Electrodynamics, which is free from screening of static magnetic
field. However, the study of magnetic screening is beyond 
the scope of HTL approximation 
as the transverse component of the gluon self energy  
vanishes in the static limit in this framework.  
Magnetic screening is relevant if any physical quantity is sensitive
to the scale $g_s^2T$, at which all the loop contributions are of the same order
\cite{asmilga} and hence the perturbation theory breaks down~\cite{alinde}.
The production of soft photons ($E \leq g_sT$) 
from QGP is non-perturbative because it 
is sensitive to the magnetic screening mass of the gluons~\cite{agz} and
consequently the soft photon emission rate is poorly known.
Therefore, in the present work we consider only the
production of hard photons ($E\ge T$) which is insensitive
to the scale $g_s^2T$. For such cases (hard photon emission) the 
infra-red divergences could be eliminated within the framework of HTL  
as discussed below.

The theory of  HTL begins with the observation that at non-zero
temperature there are two energy scales -  one associated with the
temperature $T$, referred to as the hard scale and the other connected 
with the fermionic mass $\sim g_sT$ ($g_s<<1$), induced by the temperature,
known as the soft scale.  A momentum $p^\mu$ appearing in the self 
energy diagram of
photon would be called soft (hard) if both the temporal and the 
spatial components are $\sim g_sT$ (any component is $\sim T$).
If any physical quantity is sensitive to the soft scale then
HTL resummation becomes essential, {\it i.e.} in such cases
the correlation function has to be expanded in terms of the
effective vertices and propagators, where the effective quantities
are the corresponding bare quantities plus the high temperature
limit of one loop corrections.

The notion of HTL can be clearly demonstrated 
in massless $\phi^4$ theory in the following way. Consider the Lagrangian density
\be
{\cal L}=\frac{1}{2}(\partial\phi)^2-g^2\phi^4.
\ee
The  thermal mass (self energy)
resulting from the one loop tadpole diagram in this model is
$m_{\s th}^2\sim g^2T^2$. At soft momentum scale ($p^\mu\sim gT$)
the inverse of the bare propagator goes as $\sim g^2T^2$..
Thus, the one loop (tadpole) correction is as large as the
tree amplitude. Therefore, this tadpole is a HTL by definition. 
Braaten and Pisarski \cite{braaten} have argued that these 
HTL contributions should be taken into account consistently  
by re-ordering the perturbation series in terms of effective 
vertices and propagators.
Therefore, according to their prescription we have
\be
{\cal L}=\frac{1}{2}(\partial\phi)^2-g^2\phi^4 - \frac{1}{2}m_{\s th}^2\phi^2
+ \frac{1}{2}m_{\s th}^2\phi^2={\cal L}_{\s eff} +{\cal L}_{\s ct},
\ee
where ${\cal L}_{\s ct}=m_{\s th}^2\phi^2/2$ is the counter term which should
be treated in the same footing as the $\phi^4$ term.
${\cal L}_{\s ct}$ has been introduced in order 
to avoid thermal corrections at higher order which has already been
included in the tree level. With the counter term the Lagrangian 
remains unchanged, so the effective theory is a mere re-ordering
of the perturbative expansion. A similar exercise has to be carried 
out in gauge theory keeping in mind that an addition and subtraction of local mass
terms will violate gauge invariance. 
The effective action for hot gauge
theories have been derived in Refs.~\cite{taylor,bpea,en,vpn,flechsig},
whereas the authors of Refs.~\cite{bi,kelly} follow the classical kinetic 
theory approach for the derivation of the HTL contributions. It has 
been shown in Ref.~\cite{nairvp} that the contribution of HTL to the
energy of the QGP is positive. The counter term required to avoid double
counting in evaluating the virtual photon production from QGP  in the
two-loop approximation has been derived in Ref.~\cite{lapth} recently.
%%%%%%%%%%%%%% Figure 2 %%%%%%%%%%%%%%%%%%%%%%%%%%%%%
\begin{figure}
\centerline{\psfig{figure=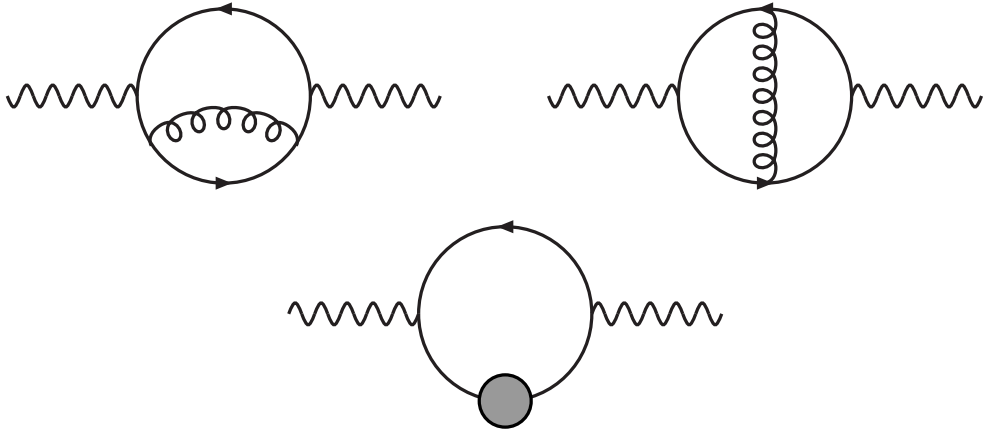,height=5.7cm,width=13cm}}
\caption{Two loop contribution to the photon self energy. A diagram
interchanging the blob in the
internal line of the third diagram should also be considered.}
\label{loop1}
\end{figure}
%%%%%%%%%%%%%%End fig 2%%%%%%%%%%%%%%%%%%%%%%%%%%%%%

\begin{figure}
\centerline{\psfig{figure=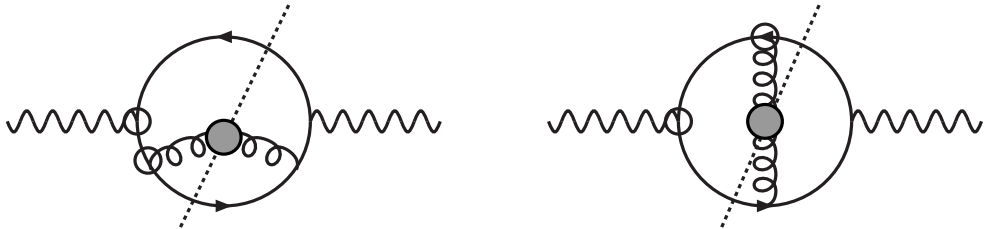,height=3cm,width=13cm}}
\caption{Two loop photon diagram relevant for bremsstrahlung processes.
The blob on the gluon (spiral line) indicates effective gluon propagator.
The circle on the vertices represent those required
to evaluate the imaginary part of the photon self energy in the
framework of thermal cutting rules (see Refs.~(\protect\cite{aurenche})
and also~(\protect\cite{gelis})).
}
\label{loop2}
\end{figure}

The photon emission from Compton and annihilation processes can be 
calculated from the imaginary parts of the first two diagrams in Fig.~(\ref{loop1}).
Since these processes involve exchange of massless quarks in the $t/u$
channels the rate becomes infrared divergent. One then 
obtains the hard contribution by introducing a lower
cut-off to render the integrals finite.
In doing so, some part of the
phase space is left out and the rate becomes cut-off dependent.
The photon rate from this (soft) part of the phase space is then 
handled using HTL resummation technique.
The application of HTL to hard photon emission rate was first 
performed in Refs.~\cite{kapusta,baier}. For hard photon
emission, one of the (soft) quark propagators in the photon
self energy diagram should be replaced by effective 
quark propagators (third diagram in Fig.~(\ref{loop1})), 
which consists of the bare propagator and
the high temperature limit of one loop corrections~\cite{klimov,hweldon}.
When the hard and the soft contributions are added,
the emission rate becomes finite because of the Landau damping
of the exchanged quark in the thermal bath and the cut-off scale
is canceled. 
The rate of hard photon emission is then obtained as~\cite{kapusta}
\be
E\frac{dR_\gamma^{QGP}}{d^3q}=\frac{5}{9}\frac{\alpha\alpha_s}{2\pi^2}
T^2\,e^{-E/T}\ln(2.912E/g_s^2T).
\ee
where $\alpha_s$ is the strong coupling constant.
Recently, the bremsstrahlung contribution to photon emission 
rate has been computed~\cite{aurenche} 
by evaluating the photon self energy in two loop HTL approximation.
The physical processes arising from two loop 
contribution (Fig.~(\ref{loop2})) 
are the bremsstrahlung of quarks, antiquarks and
quark anti-quark annihilation with scattering in the thermal bath. 
The rate of 
photon production due to bremsstrahlung process for a two flavour
thermal system with $E>T$ is given by~\cite{aurenche}
\be
E\frac{dR_\gamma^{QGP}}{d^3q}=\frac{40}{9\pi^5}\,\alpha\alpha_s
T^2\,e^{-E/T}\left(J_T-J_L\right)\ln 2,
\ee
and the rate due to $q-\bar q$ annihilation with scattering 
in the thermal bath is given by,
\be
E\frac{dR_\gamma^{QGP}}{d^3q}=\frac{40}{27\pi^5}\,\alpha\alpha_s
ET\,e^{-E/T}\left(J_T-J_L\right),
\ee
where $J_T\approx 4.45$ and $J_L\approx -4.26$.
The most important implication of this work is that the two loop
contribution is of the same order of magnitude as those evaluated
at one loop~\cite{kapusta,baier} due to the larger size of the
available phase space. 
%%%%%%%%%%%%%% Figure 3%%%%%%%%%%%%%%%%%%%%%%%%%%%%%
%%%%%%%%%%%%%%%%%%%%%End fig 3%%%%%%%%%%%%%%%%%%%%%%
In case of soft thermal photon ($E\sim g_sT$) emission rate,
all the vertices and the propagators has to be replaced
by the corresponding effective quantities. It has been
shown~\cite{baier2,aurenche1} that the result is divergent
due to the exchange of massless quarks introduced through the HTL 
effective vertices itself. 
However, such collinear singularities
for light-like external momentum could be removed with
an improved action~\cite{flechsig}.
It is also shown that such infrared singularities could be
removed (because of KLN (Kinoshita - Lee - Nauenberg) theorem~\cite{kln1,kln2}) 
by including appropriate diagrams and summing
over all degenerate initial and final states~\cite{niegawa1,niegawa2}, but
the rate is non-perturbative because it is sensitive 
to the scale $g_s^2T$~\cite{agz}. 

In this work, emission rate of hard photon 
is considered, which is well under control within HTL resummation. 
However, there
are important issues in hot gauge theories which cannot be addressed
within the HTL resummation method. For example, (i) HTL resummation is based on the
weak coupling limit ($g_s<<1$) to distinguish between hard ($T$) and soft
momentum scale ($g_sT$) but such a limit may not be 
realized in URHIC even for
the highest energy to be available at the CERN LHC in the near future. Extrapolation
of results obtained in HTL approximation to higher values of coupling constant
will be demonstrated in section 8.2 through photon spectra, (ii) it 
cannot cure the infra-red divergence problem that arises in the damping rate of fast 
fermions, (iii)it cannot remove the mass shell singularities in the  soft
photon (real) emission rate, (iv) the next to leading order correction to the Debye
mass diverges unless one includes magnetic screening, which is beyond the
scope of HTL approximation and finally (v) HTL works for a system in equilibrium 
; extension of the formalism to non-equilibrium processes is still in the
early stages of development. Results from other methods  
such as ladder approximation~\cite{cornwall}, 
renormalization group equation~\cite{km} etc. will be  very important in these
cases. For further discussions on 
the successes and limitations of HTL resummation technique and
other methods we refer to Refs.~\cite{rkobes} and ~\cite{banff}.

\subsection{Photon emission from hot hadronic gas}

To evaluate the photon emission rate from a hadronic gas
we  model the system as consisting of $\pi$, $\rho$, $\omega$
and $\eta$. The relevant vertices for the reactions 
$\pi\,\pi\,\ra\,\rho\,\gamma$ and $\pi\,\rho\,\ra\,\pi\,\gamma$
and the decay $\rho\,\ra\,\pi\,\pi\,\gamma$
are obtained from the following Lagrangian:
\be
{\cal L} = -g_{\rho \pi \pi}{\vec {\rho}}^{\mu}\cdot
({\vec \pi}\times\partial_{\mu}{\vec \pi}) - eJ^{\mu}A_{\mu} + \frac{e}{2}
F^{\mu \nu}\,({\vec \rho}_{\mu}\,\times\,{\vec \rho}_{\nu})_3,
\label{photlag}
\ee
where $F_{\mu \nu} = \partial_{\mu}A_{\nu}-\partial_{\nu}A_{\mu}$, is the
field tensor for electromagnetic field
and $J^{\mu}$ is the hadronic part of the electromagnetic
current given by
\be
J^{\mu} = ({\vec \rho}_{\nu}\times{\vec \varrho^{\nu \mu}})_3 + (
{\vec \pi}\times(\partial^{\mu}\vec \pi+g_{\rho \pi \pi}{\vec \pi}\times{\vec
\rho}^{\mu}))_3,
\label{jmu}
\ee
with ${\vec \varrho_{\mu \nu}} = \partial_{\mu}{\vec \rho}_{\nu}-\partial_{\nu}
{\vec \rho}_{\mu}-g_{\rho \pi \pi}(\vec \rho_{\mu}\times\vec \rho_{\nu})$.
$\vec\pi$, $\vec\rho$ and $A^\mu$ represent the $\pi$, $\rho$ and photon
fields respectively and the arrows represent vectors in isospin space.
$g_{\rho\pi\pi}$ denotes the
coupling strength of the $\rho-\pi-\pi$ vertex, fixed from the
observed decay width $\rho\ra\pi\pi$.
The invariant amplitudes for all these reactions have been listed in 
the appendix of Ref.~\cite{sourav}. 

For the sake of completeness we have also considered the photon 
production due to the reactions $\pi\,\eta\,\rightarrow\,\pi\,\gamma$, 
$\pi\,\pi\,\rightarrow\,\eta\,\gamma$ and the decay 
$\omega\,\ra\,\pi\,\gamma$ using the following interaction~\cite{GSW}:
\be
{\cal L} =
\frac{g_{\rho \rho \eta}}{m_{\eta}}\,
\epsilon_{\mu \nu \alpha \beta}\partial^{\mu}{\rho}^{\nu}\partial^{\alpha}
\rho^{\beta}\eta
+\frac{g_{\omega \rho \pi}}{m_{\pi}}\,
\epsilon_{\mu \nu \alpha \beta}\partial^{\mu}{\omega}^{\nu}\partial^{\alpha}
\rho^{\beta}\pi
+\frac{em_{\rho}^2}{g_{\rho}}A_{\mu}\rho^{\mu}
\label{etaro}
\ee
where $\epsilon_{\mu \nu \alpha \beta}$ is the 
totally antisymmetric Levi-Civita tensor.
The last term in the above Lagrangian is written down on the basis
of Vector Meson Dominance (VMD)~\cite{sakurai}.
The invariant amplitudes for the reactions
are given in Ref.~\cite{npa99}. The values of $g_{\rho\rho\eta}$ and
$g_{\omega\rho\pi}$ are fixed from the observed decays, $\rho\,\ra\,\eta\,\gamma$
and $\omega\,\ra\,\pi\,\gamma$ respectively~\cite{npa99}.
The constant $g_{\rho}$ is determined from the 
decay, $\rho^0\ra\,e^+e^-$. 

The importance of the role of $a_1$ as an intermediary meson in the
process $\pi\,\rho\,\rightarrow\,\pi\,\gamma$ was first emphasized 
in Refs.~\cite{xiong,song}.
Recently it has been shown~\cite{halasz} that the role of intermediary
$a_1$ in this process
is less important than thought earlier~\cite{xiong,song}.
The photon production rate obtained in Ref.~\cite{kim} is similar to that in Ref.~\cite{halasz}. 
In this article we use the
following interaction Lagrangian for the $\pi\rho a_1$ and 
$\pi a_1\gamma$ vertices~\cite{kim,rudaz,alam}:
\bea
{\cal L}&=& 
\frac{g_\rho^2f_\pi}{Z_\pi}
\left[(2c+Z_\pi)\vec \pi\cdot\vec \rho_\mu\times\,\vec a^\mu+
\frac{1}{2m_{a_1}^2}
\vec \pi\cdot
(\partial_{\mu}{\vec \rho}_{\nu}-\partial_{\nu}{\vec \rho}_{\mu})
\times\,
(\partial^{\mu}{\vec a}^{\nu}-\partial^{\nu}{\vec a}^{\mu})
\right.\nonumber\\
&&\left. +\frac{\kappa_6\,Z_\pi}{m_\rho^2}
\partial^\mu\vec \pi\cdot
(\partial_{\mu}{\vec \rho}_{\nu}-\partial_{\nu}{\vec \rho}_{\mu})
\times\,\vec a^\nu\right]\nonumber\\
&&+\frac{eg_\rho\kappa_6f_\pi}{m_\rho^2}F^{\mu\nu}
( \partial_{\mu}{\vec a}_{\nu}-\partial_{\nu}{\vec a}_{\mu}\times\vec \pi)_3
\label{aro}
\eea
where $a_\mu$ corresponds to the $a_1$ field,
$Z_\pi$ is the renormalization constant for pion fields and $f_\pi(=93$ MeV) 
is the pion decay constant. The interaction terms with coefficients
$c$ and $\kappa_6$ are introduced to improve the phenomenology of the model.
Following values of the parameters,  $m_{a_1}=1260$ MeV, $g_\rho=5.04$,
$c=-0.12$, $Z_{\pi}=0.17$ and
$\kappa_6=1.25$~\cite{rudaz,kim}, reproduce 
the width of the $a_1$ meson in vacuum. 

Most of the photon producing reactions under consideration involves
unstable particles (e.g. $\rho$ and $\omega$) in the external lines
(initial or final channels). According to Eq.~(\ref{rktp}) we need to
know the occupation probability of unstable particles in the thermal
bath. So what is the appropriate phase space density of an unstable
particle in this case?

The density of a stable hadron of mass $m$ in a thermal bath is completely
determined by the temperature, chemical potential and 
the statistics obeyed by the species  through the following 
equation:
\be
\frac{dN}{d^3x\,d^3k\,ds}=\frac{\cal N}{(2\,\pi)^3}
\frac{1}{\exp(k_0-\mu)/T\pm\,1}\,
\delta(s-m^2)
\label{stable}
\ee
where ${\cal N}$ is the statistical degeneracy, $k_0=\sqrt{\vec k^2\,+s}$ 
is the energy of the particle in the rest frame of the thermal
bath and $\mu$ is the chemical potential.  
The question we would like to ask now - How 
the equation~(\ref{stable}) will be modified if the 
particle decays within the thermal bath? 
This problem has been addressed by Weldon~\cite{weldon93} through 
the generalization of Breit-Wigner formula at finite 
temperature and density.
The distribution 
of an unstable particle in a thermal bath is given by~\cite{weldon93}
\be
\frac{dN}{d^3x\,d^3k\,ds}=\frac{\cal N}{(2\,\pi)^3}
\frac{1}{\exp(k_0-\mu)/T\pm\,1}\,P(s)
\label{unstable}
\ee
where $P(s)$ is the spectral function~\cite{bellac,zubarev} 
and can be calculated from the effective thermal propagator. It is  given by
\be
P(s)=\left[\frac{1}{\pi}
\frac{{\s Im}\Pi}{(s-m_V^2+{\s Re}\Pi)^2+({\s Im}\Pi)^2}\right]
\ee
Equation~(\ref{unstable}) indicates that to obtain 
realistic results for the photon production through a 
reaction involving unstable particle in the external 
line the finite width of the particle should be taken
into account by introducing the spectral representation
of the corresponding particle and integrating over $s$.
This is done in our calculation for the unstable vector mesons
appearing in the external line in reactions for photon production.
In the case of an unstable particle appearing in the internal
line the finite width of the particle is taken into account
through effective propagators.
However, the effects of the finite width of $\rho$ on the photon
spectra is negligible but it affects the dilepton spectra
substantially as will be shown in section 8. 
It is interesting to  note that the spectral function reduces to a 
Dirac delta function $\delta(s-m_V^2+{\s Re}\Pi)$
in the limit ${\s Im}\Pi\,\ra\,0$, i.e. when the particle is stable. 
In the calculation of the imaginary part of the self energy ${\s Im}\Pi$
of $\rho$ say, one must in principle, include all the processes 
which can create or annihilate a $\rho$ in the thermal bath. 
However, within the ambit of the model adopted in the present work
we have seen that the most dominant contribution to 
${\s Im}\Pi$ comes from the $\rho-\pi-\pi$ interaction
in the temperature range of our interest. 

\subsection{Dilepton emission from hot hadronic gas and QGP}

In order to express the dilepton emission rate from hadronic matter
in terms of the retarded current correlation function we use Eqs.~(\ref{red1})
and (\ref{red2}) in Eq.~(\ref{drretcor}) to obtain,
\be
\frac{dR}{d^4q}=-\frac{\alpha}{12\pi^4\,q^2}(1+\frac{2m^2}{q^2})
\sqrt{1-\frac{4m^2}{q^2}}{\s Im}W^{R\mu}_\mu\,f_{BE}(q_0)
\label{drcor1}
\ee
where $W_{\mn}^R$ is the retarded current correlation function.
The parametrized form of the electromagnetic current correlation function in the
$\rho$ and $\omega$ channels will be discussed in detail in section 5.2. 
Now instead of using the current correlation function directly in the above
equation one can use vector meson dominance (VMD) to obtain the dilepton
yield from ($\pi^+\pi^-\,\ra\,e^+e^-$) which is known to be the most dominant
source of dilepton production. In the low mass region one should also add 
the contributions  from the decay of vector mesons
such as $\rho$ and $\omega$. This is usually
done in the literature. 
In order to make a comparative study 
we state briefly how the emission rate from pion annihilation can be
derived from Eq.~(\ref{drcor1}). VMD relates the hadronic electromagnetic
current to the vector meson field through field current identity as
\be
J^h_\mu=\sum_{V}\, \frac{e}{g_V}m_V^2 V_\mu
\ee
where, $V$ stands for the vector fields $\rho$, $\omega$, $\phi$.
We shall keep only $\rho$ meson in the following.
The electromagnetic current correlator can then be expressed in terms of
the propagator of the vector particle in the following way:
\be
{\s Im}W_{\mn}^R=-\frac{e^2m_{\rho}^4}{g_\rho^2}{\s Im}D_{\mn}^{\rho R}
\ee
where
\bea
{\s Im}D_{\mn}^{\rho R}& = &A_{\mn}\left[\frac{{\s Im}\Pi_T^{\rho R}}{(q^2-m_\rho^2+{\s Re}
\Pi_T^{\rho R})^2+[{\s Im}\Pi_T^{\rho R}]^2}\right]\nonumber\\
&&+B_{\mn}\left[\frac{{\s Im}\Pi_L^{\rho R}}{(q^2-m_\rho^2+{\s Re}
\Pi_L^{\rho R})^2+[{\s Im}\Pi_L^{\rho R}]^2}\right].
\eea
Recall that $A_{\mn}$ and $B_{\mn}$ are the transverse and longitudinal 
projection operators whose explicit expressions are given in the appendix.
In the approximation $\Pi_T^{\rho R}=\Pi_L^{\rho R}=\Pi^{\rho R}$
we can define the $\rho$ spectral function in VMD as
\be
{\s Im}W_L^R=\frac{e^2 m_\rho^4}{g_\rho^2 q^2}
\left[\frac{{\s Im}\Pi^{\rho R}}{(q^2-m_\rho^2+{\s Re}
\Pi^{\rho R})^2+[{\s Im}\Pi^{\rho R}]^2}\right].
\label{walcor}
\ee
Using Eq.~(\ref{drcor1})
we obtain the dilepton rate from pion annihilation as
\bea
\frac{dR}{dM}&=&\frac{2\alpha^2}{\pi^2}\frac{m_\rho^4}{g_\rho^2}\frac{1}{M}
\left(1+\frac{2m^2}{M^2}\right)\sqrt{1-\frac{4m^2}{M^2}}\nonumber\\
&\times&\int e^{-M_T\cosh y/T}M_TdM_Tdy
\left[\frac{{\s Im}\Pi^{\rho R}}{(M^2-m_\rho^2+{\s Re}\Pi^{\rho R})^2+
[{\s Im}\Pi^{\rho R}]^2}\right]
\eea
in the Boltzmann approximation.
This on simplification gives
\be
\frac{dR}{dM} = \frac{\sigma(M)}{(2\pi)^4}\,M^4\,T\,\sum_n
\,\frac{K_1(nM/T)}{n}\,(1-4m_\pi^2/M^2),
\label{pipi0}
\ee
where $K_1$ is the modified Bessel function, and $M$ is the invariant mass
of the lepton pair and
$\sigma(M)$ is the cross-section for the pion 
annihilation given by
\be
\sigma(M) = \frac{4\,\pi\,\alpha^2}{3\,M^2}\,\sqrt{1-4m_\pi^2/M^2}\,
\sqrt{1-4m^2/M^2}\,(1+2m^2/M^2)\,|F_\pi(M)|^2,
\ee
where
\be
|F_\pi(M)|^2=\frac{m_\rho^4}
{(M^2-m_\rho^2+{\s Re}\Pi^{\rho R})^2+
({\s Im}\Pi^{\rho R})^2}
\ee

In the same way, the invariant mass distribution of 
lepton pairs from the vector meson decays 
is obtained using Eq.~(\ref{wel1}), as 
\bea
\frac{dR}{dM}&=&\frac{2J+1}{\pi^2}\,
M^2T\,\sum_n\,\frac{K_1(nM/T)}{n}\nonumber\\
&&\times\,\frac{M\Gamma_{{\s tot}}
/\pi}{(M^2-m_V^2+{\s Re}\Pi^{\rho R})^2+M^2\Gamma_{{\s tot}}^2}
M\Gamma_{V\,\ra\,e^+\,e^-}^{{\s vac}}
\label{decay0},
\eea
where $\Gamma_{{\s tot}}$ is the width of the vector meson
in the medium and $\Gamma^{\s vac}_{V\rightarrow e^+e^-}$ is the
partial width for the
leptonic decay mode for the off-shell vector mesons in vacuum given by
\be
\Gamma_{V\,\ra\,e^+\,e^-}^{{\s vac}}=
\frac{4\pi\alpha^2M}{3g_\rho^2}
\sqrt{1-4m^2/M^2}\,(1+2m^2/M^2)
\ee
where $m$ is the mass of the electron.

We have considered quark anti-quark annihilation 
for the evaluation of dilepton emission rate from QGP~\cite{janepr}. 
The dilepton rate is given by
\be
\frac{dR}{dM} = \frac{\sigma_{q\bar q}(M)}{(2\pi)^4}\,M^4\,T\,\sum_n
\,\frac{K_1(nM/T)}{n}
\ee
with the  cross section 
\be
\sigma_{q\bar{q}\rightarrow e^+e^-}=\frac{80\pi}{9}\frac{\alpha^2}{M^2}
\sqrt{\left(1-\frac{4m^2}{M^2}\right)}
\,\left(1+\frac{2m^2}{M^2}\right).
\ee

%%%%%%%%%%%%%%%%%%%%%%%%%%%% mass.tex %%%%%%%%%%%%%%%%%%%%%%%%
\setcounter{equation}{0}
\def\theequation{4.\arabic{equation}}

\section{Hadronic Properties at Finite Temperature}
    As emphasized earlier, the photon and dilepton emission rates
are related to the imaginary part of the photon self energy in the medium.
In this section we will study the in-medium modifications of the
particles appearing in the internal thermal loop of the photon
self energy diagram.
Here the hadronic medium consists of
mesons and baryons at a finite temperature. Due to
the interactions with real and virtual excitations, the properties of
these hadrons are expected to get modified. As a result the 
propagators appearing in the photon self energy undergo modifications.
The subject of the present section is to discuss how one incorporates 
these changes in the framework of Thermal Field 
Theory. A brief discussion of the thermal propagators is given in the
appendix.

\subsection{The Walecka model - nucleon mass}

  Before discussing the vector meson masses in the medium let us see how the 
nucleon properties are modified in matter at finite temperature. Nuclear
matter is studied using the Quantum Hadrodynamics (QHD) model~\cite{vol16}
 in which the
nucleons interact through the exchange of scalar $\sigma$ and the vector $\omega$ 
mesons. The interaction in this model is described by the Lagrangian
\be
{\cal L}_I = -g_{\wnn}\,{\bar N}\gamma_{\mu}\,N\,\omega^{\mu}+
g_{\snn}\,{\bar N}\,\sigma\,N,
\label{lqhd}
\ee
where $N(x)$, $\sigma(x)$, and $\omega(x)$ are the nucleon, $\sigma$, and
$\omega$ meson fields respectively. The $\sigma (\omega)$ field couples to the
scalar (vector) current of the nucleon with the coupling constant $g_{\snn} 
(g_{\wnn})$ which will be specified later. 

The free nucleon propagator at finite temperature and density in
general has four components (see appendix). The time-ordered $\it i.e.$
the (11)-component is physically relevant for our purposes and we will
denote this as $G^0(p)$. So we have 
\bea
G^0(p)&\equiv&G^{0(11)}(p)\nonumber\\
&=&(p\sls+M_N)\left[\frac{1}{p^2-M_N^2+i\eps}
+2\pi i\delta(p^2-M_N^2)\eta(p.u)\right]\nonumber\\
&\equiv&G_F^0(p) + G_D^0(p),
\eea 
where the first term ($G_F^0$) describes the free propagation of 
nucleon-antinucleon pairs and the second term ($G_D^0$) allows for the on-shell
propagation of particle-hole pairs. $M_N$ in the above equation is the free
nucleon mass.

\begin{figure}
\centerline{\psfig{figure=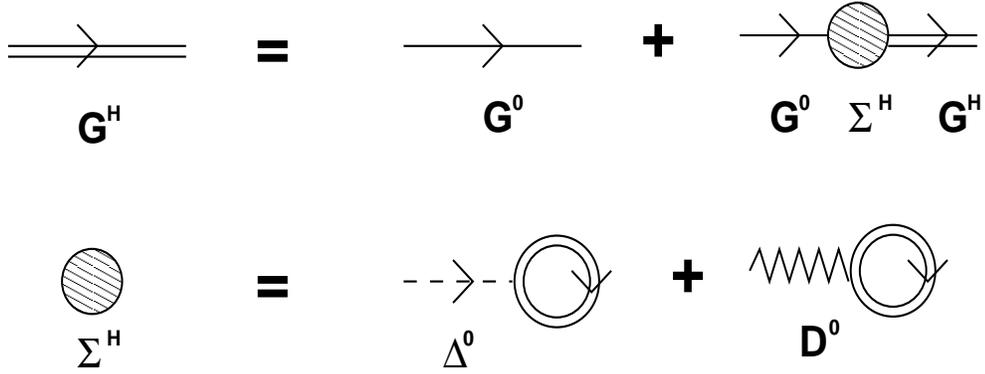,width=13cm,height=5cm,angle=-90}
}
\caption{Diagrammatic representation of Dyson-Schwinger equation for nucleons
in RHA }
\label{rhafig}
\end{figure}

 The effective mass of the nucleon in matter at finite temperature in the
presence of interaction described by Eq.~(\ref{lqhd}) will appear as a pole of
the effective nucleon propagator. 
In the Relativistic Hartree Approximation (RHA)~\cite{vol16,chin} one obtains 
the 
effective propagator by summing up scalar and vector tadpole diagrams 
self-consistently i.e. by using the interacting propagators to 
determine the self energy. The effective propagator also called 
Hartree propagator is given by (see Fig.~(\ref{rhafig}))
%%%%%%%%%%%%%%%%Figure 4%%%%%%%%%%%%%%%%%
%%%%%%%%%%%%%%%%End figure 4%%%%%%%%%%%%%%%%%
\be
G^H(p) = G^0(p) + G^0(p)\Sigma^H(p)G^H(p)
\label{rhadyson}
\ee 
where $\Sigma^H(p)$ is the nucleon self energy which contains contributions
from both scalar ($\Sigma_s$) and vector ($\Sigma_v^{\mu}$) tadpole
diagrams~\cite{vol16,chin} and is given by
\be
\Sigma^H = \Sigma_s^H - \gamma^{\mu}\Sigma_{\mu v}^H,
\label{sigma}
\ee
where
\be
\Sigma_s^H = i\frac{g_{\snn}^2}{m_{\sigma}^2}\,\int\,\frac{d^4p}{(2\pi)^4}\,
{\rm {Tr}}[G^H(p)] 
\label{sigmas1}
\ee 
and,
\be
\Sigma_{\mu v}^H = i\frac{g_{\wnn}^2}{m_\omega^2}\,\int\,\frac{d^4p}{(2\pi)^4}\,
{\rm {Tr}}[\gamma_{\mu}\,G^H(p)]
\label{sigmav1}
\ee
Here, $m_\sigma\,\,(m_\omega)$ is the mass of the neutral scalar (vector) meson.
The solution of Eq.~(\ref{rhadyson}) now reads,
\bea
G^H(p)&=&(\bar p\sls+M_N^{\ast})\left[\frac{1}{\bar p^2-M_N^{\ast 2}+i\eps}
+2\pi i\delta(\bar p^2-M_N^{\ast 2})\eta(\bar p.u)\right]\nonumber\\ 
&\equiv&G_F^H(p) + G_D^H(p)
\label{gH}
\eea
One observes that the pole structure of the
effective nucleon propagator in RHA resembles that of the free
propagator with shifted mass and four-momentum {\it i.e.}
 $\bar p=p+\Sigma_v^H$
and $M_N^\ast=M_N+\Sigma_s^H$, is the effective mass. 
Using $G_D^H$ in place of the full Hartree 
propagator in Eqs.~(\ref{sigmas1}) and (\ref{sigmav1}) defines the Mean
Field Theory (MFT) values of the self energies. This is equivalent to 
solving the meson field equations with the replacement of the meson
field operators by their expectation values which become classical
fields i.e. $\sigma \ra \langle\sigma\rangle$ and 
$\omega \ra \langle\omega\rangle$. This yields $\langle\sigma\rangle
=g_{\snn}\,\rho_s/m_\sigma^2$ and $\langle\omega^\mu\rangle = 
g_{\wnn}\,\delta^{\mu 0}\rho_B/m_\omega^2$
which indicate that the nuclear ground state contains scalar and vector
meson condensates generated by baryon sources. The spatial part of
the $\omega$ condensate vanishes due to rotational symmetry in infinite
nuclear medium. These condensates 
are related to the scalar and vector self energies generated by summing
tadpole diagrams in QHD as,
$\Sigma_s=-g_{\snn}\langle\sigma\rangle$ and
$\Sigma_v^0=-g_{\wnn}\langle\omega^0\rangle$. 
The mean field approximation is thus to neglect the fluctuations in the
meson fields which themselves are generated by the nucleons. 

RHA is obtained when one includes the vacuum fluctuation corrections 
to the MFT results. This amounts to the inclusion of the Dirac
part of the propagator $G_F^H$ in the calculation of the self energies.
Summing over the vacuum tadpoles  results in a sum over all occupied states
in the negative energy sea of nucleons. Vacuum or quantum fluctuations,
as these are called, form an essential ingredient in a relativistic
theory of many particle systems. Since there are infinite number of
negative energy states in the vacuum one expects that the vacuum contribution
to the self energy is infinite.

Let us now find the Hartree self energy of the nucleon with the full
nucleon propagator which consists a medium and a vacuum part.
The vector part of the self energy is obtained from Eq.~(\ref{sigmav1}) as
\be
\Sigma_v^{H\mu}=8i\frac{g_{\wnn}^2}{m_\omega^2}\int \frac{d^4p}{(2\pi)^4}\,
\frac{\bar p^\mu}{\bar p^2-M_N^{\ast 2}+i\eps}-
\frac{g_{\wnn}^2}{m_\omega^2}\delta^{\mu 0}
\rho_B.
\label{sigmav2}
\ee
The first term of this equation appears to be divergent. The usual
procedure is to regularize the integral in $n$ dimensions by dimensional
regularization to render the integral finite. One can then shift the
integration variable from $p$ to $\bar p$. The resulting integral vanishes
by symmetric integration. 
The vector self energy then reduces to
$\Sigma_v^{H\mu}=-{g_{\wnn}^2}\delta^{\mu 0}\rho_B/{m_\omega^2}$ and thus gives
rise to a an effective chemical potential, $\mu^\ast=\mu-{g_{\wnn}^2\,\rho_B}/
{m_\omega^2}$.
The scalar part of the self energy follows from Eq.~(\ref{sigmas1}):
\bea
\Sigma_s^H&=&8i\frac{g_{\snn}^2}{m_\sigma^2}\int\frac{d^4p}{(2\pi)^4}
\,\frac{M_N^{\ast 2}}{\bar p^2-M_N^{\ast 2} +
i\eps}- \frac{4g_{\snn}^2}{m_\sigma^2}
\int\frac{d^3p}{(2\pi)^3}\frac{M_N^{\ast}}{E^\ast}\,\nonumber\\
&&\times\,\left[\frac{}{}f_{FD}(\mu^{\ast},T)+{\bar f}_{FD}(\mu^{\ast},T)\right]
\label{sigmas2}
\eea
where
\bea
f_{FD}(\mu^{\ast},T)& = &\frac{1}{{\rm exp}[(E^{\ast}-\mu^{\ast})/T]+1}\nonumber\\
{\bar f}_{FD}(\mu^{\ast},T)& = &\frac{1}{{\rm exp}[(E^{\ast}+\mu^{\ast})/T]+1}
\nonumber\\
E^{\ast} & = & \sqrt{({\vec p}^2+M_N^{\ast 2})}\nonumber\\
\eea
Here $\rho_B$ is the baryon density (in the present 
work we will take $\rho_B=0$) of the medium and is given by
\be
\rho_B = \frac{4}{(2\pi)^3}\int\,d^3p\,
[f_{FD}(\mu^{\ast},T)-{\bar f}_ {FD}(\mu^{\ast},T)].
\label{bden}
\ee
The first term in Eq.~(\ref{sigmas2}), to be denoted by $\Sigma_s^{(1)}$,
 represents the contribution to the
scalar self energy from the filled Dirac sea and is ultraviolet divergent.
We will now proceed to renormalize this divergent contribution. The first step
is to isolate the divergences through dimensional regularization. 
This gives
\bea
\Sigma_s^{(1)}&=&-\frac{g_{\snn}^2}{m_\sigma^2}\frac{\Gamma(2-n/2)}{2\pi^2}
M_N^{\ast 3}
\nonumber\\
&=&-\frac{g_{\snn}^2}{m_\sigma^2}\frac{\Gamma(2-n/2)}{2\pi^2}
(M_N^3+3M_N^2\Sigma_s^H+ 3M_N{\Sigma_s^H}^2+{\Sigma_s^H}^3)
\eea
since $M_N^{\ast}=M_N+\Sigma_s^H$.
The divergence in $\Sigma_s^{(1)}$ now appears in the pole of the 
$\Gamma$-function for physical dimension $n=4$.
The counter terms needed to remove the divergent contributions
from the loop corrections to the measurable amplitudes are
\be
{\cal L}_{CT}=\sum_{n=1}^4\alpha_n\,\sigma^{n}/n!.
\label{lCT}
\ee
Including the contributions from the counter terms the renormalized
self energy becomes
\be
\Sigma_s^{(1) ren}=\Sigma_s^{(1)}+\Sigma_s^{CTC}, 
\ee
where
\be
\Sigma_s^{CTC}=\sum_{n=0}^3\frac{1}{n!}\left(\frac{-g_{\snn}}{m_\sigma^2}\right)
\left(\frac{-\Sigma_s^H}{g_{\snn}}\right)^n\alpha_{n+1}.
\ee
The coefficients ($\alpha_i$) are fixed by defining a set of
renormalization conditions. Since the scalar density 
$\langle\bar\psi\psi\rangle$ is not a conserved quantity the
tadpole diagrams appear in the self energy. The tadpole contribution
must vanish in normal vacuum (free space) i.e. $\langle\sigma\rangle_0=0$. 
This is ensured by the term $\alpha_1\sigma$ in ${\cal L}_{CT}$.
$\alpha_2\sigma^2$ is the meson mass counter term which ensures that
$m_\sigma$ is the physical (measured) mass. Since the original Lagrangian
of QHD~\cite{vol16} does not contain $\sigma^3$ and $\sigma^4$ terms,
at the tree level, three and four point meson amplitudes must vanish. The
last two counter terms in Eq.~(\ref{lCT}) are chosen to maintain this condition
at zero external momenta for the $\sigma$ meson when nucleon loop corrections are
included. We thus have
\be
\alpha_n=-i(-g_{\snn})^n(n-1)!\int\frac{d^4p}{(2\pi)^4}{\rm Tr}[G_F^0(p)^n].
\ee

Consequently the effective nucleon mass reads
\bea
\Sigma_s^H & = &M_N^{\ast} - M_N\nonumber\\
&=& -\frac{4g_{\snn}^2}{m_\sigma^2}\int\,\frac{d^3p}{(2\pi)^3}\,
\frac{M_N^{\ast}}{E^{\ast}}
\left[\frac{}{}f_{FD}(\mu^{\ast},T)+{\bar f}_ {FD}(\mu^{\ast},T)
\right]\nonumber\\
           & + & \frac{g_{\snn}^2}{m_\sigma^2}\,\frac{1}{\pi^2}\left[M_N^{\ast 3}
{\rm ln}\left(\frac{M_N^{\ast}}{M_N}\right)-
M_N^2(M_N^{\ast}-M_N)\right.\nonumber\\
           & - & \left.\frac{5}{2}M_N(M_N^{\ast}-M_N)^2-
\frac{11}{6}(M_N^{\ast}-M_N)^3
\right].
\label{nmass}
\eea
The solution of this equation gives the effective nucleon mass $M_N^\ast$ as
a function of temperature and baryon density. At zero baryon density it can
be parametrized as
\be
M_N^\ast=M_N\left[1-0.0264\left(\frac{T ({\s GeV})}{0.16}\right)^{8.94}\right].
\ee
\subsection{The vector meson mass}

In a medium meson properties get modified due to its coupling to nuclear
excitations. This modification is contained in the meson self energy 
which appears in the Dyson-Schwinger equation for the effective propagator
in the medium. The interaction vertices are provided by the Lagrangian
\be
{\cal L}_{VNN} = g_{VNN}\,\left({\bar N}\gamma_{\mu}
\tau^a N{V}_{a}^{\mu} - \frac{\kappa_V}{2M_N}{\bar N}
\sigma_{\mu \nu}\tau^a N\partial^{\nu}V_{a}^{\mu}\right),
\label{lagVNN}
\ee
where $V_a^{\mu} = \{\omega^{\mu},{\vec {\rho}}^{\mu}\}$,
$N$ is the nucleon field
and $\tau_a=\{1,{\vec {\tau}}\}$. $\vec \tau$ are the Pauli matrices,
$\kappa_V$ is the nucleon-vector meson tensor coupling constant will
be specified later.

The lowest order contribution to the vector meson self energy is expressed
in terms of the self-consistent nucleon propagator described 
in Eq.~(\ref {gH}). This is given by
\be
\Pi^{\mn}(k)=-2ig_{VNN}^2\int\frac{d^4p}{(2\pi)^4}{\rm Tr}\left[\Gamma^\mu(k)\,
G^H(p)\Gamma^\nu(-k)\,G^H(p+k)\right],
\ee
where $\Gamma^\mu$ represents the meson-nucleon vertex function obtained from
Eq.(~\ref{lagVNN}) and is given by
\bea
\Gamma^\mu(k)&=&\gamma^\mu;~~~~~~~~~~~~~~~~~~~~~~~~~~{\rm for}~~ \omega\nonumber\\
\Gamma^\mu(k)&=&\gamma^\mu+i\frac{\kappa_{\rho}}{2M_N}\sigma^{\mu\alpha}k_\alpha;
~~~~~~~~ {\rm for}~~\rho
\label{vtxfn}
\eea
where $\sigma^{\mu\alpha}=\frac{i}{2}[\gamma^\mu,\gamma^\alpha]$.
The vector meson self energy can be written as a sum of two parts
\be
\Pi^{\mn}(k)=\Pi_F^{\mn}(k)+\Pi_D^{\mn}(k),
\ee
where
\bea
\Pi_F^{\mn}(k)&=&-2ig_{VNN}^2\int\frac{d^4p}{(2\pi)^4}{\rm Tr}\left[\Gamma^\mu(k)\,
G_F^H(p)\Gamma^\nu(-k)\,G_F^H(p+k)\right]\nonumber\\
\Pi_D^{\mn}(k)&=&-2ig_{VNN}^2\int\frac{d^4p}{(2\pi)^4}{\rm Tr}\left[
\Gamma^\mu(k)\,G_F^H(p)\Gamma^\nu(-k)\,G_D^H(p+k)\right.\nonumber\\
&&+\Gamma^\mu(k)\,G_D^H(p)\Gamma^\nu(-k)\,G_F^H(p+k)\nonumber\\
&&+\left.\Gamma^\mu(k)\,G_D^H(p)\Gamma^\nu(-k)\,G_D^H(p+k)\right].
\label{fullpi}
\eea
$\Pi_F^{\mn}$ is the vacuum polarization. This is a bilinear function of 
$G_F^H$ and hence describes the correction to the meson propagators 
due to coupling to $N\bar N$ excitations. The $N\bar N$ pairs can
be excited only if the four-momentum carried by the mesons
is in the time-like region $(k^2>0)$. Hence the shift in the mass of the 
vector mesons due to vacuum polarization is caused by processes like
$V\ra N\bar N \ra V$ where $N$ represents nucleons in the modified Dirac
sea having an effective mass $M_N^\ast$, smaller than what it would be in
free space. We have seen that $\Pi_F^{\mn}$ causes a substantial 
negative shift in the masses of vector mesons.
>From Eq.~(\ref{fullpi}) we have
\be
\Pi_F^{\mn}(k)=-2ig_{VNN}^2\int\frac{d^4p}{(2\pi)^4}\frac{
{\rm Tr}[\Gamma^\mu(p\sls +M_N^\ast)\Gamma^\nu(p\sls+k\sls+M_N^\ast)]}
{(p^2-M_N^{\ast 2})[(p+k)^2-M_N^{\ast 2}]}.
\ee
>From naive power 
counting it can be seen that this part of the self energy is ultraviolet
divergent and has to be renormalized. A few comments about renormalizability
of the interaction given by Eq.~(\ref{lagVNN}) is in order here.
At very large momenta the propagator for massless
boson $\sim O(k^{-2})$, whereas for massive vector bosons it goes as $\sim O(1)$.
This poses severe problems to the renormalizability of the theory with
massive vector bosons.
However, in a gauge theory with spontaneous 
symmetry breaking the vector gauge bosons acquire mass in such a way
that the renormalizability of the theory is always preserved. The theory
which involves neutral massive vector bosons coupled to a
conserved current is also renormalizable. This is because in a physical
process the propagator
$\bar D_0^{\mn}=(-g^{\mn}+p^\mu p^\nu/m^2)/(p^2-m^2+i\eps)$ appears 
between two conserved currents $J_\mu$ and $J_\nu$ and the offending
term $p^\mu p^\nu/m^2$ does not contribute because of current conservation
$(p_\mu J^\mu = 0)$, making the theory renormalizable. This is the case
for the $\omega$ meson which we shall consider first 
(see Refs.~\cite{jean,caillon}).
The counter term required in this case is
\be 
{\cal L}_{VNN}^{CT}=-\frac{1}{4}\zeta V^{\mn}\,V_{\mn}.
\ee
We use dimensional regularization to separate the divergent and the
finite parts. The divergences now appear as a pole in the gamma function
at the physical dimension $n=4$. The renormalized vacuum polarization
tensor for the $\omega$ is then given by
\be
\Pi_F^{\mn}(k)=(g^{\mn}-k^\mu k^\nu/k^2)\Pi_F^{ren}(k^2),
\ee
where
\bea
\Pi_F^{ren}(k^2)&=&\frac{g_{\omega NN}^2}{\pi^2}\left\{\Gamma(2-n/2)\int_0^1\,
dz\,z(1-z)\right.\nonumber\\
&&-\left.\int_0^1\,dz\,z(1-z)\ln[M_N^{\ast 2}-k^2z(1-z)]\right\}-\zeta
\eea
in which the counter term contribution 
\be
\Pi_F^{\mn CTC}=-\zeta(g^{\mn}-k^\mu k^\nu/k^2)
\ee
has been included. $\zeta$ is now determined by the renormalization condition
\be
\Pi_F^{ren}(k^2)|_{M_N^\ast\ra M_N}=0. 
\ee
Finally, we arrive at
\bea
\Pi_F^\omega (k^2)&=&\frac{1}{3}{\rm Re}(\Pi_F^{ren})^\mu_\mu\nonumber\\
&=&-\frac{g_{\omega NN}^2}{\pi^2}k^2\int_0^1\,dz\,z(1-z)\ln\left[
\frac{M_N^{\ast 2}-k^2z(1-z)}{M_N^2-k^2z(1-z)}\right].
\eea

Renormalization of the vacuum self energy for the $\rho$ meson presents
additional problems because of the tensor interaction. A 
phenomenological subtraction procedure, as described in 
Refs.~\cite{shiomi,hatsuda2}
is used to arrive at the following expressions:
\be
\Pi_F^\rho (k^2)=-\frac{g_{\rho NN}^2}{\pi^2}
k^2\,\left[I_1+M_N^{\ast}\frac{\kappa_\rho}{2M_N}I_2+\frac{1}{2}\,(\frac{\kappa_\rho}
{2M_N})^2\,(k^2I_1+M_N^{\ast 2}I_2)\right],
\label{pik2}
\ee
where
\be
I_1=\int_{0}^{1}\,dz\,z(1-z)\,\ln\left[\frac{M_N^{\ast 2}-k^2\,z(1-z)}
{M_N^2-k^2\,z(1-z)}\right],
\ee
\be
I_2=\int_{0}^{1}\,dz\,\ln\left[\frac{M_N^{\ast 2}-k^2\,z(1-z)}
{M_N^2-k^2\,z(1-z)}\right].
\ee

The medium dependent part of the polarization, $\Pi_D^{\mn}$, describes
the coupling of the vector mesons to particle-hole excitations. It
contains at least one on-shell nucleon propagator which provides
a natural ultraviolet cutoff in the loop momenta. This part of the
self energy leads to an increased effective mass of the vector mesons
in the medium. 

The change in the hadronic mass in the medium 
can be understood from 
the following phenomenological
arguments~\cite{eletsky}. 
Consider the propagation of a vector meson in a nuclear medium. 
The attenuation of the amplitude at a distance $z$, in a Fermi gas 
approximation, is given by $e^{-n\sigma z}$,  where $n$ is the density 
of nucleons and $\sigma$ is the meson-nucleon interaction cross section. 
The optical theorem relates $\sigma$ to the imaginary part of the
forward scattering amplitude; $\sigma=4\pi{\rm Im}{\cal F}(E)/k$. It then follows that
the meson wave function $\psi \sim \exp[2\pi inz{\cal F}(E)/k]$. In terms of an
effective mass $(m_{\rm eff}=m+\D m)$, the propagation can also be described by
$\psi\sim \exp[i\sqrt{E^2-m_{\s eff}^2}z]$. Comparing the arguments of
the exponential we get
\be
\D m=-\frac{2\pi n k}{m}{\rm Re}{\cal F}(E).
\ee
This relation clearly shows that the enhancement or reduction
of hadronic masses depends on the sign of ${\s Re}{\cal F}(E)$.
 
In a hot and dense medium because of Lorentz invariance and current
conservation the general structure of the polarization tensor takes the form
\be
\Pi^{\mu \nu} = \Pi_T(k_0,|\vec k|)A^{\mu \nu}+\Pi_L(k_0,|\vec k|)B^{\mu \nu}
\ee
where the two Lorentz invariant functions $\Pi_T$ and $\Pi_L$ are 
obtained by contraction:
\bea
\Pi_L&=&-\frac{k^2}{|\vec k|^2}u^{\mu}u^{\nu}\Pi_{\mu \nu}\nonumber\\
\Pi_T&=&\frac{1}{2}(\Pi_{\mu}^{\mu}-\Pi_L)
\eea
$u_{\mu}$ is the four velocity if the thermal bath.

In the case of the vector meson interacting with real particle-hole
excitations in the nuclear medium these are given by
\bea
\Pi_{\mu \nu}^D &=& -2ig_{VNN}^2\,\int\,\frac{d^4p}{(2\pi)^4}\,{\rm{Tr}}\left[
\frac{}{}\Gamma^{\mu}(k)G_F(p)\Gamma^{\nu}(-k)G_D(p+k)+(F\leftrightarrow D)
\right]\nonumber\\
&=&(\Pi^{D,v}+\Pi^{D,vt}+\Pi^{D,t})_{\mu \nu}
\eea
with
\bea
(\Pi^{D,v})_{\mu}^{\mu} &=& \frac{g_{VNN}^2}
{2\pi^2}\,\frac{1}
{|\vec k|}\,\int\,\frac{pdp}{\omega_p}\,\left[(k^2+2M_N^{\ast 2})
\ln\left\{
\frac{(k^2+2|\vec p||\vec k|)^2-4k_0^2\omega_p^2}{(k^2-2|\vec p||\vec k|)^2-
4k_0^2\omega_p^2}\right\}\right.\nonumber\\
&&-\left.\frac{}{}8|\vec p||\vec k|\right]
\left[\frac{}{}f_{FD}(\mu^{\ast},T)+{\bar f}_{FD}(\mu^{\ast},T)\right]
\eea
\bea
(\Pi^{D,vt})_{\mu}^{\mu} &=& \frac{3g_{VNN}^2}{\pi^2}\,
M_N^{\ast}\left(\frac{\kappa_V}{2M_N}\right)\,\frac{k^2}{|\vec k|}\,\int\,
\frac{pdp}{\omega_p}\,
\ln\left\{
\frac{(k^2+2|\vec p||\vec k|)^2-4k_0^2\omega_p^2}{(k^2-2|\vec p||\vec k|)^2-
4k_0^2\omega_p^2}\right\}\nonumber\\
&&\times
\left[\frac{}{}f_{FD}(\mu^{\ast},T)+{\bar f}_{FD}(\mu^{\ast},T)\right]
\eea
\bea
(\Pi^{D,t})_{\mu}^{\mu}&=& \frac{g_{VNN}^2}{4\pi^2}\,
\left(\frac{\kappa_V}{2M_N}\right)^2\frac{k^2}
{|\vec k|}\,\int\,\frac{pdp}{\omega_p}\,\left[\frac{}{}(k^2+8M_N^{\ast 2})
\right.\nonumber\\
&&\times\left.
\ln\left\{
\frac{(k^2+2|\vec p||\vec k|)^2-4k_0^2\omega_p^2}{(k^2-2|\vec p||\vec k|)^2-
4k_0^2\omega_p^2}\right\}
-4|\vec p||\vec k|\right]\nonumber\\
&&\times\left[\frac{}{}f_{FD}(\mu^{\ast},T)+{\bar f}_{FD}(\mu^{\ast},T)\right]
\eea
The longitudinal component of the polarization tensor is given by
\be
\Pi_L^D=\Pi_L^{D,v}+\Pi_L^{D,vt}+\Pi_L^{D,t}
\ee
with
\bea
\Pi_L^{D,v}&=&-\frac{g_{VNN}^2}{4\pi^2}\,
\frac{k^2}{|\vec k|^3}
\int\,\frac{pdp}{\omega_p}\,\left[\frac{}{}\{(k_0-2\omega_p)^2-|\vec k|^2\}
\ln{\frac{k^2-2k_0\omega_p+2|\vec p||\vec k|}{k^2-2k_0\omega_p-2|\vec p||\vec k|}}
\right.\nonumber\\
&&+\left.\{(k_0+2\omega_p)^2-|\vec k|^2\}
\ln{\frac{k^2+2k_0\omega_p+2|\vec p||\vec k|}{k^2+2k_0\omega_p-2|\vec p||\vec k|}}
-8|\vec p||\vec k|\right]
\nonumber\\
&&\times\left[\frac{}{}f_{FD}(\mu^{\ast},T)+
{\bar f}_{FD}(\mu^{\ast},T)\right]
\eea
whereas,
\bea
\Pi_L^{D,vt} &=&\frac{g_{VNN}^2}{\pi^2}\,
M_N^{\ast}\left(\frac{\kappa_V}{2M_N}\right)
\frac{k^2}{|\vec k|}
\int\,\frac{pdp}{\omega_p}\,
\ln\left\{
\frac{(k^2+2|\vec p||\vec k|)^2-4k_0^2\omega_p^2}{(k^2-2|\vec p||\vec k|)^2-
4k_0^2\omega_p^2}\right\}\nonumber\\
&&\times\left[\frac{}{}f_{FD}(\mu^{\ast},T)+{\bar f}_{FD}(\mu^{\ast},T)\right]
\eea
and finally,
\bea
\Pi_L^{D,t}&=&-\frac{g_{VNN}^2}{2\pi^2}\,
\left(\frac{\kappa_V}{2M_N}\right)^2
\frac{k^2}{|\vec k|}
\int\,\frac{pdp}{\omega_p}\,
\left[\left\{2|\vec p|^2-\frac{k^2}{2}-\frac{(k^2-2k_0\omega_p)^2}{2|\vec k|^2}
\right\}
\right.\nonumber\\
&&\times\left.\ln{\frac{k^2-2k_0\omega_p+2|\vec p||\vec k|}
{k^2-2k_0\omega_p-2|\vec p||\vec k|}}
+\left\{2|\vec p|^2-k^2-\frac{(k^2+2k_0\omega_p)^2}{|\vec k|^2}\right\}
\right.\nonumber\\
&&\times\left.\ln{\frac{k^2+2k_0\omega_p+2|\vec p||\vec k|}
{k^2+2k_0\omega_p-2|\vec p||\vec k|}}
-\frac{4|\vec p|k_0^2}{|\vec k|}\right]\nonumber\\
&&\times\left[\frac{}{}f_{FD}(\mu^{\ast},T)+
{\bar f}_{FD}(\mu^{\ast},T)\right]
\eea
In the above the superscripts `$v$', `$vt$' and `$t$' represent
the vector-vector, vector-tensor and tensor-tensor components respectively
arising from the product of vector and tensor terms in Eq.~(\ref{vtxfn}).
The dispersion relation for the longitudinal (transverse) mode now reads 
\be
k_0^2-|\vec k|^2-m_V^2+{{\rm Re}}\Pi_{L(T)}^D(k_0,{\vec k})+{\rm {Re}}
\Pi^F(k^2) = 0
\label{disp}
\ee
%%%%%%%%%%%%%%% END OF PI %%%%%%%%%%%%%%%%%%%%%%%%%%%%%%%%%%%%%%%%%%%%%%

Usually the physical mass $(m_V^{\ast})$ is defined as the lowest
zero of the above equation in the limit ${\vec k}\ra 0$. 
In this limit
$\Pi_T^D = \Pi_L^D = \Pi^D$, and we have,
\be
\frac{1}{3}\Pi_{\mu}^{\mu}=\Pi= \Pi^D+\Pi^F
\ee
where
\be
\Pi^D(k_0,{\vec k}\ra 0) =  
-\frac{4g_{VNN}^2}{\pi^2}\,\int\,p^2dp\,F(|\vec p|,M_N^{\ast})\,
[\frac{}{}f_{FD}(\mu^{\ast},T)+{\bar f}_{FD}(\mu^{\ast},T)]
\ee
with
\bea
F(|\vec p|,M_N^{\ast})&=&\frac{1}{\omega_p(4\omega_p^2-k_0^2)}
\,\left[\frac{2}{3}(2|\vec p|^2+3M_N^{\ast 2})+k_0^2\left\{2M_N^{\ast}
(\frac{\kappa_V}{2M_N})\right.\right.\nonumber\\
&&+\left.\left.\,\frac{2}{3}(\frac{\kappa_V}{2M_N})^2(|\vec p|^2
+3M_N^{\ast 2})\right\}\right]
\eea
where $\omega_p^2={\vec p}^2+M_N^{\ast 2}$.

The effective mass of the vector meson is then obtained by solving the
equation:

\be
k_0^2 - m_V^2 + {\rm {Re}}\Pi = 0.
\label{mass}
\ee
The effective masses (denoted by asterix) take the following parametrized forms:
\bea
m_\rho^\ast&=&m_\rho\left[1-
0.127\left(\frac{T({\s GeV})}{0.16}\right)^{5.24}\right]\nonumber\\
m_\omega^\ast&=&m_\omega\left[1-
0.0438\left(\frac{T({\s GeV})}{0.16}\right)^{7.09}\right].
\label{pmasswal}
\eea
The effective mass of the $a_1$ meson, 
$m_{a_1}^\ast$ has been estimated from $m_{\rho}^\ast$ by using Weinberg's
sum rule~\cite{weinberg}.

One finds reference to two other kinds of masses in the literature. 
The invariant mass is defined
as the lowest order zero of Eq.~(\ref{disp}) with $\Pi^D$ neglected. 
Again, the screening mass of a vector
meson is obtained from the pure imaginary zero of the quantity
on the left hand side of the same equation with $k_0=0$.
These two masses are different because of the non-analyticity
of the polarization tensor at the origin {\it i.e.} at
($k_0,\vec{k})=(0,\vec{0}$).

%%%%%%%%%%%%%%%%%%%%% chiral models %%%%%%%%%%%%%%%%%%%%%%%%%%%%%%%%%%%%%
\setcounter{equation}{0}
\def\theequation{5.\arabic{equation}}

\section{Models with Chiral Symmetry}

In this section we will discuss finite temperature effects
on the vector meson properties for those models which respect
chiral symmetry. The effects of in-medium properties of vector mesons
on the electromagnetic ejectiles will be presented in section 8.

\subsection{The gauged linear sigma model}
The linear sigma model (LSM) is a beautiful tool to describe
the low energy dynamics of pions, because it shows 
explicitly how the spontaneous symmetry breaking (SSB) 
of global chiral symmetry (SU(2)$\bigotimes$SU(2)) by the isosinglet 
$\sigma$ field generates pions, the Nambu-Goldstone (NG) bosons.
However, there are reservations about the description of the $\sigma$ meson as
a well-defined degree of freedom because of its large decay width which is 
comparable to its mass.    
But, it has been argued by Hatsuda and Kunihiro~\cite{plb185} 
(see also Ref.~\cite{raja}) that in the limit of
chiral symmetry restoration, the decay of $\sigma$ to
two pion state should be disallowed as $\sigma$ and $\pi$
become degenerate in mass in this limit. They have explicitly shown that 
the width of $\sigma$ due to $\sigma\ra 2\pi$ decay vanishes as 
$T\ra T_\chi$, where $T_\chi$ is the critical temperature for
chiral transition (it is still not known whether the critical
temperature for deconfinement and chiral transition are the same or 
not, in the present work no distinction is made between them).

The properties of vector mesons at finite temperature
within the ambit of gauged 
LSM are studied by Pisarski~\cite{pisarski2,rdp}.
In the following we will discuss the main results of his work. 

The simplest version of LSM contains isosinglet $\sigma$ field and 
isotriplet pion field and respects the charge conjugation,
parity and time reversal symmetry (CPT)~\cite{pisarski2}. The Lagrangian obeying
these constraints is
\be
{\cal L}_{LSM}=tr\mid\partial_{\mu}\Phi\mid^2 +\mu^2\,tr \mid\Phi\mid^2
+\frac{1}{2}\lambda\,tr(\mid\Phi\mid^2)^2-h\,tr(\Phi)
\label{lsm}
\ee
where $\Phi$ is defined as
\be
\Phi=\frac{1}{2}(\sigma + i\vec{\pi}\cdot\,\vec{\tau})
\ee
with $\vec{\tau}$ being the Pauli matrices. The non-zero value of $h$ ensures
that the pions are massive and consequently 
PCAC (partially conserved axial current) relation is satisfied. 
Note that $\mid\Phi\mid^2=\sigma^2+\vec\pi^2$ is chirally invariant and
elimination of the $\sigma$ field by imposing the condition 
$\sigma^2+\vec\pi^2=f_\pi^2$  results in the Non-Linear Sigma Model,
which will be discussed in the next section.

In the gauged LSM (see Refs. ~\cite{pisarski2} and
\cite{gg} for detail discussions) one introduces the vectors and 
their chiral partners (axial
vector) through left and right handed fields as follows,
\be
V^{\mu}_l=(\vec\rho^{\mu}+\vec a^{\mu})\cdot\vec t +(\omega^\mu+f_1^\mu)
\ee
\be
V^{\mu}_r=(\vec\rho^{\mu}-\vec a^{\mu})\cdot\vec t +(\omega^\mu-f_1^\mu)
\ee
where $a_1$ and $f_1$ are the chiral partners of the $\rho$ and $\omega$
mesons respectively and $\vec{t}=\vec{\tau}/2$.

The inclusion of vector (axial) mesons would increase the number
of possible couplings and hence the number of arbitrary parameters
become large. How to include vector meson in LSM with minimal 
coupling to the matter fields ($\pi$ and $\sigma$)?
To perform this, following Kroll et al~\cite{kroll},
one assumes that the Lagrangian and its chiral transformation
properties should be such that the current generated by the chiral transformation
is proportional to the vector field itself, which leads to the relations
known as field current identities. This requirement automatically leads to
the idea of VMD due to Sakurai~\cite{sakurai}. The field current
identity is achieved by promoting the SU(2) global chiral symmetry
of vector fields to a local gauge symmetry as was done for the
first time by Yang and Mills~\cite{yang} for the isospin symmetry.
The Lagrangian for the vector field reads,
\be
{\cal L}_{lr}=\frac{1}{2}tr\mid\,F^{\mn}_l\,\mid^2 
+\frac{1}{2}tr\mid\,F^{\mn}_r\,\mid^2 
+m_0^2\,tr[(V^{\mu}_l)^2+(V^{\mu}_r)^2]
\label{lr}
\ee
where $F^{\mn}_{l,r}=\partial^\mu\,V^{\nu}_{l,r}
-\partial^\nu\,V^{\mu}_{l,r}-ig\left[V^{\mu}_{l,r},V^{\nu}_{l,r}\right]$.
In the above Lagrangian the kinetic term for the gauge fields
remains invariant under the transformation and the field-current
identity is obtained through Gell-Mann Levy theorem 
from the mass term of the gauge fields as
\be
J^{\mu}_{l,r}=\frac{m_0^2}{g}V^{\mu}_{l,r}
\ee
We note that chiral symmetry is a global one in QCD, therefore, the
local symmetry has to be broken and this is achieved precisely by the
mass term of the vector fields in the Lagrangian.
Next one has to introduce the interaction of matter fields ($\pi$ and $\sigma$)
with the gauge fields preserving the field current identity.
Noting that the ordinary derivatives occurring in Eq.~(\ref{lsm})
spoils the field-current identity, we introduce the required interactions
consistent with the gauge principle {\it i.e.} by replacing the partial
derivatives by covariant derivatives:
\be
D^\mu\Phi=\partial^\mu\Phi-ig(V^{\mu}_l\Phi-\Phi\,V^{\mu}_r)
\label{covar}
\ee
Finally the resulting Lagrangian density for the gauged LSM is 
obtained from Eqs.~(\ref{covar}), ~(\ref{lsm}) and ~(\ref{lr}) as
\be
{\cal L}_{glsm}=tr\mid\,D_{\mu}\Phi\mid^2 +\mu^2\,tr \mid\Phi\mid^2
+\frac{1}{2}\lambda\,tr(\mid\Phi\mid^2)^2-h\,tr(\Phi)
+{\cal L}_{lr}
\label{glsm}
\ee
Expanding the kinetic term for the matter field one finds
\be
tr\mid\,D_{\mu}\Phi\mid^2=\frac{1}{2}
\left[(\partial_{\mu}\sigma+g\vec a\cdot\vec\pi)^2+
(\partial_{\mu}\pi+g\vec \rho_\mu\times\vec\pi-ga^{\mu}\sigma)^2
+g^2(\sigma^2+\pi^2)(f_1^\mu)^2\right]
\label{keeq}
\ee
The above equation indicates that (i) a shift in the $\sigma$ 
field ($\sigma\ra\sigma_0+\sigma$)
gives rise to mixing between $\pi$ and $a_1$ fields 
(a term $\sim\,g\sigma_0\partial_\mu\pi\cdot\,a_1^\mu$ arises from the
second term of the r.h.s. of Eq.~(\ref{keeq})), which has  
to be eliminated by an appropriate shift in the $a_1$ field, (ii)there
is no interaction term involving $\omega$, which can come into
the picture through anomaly and (iii) the kinetic term for pion gets modified
because of the shift in the $a_1$ field. Thus to get back the
cannonical form of this term one has to renormalize the
pion field $\pi\ra\pi/\sqrt{Z_\pi}$, where $Z_\pi=m_\rho^2/m_{a_1}^2$; 
$Z_\pi{^2}=1/2$ gives the Kawarabayashi - Suzuki 
- Riazuddin- Fayyazuddin (KSRF) relation~\cite{ks,rf}.
After some algebra one gets~\cite{rdp},
$m_{\pi}^2=h/(Z_{\pi}\sigma_0)$, $m_{\sigma}^2=h/\sigma_0+2\lambda\sigma_{0}^2$,
$f_\pi=\sqrt{Z_\pi}\sigma_0$. Taking $m_\pi=137$ MeV, $m_\sigma=600$ MeV,
$m_\rho=770$ MeV and $m_{a_1}=1260$ MeV,  
we get $\sigma_0=152$ MeV, $h=(102$ MeV)$^3$,
$\mu=412$ MeV and $\lambda=7.6$.

With these inputs
Pisarski has evaluated the thermal masses of $\rho$  and $a_1$ at 
lowest order in $g$ at low temperature. We quote the results
below~\cite{rdp},
\be
m_{\rho}^2(T)\approx m_\rho^2-\frac{g^2\pi^2T^4}{45m_\rho^2}
\left[\frac{4m_{a_1}^2(3m_\rho^2+4k^2)}{(m_{a_1}^2-m_\rho^2)^2}-3\right]
\ee
\be
m_{a_1}^2(T)\approx m_{a_1}^2+\frac{g^2\pi^2T^4}{45m_\rho^2}
\left[\frac{4m_{a_1}^2(3m_{a_1}^2+4k^2)}{(m_{a_1}^2-m_{\rho}^2)^2}
+\frac{2m_{\rho}^4}{m_{a_1}^2(m_{a_1}^2-m_{\sigma}^2)}-
\frac{m_{a_1}^2}{m_{\rho}^2}\right]
\ee

In the chiral limit ($\sigma_0$ goes to zero and many of the
couplings vanish) assuming the validity of VMD in the medium
Pisarski has showed that~\cite{rdp9505} $\rho$ and $a_1$ become
degenerate with a mass value $\sim 962$ MeV ($\rho$ mass increases).
On the other hand, if one adopts a scenario where vector meson dominance
(VMD) is not 
valid in the medium then $m_\rho(T_\chi)=m_{a_1}(T_\chi)=630$
MeV ($\rho$ mass decreases). 
However, it is important to mention at this point that the chiral
symmetry can also be realized via the Georgi limit~\cite{georgi}
where the $\rho$ meson becomes massless. Pisarski~\cite{pisarski2}
has argued that the results obtained by Georgi in the non-linear
sigma model can be translated in terms of gauged linear sigma 
model without the validity of VMD, for which there is no
unique prediction for the behaviour of $\rho$ mass at non-zero
temperature. Thus the behaviour of in-medium 
$\rho$ depends on the validity of VMD in the medium. The thermal
shift of the pole of the spectral function of $\rho$ in the 
gauged LSM are shown in Fig.~(\ref{7fig9}) of section 8. 

\subsection{The gauged non-linear sigma model}
It is well-known that the global $SU(2)_l\bigotimes\,SU(2)_r$
symmetry of 2 flavours QCD is expected to be spontaneously 
broken to the subgroup $SU(2)_V$ and the pions
appear as the N-G boson. The non-linear sigma model
with $SU(2)_l\bigotimes\,SU(2)_r/SU(2)_V$ is an effective
theory of QCD for the description of pion dynamics.
The in-medium properties of vector mesons have been studied
by Song~\cite{cs,cs93} in the framework of gauged non-linear
sigma model (NLSM)~\cite{meissner}. This model 
will be discussed only briefly
because it is very similar to the gauged LSM: the
main difference is that the $\sigma$ degree of freedom
is eliminated in NLSM by the non-linear realization of
chiral symmetry as mentioned in the previous sections.
We start with the observation that a perfectly valid 
parametrization of $\Phi$ could be,
\be
U=\exp[\frac{2i}{F_\pi}\sum_a\frac{\phi_a\tau^a}{\sqrt{2}}]\equiv
\exp[\frac{2i}{F_\pi}\phi]
\ee 
where $\phi=\phi^a\tau^a/\sqrt{2}$ is the pseudoscalar 
field and $F_\pi=\sqrt{2}f_\pi$.
The Lagrangian for the NLSM
based on the manifold 
$SU(2)_l\bigotimes\,SU(2)_r/SU(2)_V$ is given by,
\be
{\cal L}_0=\frac{f_{\pi}^2}{4}
tr\left[\partial_{\mu}U\partial^{\mu}U^{\dagger}\right]
\label{nlspi}
\ee

The vector and the axial vector fields can be introduced 
as the Yang-Mills gauge as before to minimize the 
number of arbitrary parameters in 
the model. The resulting Lagrangian
is given by,
\bea
{\cal L}_{NLSM}&=&\frac{f_{\pi}^2}{4}
tr\left[\,D_{\mu}UD^{\mu}U^{\dagger}\right]
-\frac{1}{2}tr\mid\,F^{\mn}_l\,\mid^2 
-\frac{1}{2}tr\mid\,F^{\mn}_r\,\mid^2 \nonumber\\
&&+m_0^2\,tr[(V^{\mu}_l)^2+(V^{\mu}_r)^2]
\label{nlsm}
\eea
where $V_{\mu}^{l,r}=(v_\mu\pm\,a_\mu)/2$, $v_\mu$ and $a_\mu$
denote vector and axial vector fields.
To improve the phenomenology of the model, 
the following higher dimensional terms  can be added to 
the Lagrangian~\cite{cs,brh} without spoiling the 
symmetry under consideration,
\be
{\cal L}_{6dim}=-i\xi\,tr\left[\,D_{\mu}UD_{\nu}U^{\dagger}\,F^{l,\mn}
+\,D_{\mu}U^{\dagger}D_{\nu}U\,F^{r,\mn}\right]
\label{non}
\ee
where $\xi$ is a constant determined form the
decay of vector mesons~\cite{cs}. The thermal shift of the
$\rho$-mass has been evaluated by Song with pion loop, pion
tadpole and pion-$a_1$ loop resulting from the
interaction terms of the Lagrangian given in Eqs.~(\ref{nlsm}) and (\ref{non}).
It shows negligible change in the $\rho$-mass from its vacuum value.

The effective masses of $\rho$, $a_1$ and $\omega$ at non-zero 
temperature have also been calculated by Song~\cite{cs93} with
a $SU(3)_l\,\bigotimes SU(3)_r$ symmetric Lagrangian:
\bea
{\cal L}_{NLSM}&=&\frac{f_\pi^2}{4}tr\left[\,D_{\mu}UD^{\mu}U^{\dagger}\right]
-\frac{1}{2}tr\mid\,F^{\mn}_l\,\mid^2 
-\frac{1}{2}tr\mid\,F^{\mn}_r\,\mid^2 \nonumber\\
&&+m_0^2\,tr[(V^{\mu}_l)^2+(V^{\mu}_r)^2]
+\frac{1}{4}f_\pi^2\,tr\left[M(U+U^{\dagger}-2)\right]\nonumber\\
&&-i\xi\,tr\left[\,D_{\mu}UD_{\nu}U^{\dagger}\,F^{l,\mn}
+\,D_{\mu}U^{\dagger}D_{\nu}U\,F^{r,\mn}\right]\nonumber\\
&&+\kappa\,tr\left[F_{\mn}^lU\,F^{r,\mn}U^\dagger\right]
\label{su3}
\eea
where $U$ is defined as
${\displaystyle
U=\exp[\frac{2i}{F_\pi}\sum_i\frac{\phi_a\lambda^a}{\sqrt{2}}]\equiv
\exp[\frac{2i}{F_\pi}\phi]}$ and $\lambda_a$`s are Gell-Mann matrices.  The two higher dimensional
terms with co-efficients $\xi$ and $\kappa$ are added to improve
the phenomenology. Please note that although these terms retain
the gauge invariance of the model, they  spoil the renormalizability
of the model.

The dynamics of $\omega$ is governed by the 
anomalous interaction,  also known as Wess-Zumino interaction
given by
\be
{\cal L}_{anomaly}=
\frac{3g^2}{8\pi^2\,F_\pi}\,
\epsilon_{\mu \nu \alpha \beta}\partial^{\mu}{\omega}^{\nu}
\,tr[\partial^{\alpha}
\rho^{\beta}\pi]
\label{anomaly}
\ee
This is very similar to the Gell-Mann Sharp Wagner~\cite{GSW}
interaction already encountered in section~3.2.

Song~\cite{cs93} has considered the following values 
of the parameters consistent
with the vacuum properties of the vector and axial 
vector mesons: 
$(g,\kappa,\xi)=(10.30,0.34,0.45)$ and
$(6.45,-0.29,0.06)$, referred to as set I and II 
respectively.

The calculation of thermal mass shift of the vector and axial
vector mesons with all these inputs reveal that: 
(i)with parameter set I $\rho$ and $\omega$ masses increase
with different rate and $a_1$ mass decreases,
(ii) with parameter set
II the thermal mass shift of $\rho$ and $\omega$ is negligibly small 
but $a_1$ mass decreases slightly. The modification of hadronic masses
due to thermal interactions within the ambit of the model 
discussed above are presented in Fig.~(\ref{7fig9}) of section 8 . 

\subsection{The hidden local symmetry approach}
In case of the two chiral models described above the vector 
mesons are introduced as Yang-Mills field and the mass term for the
gauge boson are put in by hand which may not be entirely satisfactory,
whereas in the hidden local symmetry
(HLS) approach the $\rho$ meson is generated as a dynamical
gauge boson of a hidden symmetry in the NLSM~\cite{bandoprl,bandopr}.
It has been explicitly shown by Bando et al that, in general,
any NLSM corresponding to the manifold $G/H$ is gauge equivalent 
to a ``linear'' model having $G_{global}\bigotimes\,H_{local}$
symmetry. Accordingly, the Lagrangian of Eq.~(\ref{nlspi})
can be written in a form that exhibits, besides 
$SU(2)_l\bigotimes\,SU(2)_r$ global, a local
$SU(2)_V$ symmetry - the hidden symmetry and $\rho$ 
meson appears as a gauge boson corresponding to this symmetry.
(The axial vector $a_1$ is not included in the minimal
version of HLS Lagrangian.) To make it more explicit, we introduce
two $SU(2)$-matrix valued variables $\xi_l(x)$ and $\xi_r(x)$
with the transformation properties~\cite{bandoprl},
\be
\xi_{l,r}(x) \ra h(x)\xi_{l,r}(x)g^\dagger_{l,r}
\ee
with 
\be
U=\xi_l^\dagger\xi_r
\label{capu}
\ee
where $h(x)\in\,[SU(2)_V]_{local}$ and 
$g_{l,r}\in\,[SU(2)_{l,r}]_{global}$.
$\xi_{l,r}$ is parametrized as
\be
\xi_{l,r}=\exp[i\Sigma(x)/f_\Sigma\mp\,i\pi/f_\pi\,]
\label{unph}
\ee
where $\pi=\pi^a\,t^a$ and  
$\Sigma=\Sigma^a\,t^a$.
We note at this point that the unwanted degrees of freedom,
$\Sigma$, which have entered in the system via
Eqs.(\ref{capu}) and (\ref{unph}) are known as ``compensator''- the would be
N-G boson  which has to be eaten up by the hidden gauge boson,
$\rho$ and these extra degrees of freedom reappear as the longitudinal
polarization of the (massive) $\rho$.  
Now we define the covariant derivative as
\be
D_\mu\xi_l=\partial_\mu\xi_l-igV_\mu\xi_l+i\xi_l\,l_\mu
\ee
and
\be
D_\mu\xi_r=\partial_\mu\xi_r-igV_\mu\xi_r+i\xi_r\,r_\mu
\ee
where $l_\mu (r_\mu)$ is the external field corresponding to
the gauging of $SU(2)_l\bigotimes SU(2)_r$ and $V_\mu$ is the gauge field
corresponding to the symmetry $[SU(2)_V]_{local}$. 
With these fields two types of 
$[SU(2)_l\bigotimes\,SU(2)_r]_{global}$$\bigotimes[SU(2)_V]_{local}$
invariants can be constructed~\cite{bandoprl,bandopr,fuji},
\be
{\cal L}_{V}=-\frac{f_{\pi}^2}{4}\,
tr\,\left[D_\mu\xi_l\cdot\xi_l^{\dagger}
+D_\mu\xi_r\cdot\xi_r^{\dagger}\right]^2
\ee
\be
{\cal L}_{A}=-\frac{f_{\pi}^2}{4}\,
tr\,\left[D_\mu\xi_l\cdot\xi_l^{\dagger}
-D_\mu\xi_r\cdot\xi_r^{\dagger}\right]^2
\ee
and a linear combination ${\cal L}={\cal L}_{A}+a{\cal L}_{V}$
is equivalent to the original Lagrangian given in Eq.~(\ref{nlspi}).
By fixing the gauge $\xi_l^\dagger=\xi_r=\exp(i\pi/f_\pi)$
(and hence eliminating the unphysical degrees of freedom,
$\Sigma$) one can show that
${\cal L}_{A}={\cal L}_{0}$, 
while ${\cal L}_{V}$ vanishes when equation of motion for
$V_\mu$ is used. So far $V_\mu$ has been treated as an auxiliary field.
It is assumed that the kinetic term for this field is
generated by quantum effects or by the QCD dynamics. The full
Lagrangian with the kinetic term is
\be
{\cal L}_{HLS}={\cal L}_{A}+a{\cal L}_{V}-\frac{1}{4}\vec\varrho_{\mn}\vec\varrho^{\mn}
\label{hlslag}
\ee
where $\vec\varrho_{\mn}$ is the non-abelian field tensor for the $\rho$ meson.
The Lagrangian of Eq.~(\ref{hlslag}) can be written as~\cite{bandoprl},
\be
{\cal L}_{HLS}=\frac{1}{2}(\partial_\mu\vec\pi)^2+
\frac{1}{2}ag\vec\rho^\mu\cdot\vec\pi\times\partial_\mu\vec\pi+
\frac{1}{2}g^2af_\pi^2\vec\rho_\mu^2
-\frac{1}{4}\vec\varrho_{\mn}\vec\varrho^{\mn}+......
\ee
The above equation implies that the mass of the $\rho$ meson  
($m_\rho^2=ag^2f_\pi^2$) is generated due to SSB via Higgs mechanism
and the unphysical N-G modes ($\Sigma$ not $\pi$) are ``eaten-up''
by the gauge boson
{\it i.e.} the three extra degrees of freedom
get converted to the three longitudinal polarization
of the massive gauge boson. For $a=2$ one recovers the KSRF relation. This 
value of $a$ also results in universal coupling of $\rho$.

Harada et al~\cite{harada} have evaluated the finite temperature
effects on the $\rho$-mass upto one loop order in the HLS approach
due to the thermal pion and $\rho$ meson interactions.
Their results reveal that at high temperature the 
reduction in $\rho$ mass due to pion loop is overwhelmed by
the increase due to thermal $\rho$ loop contribution, although
the net shift is rather small. The contribution of thermal
pions to the $\rho$ self energy in this model is different from 
other calculations because in HLS approach there is no pion
tadpole contribution. We will study the effects of the mass shift of $\rho$
in this approach to the electromagnetic probes later in
this article.

%%%%%%%%%%%%%%%%%%%%% QSR %%%%%%%%%%%%%%%%%%%%%%%%%%%%%%%%%%%%%
\setcounter{equation}{0}
\def\theequation{6.\arabic{equation}}

\section{Spectral Constraints at Finite $T$}

In the previous section we have discussed the change in the hadronic properties
using effective Lagrangian approach. However, medium modifications
can also be studied by applying QCD sum rules (QSR)~\cite{BS,hl,HKL}.
As many good reviews are available on 
the QSR at zero temperature ~\cite{shifman,reinders,ioffe,ms,narison}  
after the original work of Shifman et al~\cite{svz}, 
we, therefore, very briefly introduce the basic principles of 
QSR in vacuum and then discuss the QSR at non-zero temperature. 

\subsection{QCD sum rules at zero temperature}
The basic aim of the QCD sum rule approach is to evaluate the 
resonance parameter (mass, coupling constant etc) in low energy hadronic
physics in terms of the vacuum expectation values of quantities such
as quark and gluon condensates, $\langle 0|{\bar q}q|0 \rangle$ (indicates chiral
symmetry breaking), 
$\langle 0|G_{\mu \nu}^a G^{\mu \nu a}|0 \rangle$ (signals the break
down of scale invariance), etc,  where $q(x)$ is the
quark field and $G_{\mu \nu}^a(x)$ is the gluon field tensor. 
These condensates are non-perturbative in nature and appear as 
a power corrections to the leading logarithmic (perturbative)
behaviour. These power corrections
are far more important than higher order $\alpha_s$ 
corrections~\cite{reinders}. In the following we will discuss
how the QCD sum rule approach connects the
perturbative and non-perturbative domain and 
leads to the determination of resonance parameter 
({\it i.e.} mass and the resonance
strength) of the $\rho$ meson.

The QCD sum rule approach starts with the Wilson operator 
product expansion for the time
ordered product of two (or more) currents. 
The gluon and quark condensates
appear as higher dimensional operators in the expansion. The coefficients
of this expansion contain the short distance part and the long distance part
is contained in the vacuum expectation values. The coefficient can be 
evaluated perturbatively in terms of the parameters 
($\alpha_s$ and the quark masses) of the Lagrangian used. 
We consider the time-ordered or causal current correlator
\begin{eqnarray}
\Pi_{\mu \nu}(q)&\equiv&i\,\int\,d^4x\,e^{iq\cdot x}\,\langle 0|T\left(\frac{}{}
J_{\mu}(x)J_{\nu}(0)\right)|0
\rangle
\nonumber\\
&=&(q_{\mu}q_{\nu}-q^2g_{\mu \nu})\Pi(q^2)
\label{eq1}
\end{eqnarray}
In this article we will confine our discussions for currents
with quantum numbers $J^{PC}=1^{--}, I=1$,
\begin{equation}
J_{\mu}^{\rho} = \frac{1}{2}\,({\bar u}\gamma_{\mu} u - 
{\bar d}\gamma_{\mu}d)
\label{eq2}
\end{equation} 
where $u$ and $d$ are the quark fields for up and down quark
respectively.
The analytic structure of the correlator ($\Pi$), for spacelike $Q^2 = -q^2$,
can be expressed through a dispersion relation:
\begin{equation}
\Pi(Q^2) = \frac{1}{\pi}\,\int\,\frac{{\rm Im}\Pi(s)\,ds}{s+Q^2}
+({\rm subtraction}) 
\label{eq3}
\end{equation} 
Imaginary part of $\Pi$ is proportional to the spectral density which can be 
modeled as consisting of a conspicuous resonance and a continuum with a 
sharp threshold $\omega_0$,
\begin{equation}
{\rm Im}\Pi(s) = \pi\,\sum_{\rm Res}\,{\cal G}_R\,m_R^2\,
\delta(s-m_R^2)+\frac{1}{8\pi}\left(1+
\frac{\alpha_s}{\pi}\right)\,\theta(s-\omega_0)
\label{eq4}
\end{equation} 
with a resonance strength ${\cal G}_R$ and a pole position at $m_R^2$.

The theoretical side of the sum rule is derived from an operator product
expansion for large $Q^2 = -q^2$ (deep Euclidean region where asymptotic
freedom is realized)
as has been suggested by Shifman 
et. al~\cite{svz}. Thus we write
\begin{equation}
i\,\int\,d^4x\,e^{iq\cdot x}\,T\left(\frac{}{}J_{\mu}(x)J_{\nu}(x)\right)
= C_I(q) + \sum_{n}\,C_n(q){\cal O}_n
\label{eq5}
\end{equation} 
where $I$ is the identity operator, $C$'s are the Wilson coefficients,
and ${\cal O}_n$'s are the local gauge invariant operators constructed from the 
quark and gluon fields. The operators are ordered by their increasing
dimensions and therefore, the coefficients fall off by the corresponding 
power of $q^2$. On dimensional ground one sees that the operators of dimension
$d > 0$ leads to $1/q^d$ power corrections. However, for large $Q^2 = -q^2$
a fewer power corrections ($d = 6$) is sufficient to converge the series.
Taking the vacuum expectation value of Eq.~(\ref{eq5}) we 
obtain~\cite{reinders}
\begin{equation}
\Pi_{\mu \nu} = (q_{\mu}q_{\nu}-q^2g_{\mu \nu})\Pi
\label{eq6}
\end{equation} 
where
\begin{eqnarray}
\Pi&=&-\frac{1}{8\pi^2}\left(1+\frac{\alpha_s}{\pi}\right)\,
\ln{\frac{Q^2}{\mu^2}}+\frac{1}{2Q^4}\,\langle 0|m_u{\bar u}u+m_d{\bar d}d
|0\rangle \nonumber\\
&&+\frac{1}{24Q^4}\langle 0|\frac{\alpha_s}{\pi}G_{\mu \nu}^aG^{\mu \nu a}|0
\rangle - \frac{\pi \alpha_s}{2Q^6}\langle 0|({\bar u}\gamma_{\mu}\gamma_5
\lambda^a u-{\bar d}\gamma_{\mu}\gamma_5\lambda^a d)^2|0\rangle \nonumber\\
&&-\frac{\pi \alpha_s}{9Q^4}\,\langle 0|({\bar u}\gamma_{\mu}\lambda^a u+
{\bar d}\gamma_{\mu}\lambda^a d)\,\sum_{q=u,d,s}\,{\bar q}\gamma_{\mu}\
\lambda^a q|0 \rangle
\label{eq7}
\end{eqnarray}

The left hand side (l.h.s.) is the well known two 
point Greens function which can be 
expressed in terms of the phenomenological
parameters, characterizing the strong interaction processes, consistent
with the current under consideration, via the dispersion relation.
The right hand side (r.h.s.) has been evaluated by using OPE in the short distance (asymptotic
freedom) region. The vacuum expectation value of the higher dimensional
operator appears as a power correction to the asymptotic contribution
(the first logarithmic term in the r.h.s. of the above equation).

The sum rule therefore, becomes (modulo subtractions)
\begin{equation}
\frac{1}{\pi}\,\int\,\frac{{\rm Im}\Pi(s)\,ds}{s+Q^2} = \Pi
\label{eq8}
\end{equation} 

In Eq.~(\ref{eq8}) r.h.s. corresponds to large $Q^2$ or small distance
scale with fewer power corrections and l.h.s should be saturated by the lowest
resonance, which is a long distance phenomenon. Therefore, in order to get a
balance between the two sides we would like to have a weight function which
enhances the low $Q^2$ contribution relative to the high $Q^2$ one. This can
be done by taking additional derivative with respect to $Q^2$, and then taking
$Q^2$ and the number of derivatives $n$ to infinity, we obtain Borel 
transformed sum rule. Borel transformation is equivalent to the following
mathematical operation:
\begin{equation}
{\hat L}_M\,\frac{1}{s+Q^2} = \frac{1}{M_B^2}\,e^{-s/M^2}
\label{eq9}
\end{equation} 
where
\begin{equation}
{\hat L}_M = lim_{_{_{_{\!\!\!\!\!\!\!\!\!\!\!\!\!\!\!\!\!\!\!\!\!\!\!\!\!\!\!
\stackrel{Q^2,n\rightarrow \infty} {Q^2/n=M_B^2=
{\rm const.}}}}}}
 \frac{1}{(n-1)!}\,Q^{2n}\left(-\frac{\partial}{\partial Q^2}\right)^n
\label{eq10}
\end{equation} 
and $M_B$ is the Borel mass.
Applying Eq.~(\ref{eq9}) on the l.h.s. of Eq.~(\ref{eq8}) and Eq.~(\ref{eq10})
on the r.h.s. and expressing vacuum expectation value of four fermion  
operators in terms of two fermions, we  obtain~\cite{reinders}
\bea
\int\,e^{-s/M_B^2}\,{\rm Im}\Pi(s)\,ds&=&\frac{1}{8\pi}\,M_B^2\left[
1+\frac{\alpha_s}{\pi}+\frac{8\pi^2}{M_B^4}\langle 0| m_q{\bar q}q|0\rangle\right.
\nonumber\\
&&+\left.\frac{\pi^2}{3M_B^4}\langle0|\frac{\alpha_s}{\pi}G_{\mu \nu}^a
G^{\mu \nu a}|0\rangle\right.\nonumber\\
&&-\left.\frac{448}{81}\,\frac{\pi^3\alpha_s}{M_B^6}\,
{\langle 0|{\bar q}q|0\rangle}^2\right] 
\label{eq11}
\eea
Substituting the various values of the matrix elements as given in 
Ref.~\cite{reinders} we obtain
\begin{equation}
\int\,e^{-s/M_B^2}\,{\rm Im}\Pi(s)\,ds = \frac{1}{8\pi}\,M_B^2\left[
1+\frac{\alpha_s}{\pi}+\frac{0.04}{M_B^4}-\frac{0.03}{M_B^6}\right]
\label{eq12}
\end{equation} 
Differentiating Eq.~(\ref{eq12}) with respect to $1/M_B^2$ to obtain another sum
rule:
\begin{equation}
\int\,e^{-s/M_B^2}\,{\rm Im}\Pi(s)\,sds = \frac{1}{8\pi}\,M_B^4\left[
1+\frac{\alpha_s}{\pi}-\frac{0.04}{M_B^4}+\frac{0.06}{M_B^6}\right]
\label{eq13}
\end{equation} 
In Eqs.~\ref{eq12} and \ref{eq13} the terms $M_B^{-4}$ and $M_B^{-6}$
arise due to gluon and quark condensates respectively.
Assuming that r.h.s. of Eq.~(\ref{eq4}) is saturated by the $\rho$ resonance  
we get from Eqs.~(\ref{eq12}) and 
(\ref{eq13}) 
\begin{equation}
m_{\rho}^2 = M_B^2\,\frac{\left(1+\alpha_s/\pi\right)\,\left[1-\left(1+
\omega_0/M_B^2\right)e^{-\omega_0/M_B^2}\right]
-0.04/M_B^4+0.06/M_B^6}
{\left(1+\alpha_s/\pi\right)\,[1-
e^{-\omega_0/M_B^2}]+0.04/M_B^4-0.03/M_B^6}
\label{eq14}
\end{equation} 
The above expression still depends on the Borel mass $M_B$ and the continuum
threshold $\omega_0$. 
The value of $\omega_0$ can be inferred from the data of $e^+e^-$  
annihilation. The absolute value of $\rho$ mass is then
obtained by looking for the stability plateau {\it i.e.} choosing $M_B^2$ such
that ${\partial m_{\rho}(M_B^2)}/{\partial M_B^2} = 0$. To determine the 
resonance strength for the $\rho$ meson we keep only the 
$\rho$ resonance in the sum of Eq.~(\ref{eq4}) and substitute it in Eq.~(\ref{eq12})
to obtain, after an elementary integration,
\be
4\pi\,{\cal G}_\rho=\frac{M_B^2e^{m_\rho^2/M_B^2}}{2\pi\,m_\rho^2}
\left[1+\frac{\alpha_s}{\pi}+\frac{0.04}{M_B^4}-\frac{0.03}{M_B^6}
-(1+\alpha_s/\pi)e^{-\omega_0/M_B^2}\right]
\label{eq15}
\ee
${\cal G}_\rho$ is related to $g_\rho$ as ${\cal G}_\rho=1/g_\rho^2$.

Eqs.~(\ref{eq14}) and (\ref{eq15}) indicate how the resonance 
parameters of vector mesons can be extracted 
by using QCD sum rules in vacuum. In the next section we will
briefly discuss the QCD sum rules at non-zero temperature.

\subsection{QCD sum rule at non-zero temperature}

As mentioned earlier, 
the retarded correlator has the 
required analytic properties in a thermal system. 
QCD sum rules for vector mesons in the medium \cite{hl,HKL} start with  the
retarded current correlation function (defined as $W_{\mn}$ in section 2)
\begin{eqnarray}
\Pi_{\mu \nu}^R (q_0 , {\vec q})
=i \int d^4x e^{iqx}  \theta(x_0)\langle\,[J_{\mu}(x), J_{\nu}(0)]\,\rangle\ ,
\label{correlator}
\end{eqnarray}
where  $q^\mu \equiv (q_0 , {\vec q})$ is the four momentum, 
with the source (electromagnetic) currents $J_\mu$ defined 
in terms of the quark fields as (in units of $e$),
\be
J_{\mu}  =   {2 \over 3} \bar{u} \gamma_{\mu} u
 - {1 \over 3} \bar{d} \gamma_{\mu} d
 - {1 \over 3} \bar{s} \gamma_{\mu} s 
\ee
Defining the current in the $\rho$, $\omega$ and $\phi$ channels as
\be
J_{\mu}^{\rho} = (1/2)(\bar{u} \gamma_{\mu} u - 
\bar{d} \gamma_{\mu} d), 
\ee
\be
J_{\mu}^{\omega} = (1/2)(\bar{u} \gamma_{\mu} u + 
\bar{d} \gamma_{\mu} d), 
\ee 
and 
\be
J_{\mu}^{\phi} = \bar{s} \gamma_{\mu} s ,
\ee
one can express the electromagnetic current in terms of 
$\rho$, $\omega$ and $\phi$ fields as,
\be
 J_{\mu} = J_{\mu}^{\rho} + {1 \over 3} J_{\mu}^{\omega} -
      { 1 \over 3}  J_{\mu}^{\phi},
\label{emhad}
\ee
As discussed earlier
there are two independent
invariants in the medium; the transverse ($\Pi^R_T$) and 
the longitudinal ($\Pi^R_L$) 
components of the polarization tensor.  
In the limit  ${\vec q} \rightarrow 0$, as there is no spatial direction,
$\Pi^R_T$ and $\Pi^R_L$ becomes equal ($=\Pi^R$)
 and the trace of the retarded correlation
function can be expressed in terms of $\Pi^R$ as 
$\Pi^R\,\equiv \Pi_{\mu}^{\mu R}/(-3 q_0^2)$.
  Both the transverse and the longitudinal components
 satisfy the fixed ${\vec q}$ dispersion relation. 
 In particular, at ${\vec q} =0$, 
\begin{eqnarray}
\label{dispersion2}
{\rm Re} \Pi^{R} (q_0) =
 {1 \over \pi} {\rm P} \int_0^{\infty} du^2
{ {\rm Im} \Pi^R(u) \over u^2-q_0^2} + ({\rm subtraction}). 
\label{disp0}
\end{eqnarray}
 ${\rm Re} \Pi^R$ can be calculated using perturbation
 theory with power corrections (operator product expansion (OPE))
 in the deep Euclidean region $q_0^2 \rightarrow - \infty $.
  For example, OPE for ${\rm Re} \Pi^R(q_0)$, which is the same as 
 the OPE for the causal (Feynman) correlator  $\Pi^F(q_0)$,
 has a general form at $q_0^2 \equiv - Q^2 \rightarrow - \infty$,
\begin{eqnarray}
 {\rm Re} \Pi^R(Q^2 \rightarrow - \infty)
 =  - C_0 \ln Q^2 + \sum_{n=1}^{\infty}
 {C_n (\alpha_s(\mu^2), \ln (\mu^2/Q^2)) \over Q^{2n}} \langle {\cal O}_n
(\mu^2)
\rangle_{_T}
\
\
\ ,
\label{ope}
\end{eqnarray}
where $\mu$ is the renormalization point of the local composite operators, 
If $|Q|$ is much larger than other soft scales present in the system such 
as  $\Lambda_{QCD}$ and $T$, perturbative QCD can be used to calculate the
relevant quantities.
 $C_n$ are the c-number Wilson coefficients which are
 $T$ independent. All the medium effects are contained in the
 thermal average of the local operators ${\cal O}_n$.
 Since $\langle {\cal O}_n \rangle_{_T} \sim T^{2l}\cdot \Lambda_{QCD}^{2m}$
 with $l+m=n$ due to dimensional reasons, (\ref{ope})
 is a valid asymptotic expansion as long as
 $Q^2 \gg T^2$ and $\Lambda_{QCD}^2$. 
 The local operators ${\cal O}_n(\mu^2)$
 in the vector meson sum rule are essentially 
 the same with those 
 in the lepton-nucleon deep inelastic scattering (DIS) and
 can be characterized by their canonical dimension ($d$) and the 
twist ($\tau$=dimension-spin).   They are given in Ref.~\cite{HKL}
  up to dimension 6 operators and we will not recapitulate them here.
 For ${\vec q} \rightarrow 0$,
 Eq.~(\ref{ope}) is  an asymptotic series in $1/{Q}^2$
 or equivalently  an expansion with respect to  $d$.
  The  medium condensates $\langle {\cal O}_n (\mu^2) \rangle_{_T}$  may be
evaluated by
 low energy theorems, the parton distribution of hadrons
 and lattice QCD simulations. 

 Matching the l.h.s. and the r.h.s. of Eq.~(\ref{dispersion2}) 
 in the asymptotic region $q_0^2 \rightarrow - \infty $
 is the essential part of QSR. This procedure 
 gives constraints on the spectral integral and hence the 
 hadronic properties in the medium as well as in the vacuum.
 There are two major procedures for this matching, namely 
 the Borel sum rules (BSR)~\cite{svz}(discussed in the previous section) and the
  finite energy sum rules (FESR) \cite{kpt},
 which can be summarized as
\begin{eqnarray}
\label{sumrules}
\int_0^{\infty}
 & dq_0^2\ \ W(q_0^2)& \ [{\rm Im} \Pi^R(q_0)
 - {\rm Im} \Pi^R_{_{OPE}}(q_0) ] =0 ,\\ 
 & W(q_0^2) & = \left\{ \begin{array}{ll}
                    q_0^{2n} \ 
 \theta(\omega_0 -q_0^2) & \ \ \ \ \ \ \ \ \  ({\rm FESR}), \\ 
\nonumber
                    e^{-q_0^2/M_B^2} & \ \ \ \ \ \ \ \ \ ({\rm BSR}).
                   \end{array}     \right.
\end{eqnarray}
Here ${\rm Im} \Pi^R_{_{OPE}}(q_0)$ is a hypothetical imaginary part of
 $\Pi^R$ obtained from OPE and $M_B$ is the Borel mass.
 
We have seen earlier that in QSR in the vacuum, the spectral function
 (i.e. ${\rm Im} \Pi^R$ in  Eq.~(\ref{disp0}))
 is modeled with a  resonance pole and the 
 continuum to extract the mass and decay constant of hadrons.
 In the medium, such a simple parametrization is not always
 justified because of the thermal broadening of the spectrum and
 also because of the new spectral structure due to Landau damping and the 
 thermal mixing among mesons. 
  Therefore, the model independent constraints obtained from
 QSR are only for the weighted spectral integral.
 
 For example, the first three finite energy sum rules at finite $T$
 read \cite{HKL}
\begin{eqnarray}
I_1 & = & \int_0^{\infty}
 [{\rm Im} \Pi^R(q_0) - {\rm Im} \Pi^R_{_{OPE}}(q_0)] \ dq_0^2 = 0,
  \\
 I_2 & = & \int_0^{\infty}
 [{\rm Im} \Pi^R(q_0) - {\rm Im} \Pi^R_{_{OPE}}(q_0)]\ q_0^2 
  \ dq_0^2 = - C_2 \langle {\cal O}_2 \rangle_{_T}, \\
I_3 & = & \int_0^{\infty}
 [{\rm Im} \Pi^R(q_0) - {\rm Im} \Pi^R_{_{OPE}}(q_0)] \ q_0^4 
 \ dq_0^2 = C_3 \langle {\cal O}_3 \rangle_{_T} .
\end{eqnarray}
 Similar sum rules hold  for the axial vector
 channel (in the chiral limit)
 except that one has a different operator for ${\cal O}_3$.
 One can also generalize the above  sum rules to finite
 ${\vec q}$ \cite{KS,shlee}.
  
 Explicit forms of $C_n \langle {\cal O}_n \rangle_{_T} $ have been
 calculated as \cite{HKL}
\begin{eqnarray}
C_0 & = & - {1 \over 8 \pi} (1+ {\alpha_s \over \pi}) , \ \ \ 
 C_1  =  0 ,  \\
\label{dim4}
C_2 \langle {\cal O}_2 \rangle_{_T} & = & {1 \over 24} 
\langle {\alpha_s \over \pi} G^2 \rangle_{_T} +
 {4 \over 3} \langle {\cal S} \bar{q} i \gamma_0 D_0 q \rangle_{_T} , \\
C_3 \langle {\cal O}_3 \rangle_{_T} & = & -
 \langle {\rm scalar\,4-quark} \rangle_{_T}
 +
 {16 \over 3} \langle {\cal S} \bar{q} i \gamma_0 D_0 D_0 D_0 q \rangle_{_T} . 
\end{eqnarray}
Here we have neglected the terms proportional to the
light quark masses (chiral limit) and the quark-gluon mixed operators.
The operator ${\cal S}$ is used to make the operators symmetric and traceless.
At low $T$, one may use the soft pion theorems
 and the parton distribution of the pion to estimate the r.h.s.
 of the above equations.
 When $T$ is close to $T_c$, one has to look for a totally different
 way of estimation: the simplest approach is to assume the resonance gas
 to evaluate the r.h.s., while the direct lattice simulations
  will be the most reliable way in the future.
 An important feature of the OPE in the above is the
 appearance of local operators with Lorentz indices.  This happens because
 we are taking the rest frame of the heat bath which breaks
  covariance.  

 The sum rules $I_i$ can be used to check the 
 validity of the calculations of the spectral functions
 using effective theories of QCD.  This is in fact quite 
 useful for the spectral function at finite baryon density.
 At finite $T$, especially near the critical point,
 the behaviour of the condensates with dimension $d\ge 4$ is
 not known precisely.  Therefore, it is rather difficult to
 make a strong argument on the spectral constraints near $T_c$
 at present. The future lattice simulations of these
 condensates are highly called for.

\subsection{Parametrization of the spectral functions}

In this section we will introduce a 
 parametrization of the correlator at finite $T$.
 The parametrization
 should be consistent with the experimental
  data from $e^+e^-\rightarrow
hadrons$ processes at zero $T$, and it should
 be also consistent with the  
 high energy behaviour known from perturbative QCD at $q_0 \gg T$.

As the vector mesons appear as resonances in the electromagnetic 
correlator, using Eqs.~(\ref{correlator}) and (\ref{emhad}) we can write,  
\be
{\rm Im} \Pi_{\mu \nu}^{R} = 
 {\rm Im} \Pi_{\mu \nu}^{\rho,R} +
 {1 \over 9} {\rm Im} \Pi_{\mu \nu}^{\omega,R} +
 {1 \over 9} {\rm Im} \Pi_{\mu \nu}^{\phi,R}.
\label{rcorr}
\ee
The above equation shows that the contributions of $\omega$ 
and $\phi$ mesons to the electromagnetic probes are down by
almost an order of magnitude compared to $\rho$ meson. 

As mentioned before,
at  zero three momentum
the imaginary part of the trace of the retarded correlator
can be written in terms of its longitudinal component
as,
\be
{\rm Im} \ \Pi^R_{\mu \mu}(q_0) 
= -3 \ {\rm Im} \ \Pi^R_L (q_0)
= -3 q_0^2 \ {\rm Im} \ \tilde{\Pi}^R_L (q_0).
\label{cormm}
\ee
Thus our next task is to parametrize ${\rm Im} \tilde{\Pi}^R_L (q_0)$.
 We take a
Breit-Wigner form with an energy-dependent width for the resonance
along with a continuum:
\be
 {\rm Im} \ \tilde{\Pi}_L^{R,\rho} (q_0, {\vec q}=0)
 = f_{\rho}^2 {D_{\rho} \over (q_0^2 - m_{\rho}^2)^2
 + D_{\rho}^2 } + {1 \over 8 \pi}(1+ {\alpha_s \over \pi })
 {1 \over 1 + e^{(\omega_0 - q_0)/\delta}}.
\label{parametrho}
\ee
 At zero $T$, 
 this reduces to a relativistic generalization of the parametrization
used by Shuryak~\cite{shuryak} to fit the experimental data of
$e^+e^-\rightarrow  hadrons$. 
 Here $D_{\rho}$ is 
the imaginary part of the self-energy which should in principle
contain all the channels which can destroy or create
a $\rho$ in the thermal bath. 
Hence $D_\rho$ is given by the difference of the decay-width and the formation 
width so that 
$D_\rho=q_0 \Gamma(q_0)$.  
However, we have seen
that for a baryon free matter the most dominant
contribution to $D_\rho$ comes from the pion-loop~\cite{ja}.
For a $\rho$ meson propagating
with energy $q_0$ and three momentum $\vec q$ the $\rho$ width is given by  
\bea
\Gamma_{\rho\,\rightarrow\,\pi\,\pi}(q_0, {\vec q}) 
&=& {g_{\rho \pi \pi}^2 \over 48\pi}\,W^3(s)\,\frac{s}{q_0}\,
\left[1+\frac{2T}{W(s)\sqrt{q_0^2-s}}\right.\nonumber\\
&&\times\left.\ln\left\{\frac{1-\exp[-\frac{\beta}{2}(q_0+W(s)
\sqrt{q_0^2-s})]}
{1-\exp[-\frac{\beta}{2}(q_0-W(s)\sqrt{q_0^2-s})]}\right\}
\right]
\eea
where $s = q_0^2-{\vec q}^2$ and $W(s) = \sqrt{1-4m_{\pi}^2/s}$.
In the limit $|\vec q| \rightarrow 0$ , the above expression reduces to 
the in-medium decay width  and is given by
\be
\Gamma_{\rho\,\rightarrow\,\pi\,\pi}(q_0) = 
\frac{g_{\rho\,\pi\,\pi}^2}{48\pi}\,q_0 \,W^3(q_0)
\left[\left(1+f_{BE}(\frac{q_0}{2})\right)\,\left(1+f_{BE}(\frac{
q_0}{2})\right)-f_{BE}(\frac{q_0}{2})f_{BE}(\frac{q_0}{2})
\right]
\label{width}
\ee
with $f_{BE}(x) = [e^x -1]^{-1}$,
$\omega_0$ is the continuum threshold above which the asymptotic 
freedom is restored and
$f_\rho$ is the coupling between electromagnetic current and
the $\rho$ field defined as 
\be
\langle 0\mid J_\mu^\rho \mid \rho \rangle=f_\rho m_\rho\epsilon_\mu 
\ee
Assuming vector dominance in the medium  we obtain,
\be
g_{\rho} = m_{\rho}/ f_{\rho}
\ee

In the vacuum, the standard parameters for the $\rho$ spectral
function are given by,
$m_{\rho}=0.77$ GeV, $m_{\pi}=0.14$ GeV,
$f_{\rho}=0.141$ GeV, 
$g_{\rho}=5.46$,
$\omega_0=1.3$ GeV,
$\delta=0.2$ GeV and
$\alpha_s = 0.3$. 
The resulting spectral function for the $\rho$-meson {\em in the vacuum}
should be compared with Ref.~\cite{shifman}.

Let us now concentrate on the spectral function in the $\omega$ channel.
We again take a Breit-Wigner form along with a continuum:
\be
 {\rm Im} \ \tilde{\Pi}_L^{R, \omega} (q_0, \vec q=0)
 = f_{\omega}^2 {D_{\omega} \over (q_0^2 - m_{\omega}^2)^2
 + D_{\omega}^2 } + {1 \over 8 \pi}(1+ {\alpha_s \over \pi })
 {1 \over 1 + e^{(\omega_0 - q_0)/\delta}}
\label{parametomg}
\ee
where
$f_{\omega}$ is a coupling of the current with the $\omega$-meson
 defined as 
\be 
\langle 0 \mid J_{\mu}^{\omega} \mid \omega \rangle =
 f_{\omega} m_{\omega} \epsilon_{\mu}.
\ee
Note that $f_{\omega}$ here is defined as 
 factor 3 larger than Shuryak's definition~\cite{shuryak} .
$D_{\omega}$, which is 
 the imaginary part of the self-energy, is calculated 
using the Lagrangian density given in Eq.~(\ref{etaro}).
We have shown in earlier calculations~\cite{ja,pr} that a
substantial contribution to the $\omega$ width comes from the process
$\omega\pi\ra\pi\pi$ in a thermal bath. Consequently
\be
D_{\omega}(q_0) = 
q_0 (\Gamma_{\omega\ra 3\pi}+\Gamma_{\omega\pi\ra\pi\pi})
\ee
where
\be
\Gamma_{\omega\ra 3\pi}(q_0)=C\,\int_{w_{min}}^{w_{max}}\,dw\,
\int_{x_{min}}^{x_{max}}\,dx
\mid\,F\,\mid^2\,S
\ee
$S$ is the phase space factor for thermal equilibrium, given by
\be
S=\left[(1+f_{BE}(E_1))(1+f_{BE}(E_2))(1+f_{BE}(E_3))-
f_{BE}(E_1)f_{BE}(E_2)f_{BE}(E_3)\right]
\ee
and
\be
C=\frac{g_{\omega\rho\pi}^2\,
g_{\rho\pi\pi}^2\,q_0}{48\pi^3\,m_{\pi}^2}
\ee
The limits of integration are 
\bea
w_{\s min}&=&m_\pi,\nonumber\\
w_{\s max}&=&{(q_0^2-3m_\pi^2)}/{2\,q_0},\nonumber\\ 
x_{\s max}&=&\sqrt{{0.5\omega\,(w-w_{\s max})(w^2-m_\pi^2)}
/{(2q_0 \,w-q_0^2-m_\pi^2)}},\nonumber\\
x_{\s min}&=&-x_{\s max}, \nonumber\\
E_1&=&w,\nonumber\\
E_2&=&x+(q_0-w)/2,\nonumber\\
E_3&=&-x+(q_0-w)/2,\nonumber\\
\mid\,\vec p_i\,\mid&=&\sqrt{E_i^2-m_\pi^2},
\eea
and $\vec p_i$ is the pion 3-momentum.
The amplitude for the process is
\be
\mid F \mid^2=\mid \vec p_1 \mid^2\mid \vec p_2 \mid^2
(1-Z_0^2)H
\ee
where
\be
Z_0=\frac{\omega^2+m_\pi^2-2\omega(E_1+E_2)+2E_1E_2}
{2\mid \vec p_1\vec p_2 \mid}
\ee
and
\be
H=\sum_{i=1}^6\,h_i
\ee
with
\bea
h_1&=&\frac{1}{q_{12}^2+m_\rho^2\,\Gamma_\rho^2}\nonumber\\
h_2&=&\frac{1}{q_{13}^2+m_\rho^2\,\Gamma_\rho^2}\nonumber\\
h_3&=&\frac{1}{q_{23}^2+m_\rho^2\,\Gamma_\rho^2}\nonumber\\
h_4&=&2(q_{12}q_{13}+m_\rho^2\Gamma_\rho^2)h_1h_2\nonumber\\
h_5&=&2(q_{13}q_{23}+m_\rho^2\Gamma_\rho^2)h_2h_3\nonumber\\
h_6&=&2(q_{12}q_{23}+m_\rho^2\Gamma_\rho^2)h_1h_3\nonumber\\
q_{12}&=&(E_1+E_2)^2-\vec p_3^2-m_\rho^2\nonumber\\ 
q_{13}&=&(E_1+E_3)^2-\vec p_2^2-m_\rho^2\nonumber\\ 
q_{23}&=&(E_2+E_3)^2-\vec p_1^2-m_\rho^2 
\eea
The width for $\omega\pi\ra\pi\pi$ is calculated analogously.

In the vacuum the standard parameters for $\omega$ are as follows:
$m_{\omega} =  0.782$ GeV,  
$m_{\pi}  =  0.14$  GeV, 
$f_{\omega}  = 0.138$ GeV, 
$\omega_0  = 1.1$ GeV, 
$\delta  =  0.2$ GeV and
$\alpha_s  =  0.3$. 

In a medium at finite $T$, we simply replace
 $m_{\rho}$, $\omega_0$,$f_{\rho}$ and
$g_{\rho}$ by the corresponding effective quantities
(denoted by asterix)
$m_{\rho}^*$, $\omega_0^*$, $f_{\rho}^*$ and $g_{\rho}^*$ respectively.
Since not much is known about the critical behaviour of the 
 scalar and tensor condensates at finite $T$ in QCD sum rules we take
 a simple ansatz for in-medium quantities for their $T$-dependence.
A possible parametrization of $*$-quantities at finite $T$ is
\be
{m_{V}^* \over m_{V}}  = 
{f_{V}^* \over f_{V}} = 
{\omega_{0}^* \over \omega_{0}}  =
 \left( 1 - {T^2 \over T_c^2} \right) ^{\lambda},
\label{anst}
\ee
where $\lambda$ is a sort of {\em dynamical} critical exponent and
$V$ stands for vector mesons ($\rho$ and $\omega$).
 (Note that there is no definite reason to believe that all the in-medium
 dynamical quantities are dictated by a single exponent $\lambda$.
 This is a simplest possible ansatz.)
 Since the numerical value of $\lambda$ is not known, 
 we take two typical cases:
 $\lambda=1/6$ (BR scaling) and $1/2$ 
 (Nambu scaling)~\cite{brpr}. The effective mass of $a_1$
is estimated by using Weinberg`s sum rules~\cite{weinberg}.

 Some remarks are in order here:

\noindent
(i) Eq.~(\ref{anst}) for $m_{\rho}^*$ is not entirely consistent 
 with the low temperature theorem~\cite{mdey},
 which says there should be no $O(T^2)$ correction to the mass.
 Therefore, one cannot take the ansatz too seriously at low $T$.
 In practical applications, however,
  $T < 100$ MeV is not relevant in any way since it is below
 the freeze-out temperature. 

\noindent
(ii) Local duality constraint $I_1$ in QCD sum rules implies that
 $(f_{\rho}^*)^2 = 8 \pi^2 (1+\alpha_s/\pi) 
 (\omega_0^*)^2 + {\rm (scattering\, term)}$~\cite{HKL}.
 This  condition is slightly violated for $f_{\rho}^*$ in Eq.~(\ref{anst})
 because of the existence of the scattering term (Landau damping).
 
\noindent
(iii) The assumption of vector dominance in the medium together with 
 Eq.~(\ref{anst}) simply leads to  $g_{\rho}^* = g_{\rho}$.

Under these reservations, we will use the parametrized spectral
 functions (BR scaling and Nambu scaling) in the calculation of 
 the  lepton and photon
 productions in later sections.
 Major qualitative difference between the spectral function
 in the effective Lagrangian approaches and that in this section is
 the existence of the continuum and its medium modification 
 at finite $T$.

As mentioned in the introduction the photon and dilepton emission
is determined by the retarded correlator of the electromagnetic 
current. 
Dilepton emission involving the $\rho$ and $\omega$ mesons is thus obtained by
inserting
$e^2$ times Eqs.~(\ref{parametrho}) and (\ref{parametomg}) in Eq.~(\ref{drcor1})
using Eq.~(\ref{cormm}).

%%%%%%%%%%%%%%%%%%%%%% space-time evolution%%%%%%%%%%%%%%%%%%%%
\setcounter{equation}{0}
\def\theequation{7.\arabic{equation}}

\section{Evolution Dynamics}

As mentioned earlier, in URHIC the produced matter will either be in the form
of a hot hadronic gas or a quark gluon plasma.
So far we have talked about the rate of photon and dilepton  emission 
per unit time
from unit volume of a thermal system made up of quark matter and  hadronic matter
at a fixed temperature $T$. Our next task is to consider its evolution in
space and time. This is done using relativistic hydrodynamics. 
A basic ingredient of the hydrodynamic description of the 
collision volume is the existence of a strong interaction 
time scale,
\be
\tau_i\sim \frac{1}{\Lambda_{QCD}}\sim {\s 1 fm/c}\,\sim\,\tau_{\s formation}
\ee
In any hadronic collision the produced fragments can only interact
after a proper time $\tau_i$ has elapsed after their collisions.
Thus, there is another time scale in the problem, 
the so called transit time , which is defined as
\be
\tau_{\s transit}\sim\frac{2R_A}{\gamma_{\s cm}}
\ee
$R_A$ is the nuclear radius, $\gamma_{\s cm}$ is the Lorentz factor.
If the value of $\gamma_{\s cm}$ (which is a function of the 
collision energy)
is such that $\tau_{\s transit}<\tau_{\s formation}$ then most of the
secondaries are formed after the nuclei pass through each other.
Consequently these secondaries
will not contribute to the energy density of the 
fluid in the central region. Such a scenario may be realized at RHIC (Relativistic
Heavy Ion Collider) and
LHC (Large Hadron Collider) energies. 
This particular feature has been taken into account 
in Bjorken's hydrodynamic model~\cite{bjorken}. 

\subsection{Bjorken's hydrodynamical model} 

It has been observed experimentally that the particle spectra  for
the secondaries produced in $N-N$ collisions exhibit a central plateau in the
rapidity space. This kind of behaviour
is due to the frame independence symmetry
of the hydrodynamic expansion of the system~\cite{chiu}. 
Bjorken assumed that the same kind of
plateau will also be observed in nucleus nucleus collisions~\cite{bjorken}. 
In terms of the initial condition this means that the energy density, 
pressure etc  (all the thermodynamic quantities) will be a function of 
the initial thermalization (proper) time $\tau_i$
only and {\it will not} depend on the space time rapidity $\eta$
(defined later). This initial symmetry of the thermodynamic quantities
is preserved throughout the evolution scenario. 
If the particle rapidity density is flat
or invariant under Lorentz boosts then the entropy density ($s$)
will be independent of the rapidity. Since our discussion is 
limited to the baryon free region, there is
only one independent thermodynamic variable $T$, say. Once $s$ is
independent of Lorentz boost (rapidity) so are all the thermodynamic
quantities. 

\begin{figure}
\centerline{\psfig{figure=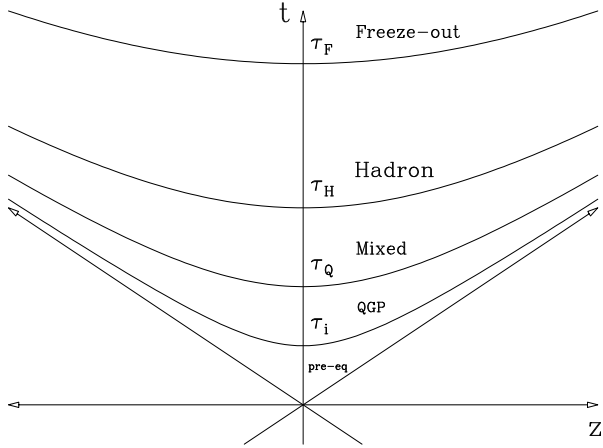,height=6cm,width=8cm}}
\caption{Space-time diagram of the collision in Bjorken hydrodynamics} 
\label{t-z}
\end{figure}

The evolution of the fluid is governed by the energy momentum
conservation equation
\be
\partial_\mu\,T^{\mn}=0
\label{eom1}
\ee
where $T^{\mn}=(\epsilon+P)u^{\mu}u^{\nu}\,+\,g^{\mn}P$ is the energy
momentum tensor for ideal fluid. For an isentropic flow the entropy 
conservation reads
\be
\partial_\mu\,s^{\mu}=0
\label{eom2}
\ee
where $s^\mu=s\,u^\mu$ is the entropy current.
Let us consider the frame independence symmetry in a two dimensional
sub-space ($t-z$ plane).
Changing the independent variables from ($t,z$) to ($\tau,\eta$) using
\be
\tau\equiv\sqrt{t^2-z^2};\,\,\,\,\,\eta\equiv\frac{1}{2}\ln\frac{t+z}{t-z}
\ee
the equation of motion Eqs.~(\ref{eom1}) and (\ref{eom2}) become
\be
\frac{\partial}{\partial\tau}\left(s\tau \cosh(y-\eta)\right)+
\frac{\partial}{\partial\eta}\left(s\sinh(y-\eta)\right)=0
\label{eom3}
\ee
\be
\frac{\partial}{\partial\tau}\left(T\tau \sinh(y-\eta)\right)+
\frac{\partial}{\partial\eta}\left(T\cosh(y-\eta)\right)=0
\label{eom4}
\ee
The independent variable $\tau$, by definition is the proper time 
of the frame which
is related to the c.m. frame by a Lorentz transformation along the $z$-axis
with velocity $z/t$. The variable $\eta$, known as the space time rapidity, 
becomes equal to the fluid rapidity $y(={1\over 2}\ln(1+v_z)/(1-v_z)$).
Putting $y=\eta$ in Eqs.~(\ref{eom3}) and (\ref{eom4}) we get
\be
\frac{\partial}{\partial\tau}\left(s\tau\right)=0
\label{enscale}
\ee
\be
\frac{\partial T}{\partial\eta}=0
\ee
These equations imply that $T$ is independent of $\eta$ and so are all the
thermodynamic quantities and $s\tau=$const.
This is the Bjorken's scaling solution. The resulting space-time
picture of the collision is shown in Fig.~(\ref{t-z}).
%%%%%%%%%%%%%% Fig. 5 %%%%%%%%%%%%%%%%%%%%%%%%%%%%%
%%%%%%%%%%%%%%%%%%%%%End Figure 5%%%%%%%%%%%%%%%%%%%%%%%%%%%%%%%

It may be noted that the above results were obtained without any specific
input from the equation of state (EOS),
it is thus a general result that one dimensional similarity flow is 
necessarily isentropic even if there is a phase transition.
For a relativistic massless
gas with statistical degeneracy $g_k$, 
$s$ and $T$ are related through the equation of state:
\be
s=4\frac{\pi^2}{90}g_kT^3
\label{sT}
\ee
Putting this expression for entropy density in the Bjorken scaling 
solution we get $T^3\tau =$const.
This is the cooling law which is extensively used to evaluate 
the signals of QGP. 
The initial temperature of the system is determined by observing that
the variation 
of temperature from its initial value  $T_i$ to final value 
$T_f$ (freeze-out temperature) with proper time ($\tau$) is governed 
by the entropy conservation (Eq.~(\ref{enscale})) 
\be
s(T)\tau=s(T_i)\tau_i
\label{entro1}
\ee
The entropy density is then expressed in terms of the observed particle (pion) multiplicity.
Using Eqs.~(\ref{sT}) and (\ref{entro1}) one gets the initial temperature as
\be
T_i^3=\frac{2\pi^4}{45\zeta(3)\pi\,R_A^2 4a_k\tau_i}\frac{dN_\pi}{dy}
\label{initemp}
\ee
where $dN_\pi/dy$ is the total pion multiplicity, $R_A$ is the radius
of the system,
$\tau_i$ is the initial thermalization time, 
and $\zeta(3)$ is the Reimann zeta function. 
$a_k=({\pi^2}/{90})\,g_k$ is the degeneracy of the produced system
and hence $k$ stands for either QGP or a hot hadronic gas. 
The rapidity density for the secondaries is obtained from\cite{heinz},
\be
\left(\frac{dN}{dy}\right)_{A-A}=
A^{\alpha}\left(\frac{dN}{dy}\right)_{p-p}
\ee
where $\alpha$ is known as the rescattering parameter.
$dN/dy\mid_{p-p}$ can be parametrized
to fit the experimental data in the central region as a function
of the centre of mass energy,$\sqrt{s}$ as
\be
\left(\frac{dN}{dy}\right)_{p-p}=0.8\ln(\sqrt{s})
\ee
The assumption of a central plateau in the rapidity distribution is not
experimentally observed in nucleus nucleus collisions at the presently available
energies. Hence the boost invariant hydrodynamics may not be a valid concept
at these energies. The concept of complete stopping in 
Landau model~\cite{landau} is not valid either at these energies. 
The physical situation
may be in between the boost invariant model of Bjorken and the Landau
model of complete stopping, which means that there may be an overlap between
the formation zone and the collision zone. 
The  Bjorken model will be used in this work
to describe the space time evolution of matter formed in URHIC.
Appropriate generalization has been made to take into account
the temperature dependent hadronic masses.

\subsection{Initial conditions and equation of state}

The set of hydrodynamic equations is not closed by itself;
the number of unknown variables exceeds the number of equations by one.
One thus needs to postulate a functional relation between any two  
variables so that the system becomes deterministic. 
The most natural course is to look for
such a relation between the pressure $P$ and the energy density $\epsilon$, as
is done in the case of thermal equilibrium. 
Under the assumption of local thermal equilibrium, this functional relation
between $P$ and $\epsilon$ is the Equation Of States (EOS).
Obviously, different EOS's will govern the hydrodynamic
flow quite differently~\cite{pasi} and as far as the search for QGP is concerned, 
the goal is to look for
distinctions in the observables due to the different EOS's (corresponding
to the novel state of QGP vis-a-vis that for the usual hadronic matter). 
It is thus imperative to understand in what respects the two EOS's differ 
and how they affect the evolution in space and time. Recently, the sensitivity
of the photon emission rate on various evolution scenarios has been studied 
in Ref.~\cite{pp}.

A physically intuitive way of understanding the role of the EOS in governing
the hydrodynamic flow lies in the fact that the velocity of sound $c_{s}^{2}
=(\partial P/\partial\epsilon)_s$ sets an intrinsic scale in the hydrodynamic
evolution. One can thus write a simple parametric form for the EOS:
$P=c_{s}^{2}(T)\epsilon$. 
Inclusion of interactions, however, may drastically alter the value of 
$c_{s}^{2}$~\cite{ah}.
In our calculation we assume the MIT bag model equation of state
for the QGP where the energy density and pressure are given by
\be
\epsilon_Q=g_Q{\pi^2T^4\over 30}+B,
\ee
and
\be
P_Q=g_Q\frac{\pi^2}{90}T^4-B.
\ee
The effective degrees of freedom in QGP, $g_Q=37$ for two flavours.
The entropy density $s_Q$ is given by $s_Q=2g_Q(\pi^2/45)T^3$. Putting
$a_k\equiv a_Q=(\pi^2/90)g_Q$ the initial temperature for a system
produced as QGP can be determined from Eq.~(\ref{initemp}).

In the hadronic phase we have to be more careful about the presence
of heavier particles and the change in their masses due to finite temperature
effects.
The ideal limit of treating the hot hadronic matter as a gas of pions 
originated from the expectation that in the framework of local 
thermalization the system would be dominated by the lowest mass hadrons 
while the higher mass resonances would be
Boltzmann suppressed. 
Indirect justification of this assumption comes from the experimental
observation in high energy collisions that most of the secondaries are pions.
Nevertheless, the temperature of the system is 
higher than
$m_{\pi}$ during a major part of the evolution and at these temperatures the
suppression of the higher mass resonances may not be complete. It may 
therefore be more realistic to include higher mass resonances in the 
hadronic sector, their relative
abundances being governed by the condition of (assumed) thermodynamic 
equilibrium.
We assume that 
the hadronic phase consists of $\pi$, $\rho$, $\omega$, $\eta$, $a_1$  
mesons and nucleons. The nucleons and heavier mesons are expected 
to play an important role in the EOS in a scenario where mass of the 
hadrons decreases with temperature. 

The energy density and pressure
for such a system of mesons and nucleons are given by
\be
\epsilon_H=\sum_{h={\s mesons}} \frac{g_h}{(2\pi)^3} 
\int d^3p\,E_h\,f_{BE}(E_h,T)
+\frac{g_N}{(2\pi)^3} 
\int d^3p\,E_N\,f_{FD}(E_N,T)
\ee
and
\be
P_H=\sum_{h={\s mesons}} \frac{g_h}{(2\pi)^3} 
\int d^3p\frac{p^2}{3\,E_h}f_{BE}(E_h,T)
+\frac{g_N}{(2\pi)^3} 
\int d^3p\frac{p^2}{3\,E_N}f_{FD}(E_N,T)
\ee
where the sum is over all the mesons under consideration and $N$ stands
for nucleons and $E_h=\sqrt{p^2 + m_h^2}$.          
The entropy density is then
\be
s_H=\frac{\epsilon_H+P_H}{T}\,\equiv\,4a_{\s{eff}}(T)\,T^3
= 4\frac{\pi^2}{90} g_{\s{eff}}(m^\ast(T),T)T^3
\label{entro}
\ee
where  $g_{\s eff}$ is the effective statistical degeneracy.
Thus, we can visualize the finite mass of the hadrons
having an effective degeneracy $g_{\s{eff}}(m^\ast(T),T)$. 
Because of the temperature dependence of the effective degeneracy
Eq.~(\ref{initemp}) has to be solved self consistently in order to
calculate the initial temperature of the system initially 
produced as a hot hadronic gas. We thus solve the equation
\be
\frac{dN_\pi}{dy}=\frac{45\zeta(3)}{2\pi^4}\pi\,R_A^2 4a_{\s{eff}}(T_i)T_i^3\tau_i
\label{dndy}
\ee
where
$a_{\s{eff}}(T_i)=({\pi^2}/{90})\,g_{\s{eff}}(m^\ast(T_i),T_i)$ .
The change in the expansion dynamics
as well as the value of the initial temperature due
to medium effects enters the calculation of the
photon emission rate through the effective statistical degeneracy.

If  the  energy/entropy density in the fireball
immediately  after  the  so-called  ``formation  time" $\tau_i$ is
sufficiently high, then the matter exists in the form of  a  QGP.
As  the  hydrodynamic expansion starts, the system begins to cool
until  the  critical  temperature  $T_c$  is  reached at a time
$\tau_Q$.  At  this
instant, the phase transition  to  the  hadronic  matter  starts.
Assuming  that  the  phase transition is a first
order one, the released latent heat maintains the temperature  of
the  system  at  the  critical temperature $T_c$, even though the
system continues to expand; the cooling due to expansion is compensated
by the latent heat liberated during the process.
Together with the possible explosive
events,  we are neglecting  the  scenarios  of  supercooling or
superheating. This process continues until all the matter has
converted  to the hadronic phase at a time $\tau_H$, the
temperature remaining constant at $T=T_c$; 
from then on,
the system continues to expand, governed by  the  EOS  of  the  hot
hadronic  matter  till the freeze-out temperature $T_f$
at the proper time $\tau_f$. Thus the
appearance of the so called mixed phase at $T=T_c$, when QGP  and
hadronic  matter  co-exist,  is a direct consequence of the first
order phase transition. Apart from the role in  QGP  diagnostics,
the  possibility of the mixed phase affects also the bulk
features of the evolution process. 

In the mixed phase, the relative proportion of QGP  and  hadronic
matter must be a function of time; initially the system consists
entirely of QGP and at the end, entirely of hot  hadronic  matter.
If we denote the fraction of the QGP
by $f_Q(\tau)$, then the entropy in the mixed phase ($s_{\s mix}$) 
can be expressed as,
\be
s_{\s mix}=f_Q(\tau)s_Q^c+f_H(\tau)s_H^c
\label{enmix}
\ee
such that at $\tau=\tau_Q$, $f_Q=1$ and at $\tau=\tau_H$, $f_H=1-f_Q=1$ 
and the
life time of the mixed phase $\tau^{\s mixed}_{\s life}$ is
$\tau_H-\tau_Q$. Here $s_Q^c$ ($s_H^c$) denotes 
the entropy density of QGP (hadronic) phase at $T_c$.
Since scaling law governing 
the variation of $s(\tau)$ must continue
to hold also in the mixed  phase, substituting
Eq.~(\ref{enmix}) in Eq.~(\ref{enscale}) we obtain for $T_i>T_c$,
\be
f_Q(\tau)=\frac{1}{r-1}\left(r\frac{\tau_Q}{\tau}-1\right)
=\frac{1}{r-1}\left(\frac{\tau_H}{\tau}-1\right)
\label{qgpfrac}
\ee
where $r$ ($=g_Q/g_{\s eff}$)
is the ratio of the degeneracy of QGP phase and the effective 
degeneracy in the hadronic phase. 
In the above equation we have used the 
relation $\tau_H=r\tau_Q$, 
obtained as a result of ($1+1$) dimensional isentropic expansion.

The quantity $f_Q$ expressed  as 
$f_Q(\tau)=(s-s_H^c)/(s_Q^c-s_H^c)$ is the volume fraction
of the QGP sector in the mixed phase and similarly 
$f_H=(s_Q^c-s)/(s_Q^c-s_H^c)$ is the volume fraction of the hadronic sector
in the mixed phase. These quantities ($f_Q(\tau)$ and $f_H(\tau)$),
will be  required to evaluate the electromagnetic probes from 
a evolution scenario, QGP$\ra$\,mixed phase$\ra$\,hadronic phase\,$\ra$ 
freeze-out, in the next section.

If $T_i=T_c$, i.e. if the system is formed in  the  mixed  phase
with a fraction $f_0$ of the QGP phase then $f_Q$ 
is given by~\cite{janepr,kkmm}
\be
f_Q(\tau)=\frac{1}{r-1}\left[(1+(r-1)f_0)\frac{\tau_i}{\tau}-1\right]
\ee
The mixed phase ends at a proper time
$\tau_H^m=(1+(r-1)f_0)\tau_i$.
In case of $s_i<s_H^c$, the value of $f_H(\tau)$ is always unity.

To make our discussion more specific, 
consider Pb + Pb collisions at CERN SPS energies. 
If we assume that the matter is formed in the QGP phase
with two flavours ($u$ and $d$), then $g_k=37$.
Taking $dN_{\pi}/dy=600$
as measured by the NA49 Collaboration~\cite{npa610} 
for Pb + Pb collisions, we obtain $T_i=185$ MeV for
$\tau_i=1$ fm/c. We have taken $T_f=130$ MeV~\cite{lkb,dumitru} 
in our calculations.
We also consider central
collisions of Pb + Pb at the RHIC
energies which correspond to about 200 GeV/A
in the centre of mass system.
The particle rapidity density in the central region is taken
as 1735 for RHIC.
The corresponding initial temperature is 
(by assuming that $\tau_i=1$ and QGP initial state) $T_i$= 265 MeV . 

%%%%%%%%%%%%%%%%%%%%%%%%%%%%%%%%%result.tex%%%%%%%%%%%%%%%%%%%%%
\setcounter{equation}{0}
\def\theequation{8.\arabic{equation}}

\section{Results}

\subsection{Hadronic properties at non-zero temperature}

In the Walecka model the effective nucleon mass 
at $T\neq 0$ has been evaluated in the Relativistic 
Hartree Approximation (RHA).
Then the $\rho$ and $\omega$ masses are computed
by evaluating their self energies due to 
$\rho-N-\bar{N}$ and $\omega-N-\bar{N}$ interactions at finite
temperature. 
The following values of the coupling constants
and masses~\cite{shiomi} have been used in our calculations: 
$\kappa_{\rho} = 6.1,~g_{\rho NN}^2 = 6.91, m_\sigma$= 458 MeV, $m_{\rho} = 
770$ MeV, $M_N = 939$ MeV, $g_{\snn}^2 = 54.3$, $\kappa_{\omega} = 0$,
and $g_{\omega NN}^2 = 102$. 
In Fig.~(\ref{7fig1}) we depict the
variation of vector meson masses as a function of temperature
in the Walecka model along with the BR and Nambu scaling scenarios.
The parametrized forms 
 of the effective masses are given in Eqs.~(\ref{pmasswal})
and (\ref{anst}).
The variation of mass in the Walecka model and BR scaling
is slower than the Nambu scaling scenario. At higher temperature
the Walecka model calculation and
the BR scaling (near $T_c$)  
tend to converge. Such a small difference in the mass
variation in the above two scenarios may not be 
visible through the photon spectra.
%\vskip 0.2cm
%%%%%%%%%%%%%% Figure 6 %%%%%%%%%%%%%%%%%%%%%%%%%%%%%
\bef
\centerline{\psfig{figure=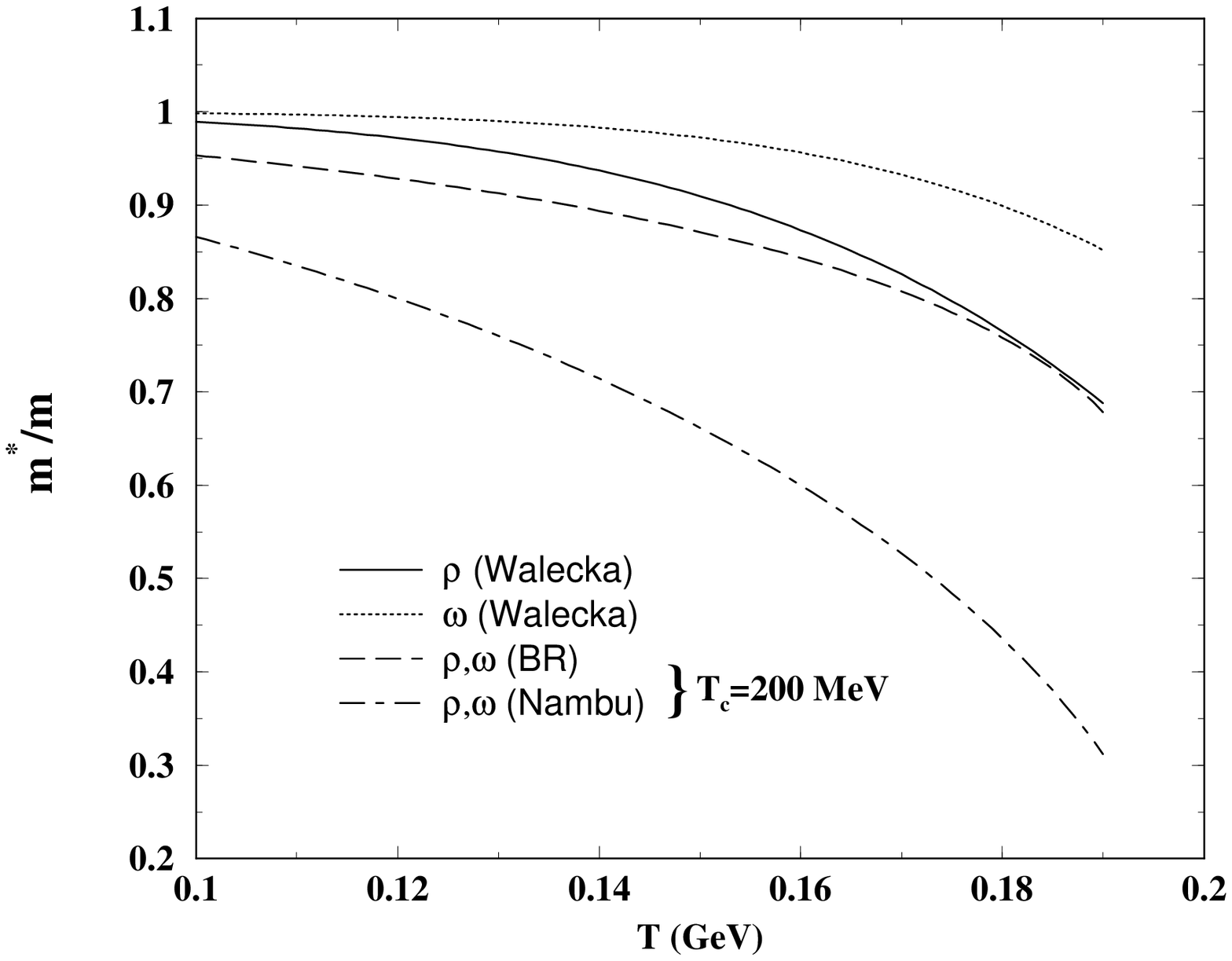,height=7cm,width=9cm}}
\caption{
Variation of vector meson mass with temperature 
for BR (long-dashed line), 
Nambu (dot-dashed line) scaling with $T_c$=200 MeV and
in the Walecka model for $\rho$ (solid line) and $\omega$
(dotted line).
}
\label{7fig1}
\eef
%%%%%%%%%%%%%%%%%%%%%%%%%%%%%%%%%%%%%%%%%%%
%\vskip 0.2cm
We also note at this point that in the Walecka model $\rho$ and $\omega$
masses show different rate of reduction~\cite{npa99} due to
different values of their coupling constants with the nucleons.
%\vskip 0.2cm
%%%%%%%%%%%%%% Figure 7 rho sp fn (Walecka model)%%%%%%%%%%%%%%%%%%%%%%%%%%%%%
\bef
\centerline{\psfig{figure=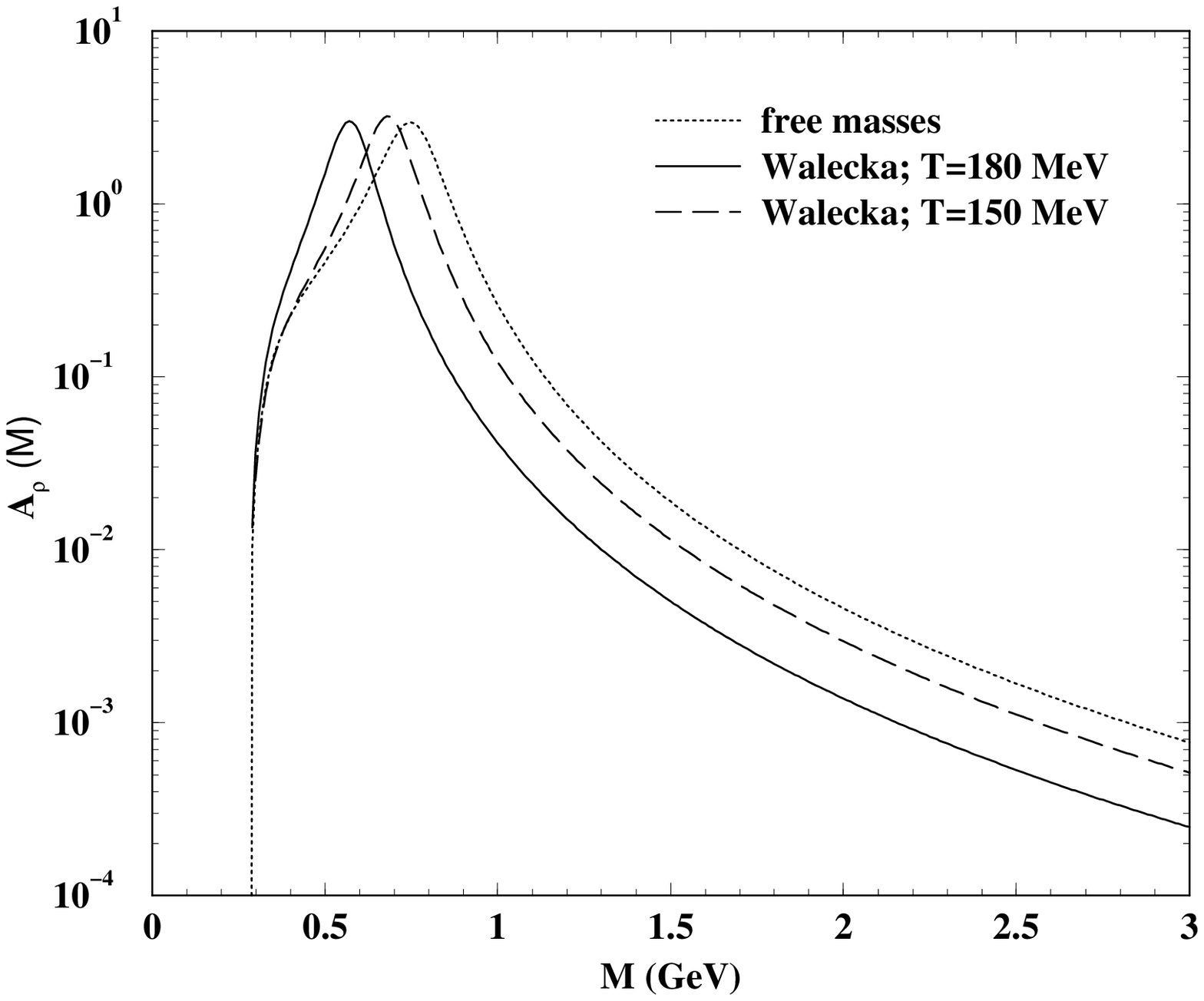,height=7cm,width=9cm}}
\caption{ Spectral function of $\rho$ meson in the Walecka model.
Solid (long dashed) line corresponds to $T=$180 MeV ($T=$150 MeV).
The spectral function in vacuum is shown by the dotted line.
}
\label{7fig2}
\eef
%%%%%%%%%%%%%%%%%%%%%%%%%%%%%%%%%%%%%%%%%%%
%\vskip 0.2cm

In Figs.~(\ref{7fig2}) and ~(\ref{7fig3}) the change in the 
$\rho$ and $\omega$ spectral functions at non-zero temperature
has been displayed. Here as well as in the following `free' mass
will indicate the physical mass in vacuum, {\it e.g.} 770 MeV
for the $\rho$ meson. $A_\rho$ (in units of $e$) is obtained
by multiplying Eq.~(\ref{walcor}) by $8\pi$. $A_\omega$ is obtained analogously.
We have used the interaction Lagrangians ~(\ref{photlag}), 
(\ref{etaro}) and (\ref{lqhd}) for this purpose. 
The shifts in both the spectral functions
towards the lower invariant mass region correspond to the reduction
of their masses due to thermal interactions (see Fig.~\ref{7fig1}). 
The broad $\omega$ peak arises due to its interaction with the
thermal pion in the heat bath. The reaction $\omega\,\pi\,\ra\,\pi\,\pi$
contributes dominantly to the survival probability of the $\omega$ 
in the medium.
%\vskip 0.2cm
%%%%%%%%%%%%%% Figure 8 omega sp fn (Walecka model)%%%%%%%%%%%%%%%%%%%%%%%%%%%%%
\bef
\centerline{\psfig{figure=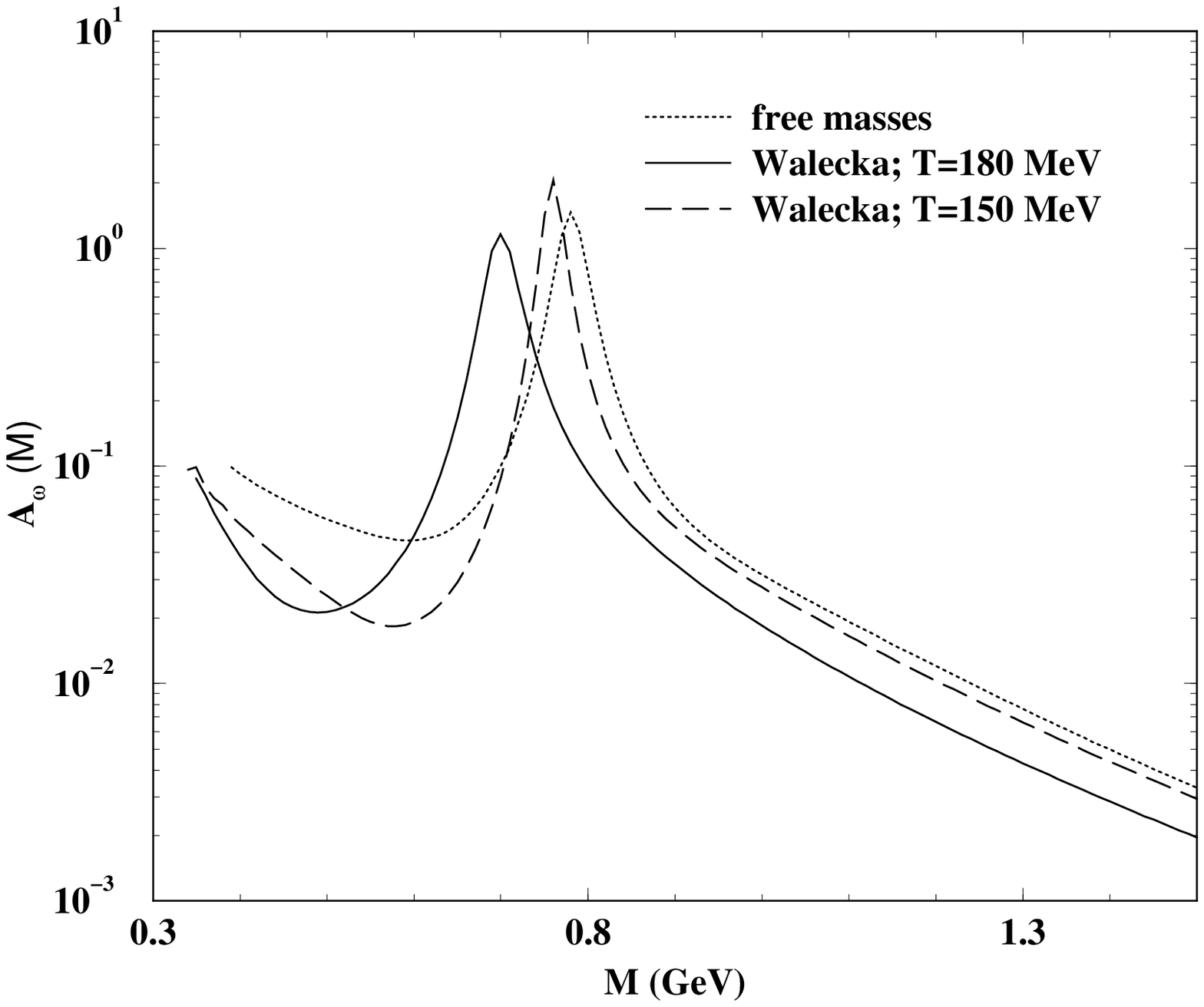,height=7cm,width=9cm}}
\caption{ Same as Fig.~\protect\ref{7fig2} for the $\omega$ meson.}
\label{7fig3}
\eef
%%%%%%%%%%%%%%%%%%%%%%%%%%%%%%%%%%%%%%%%%%%

Before we proceed further a few comments on Walecka model calculations
are in order.
In this model the major contribution to the medium effects on the $\rho$ 
 and $\omega$ mesons
arises from the nucleon-loop diagram.
For the dressing of internal lines in matter we restrict ourselves to 
the Mean Field Theory (MFT) to avoid a plethora of diagrams and to maintain
internal consistency. 
It has been shown~\cite{gale,sourav} that the change in the $\rho$ mass due
to $\rho-\pi-\pi$ interaction is negligibly small at non-zero temperature
and zero baryon density. 
Therefore the change in the $\rho$ meson mass
due to $\rho-\pi-\pi$ interaction is neglected here.
 At finite baryon density, the dynamics is more involved
 due to the   medium effects on the $\rho-\pi-\pi$ vertex,
 the pion propagator coupled with  delta-hole excitation,
 and the coupling of the $\rho$-meson  with $N^*$-hole excitations
\cite{asakawa,friman,chanfray,klingl,fp,rubw,ppllm}
 (see also the review, Ref.\cite{blrrw}.) 
 The major effect of such medium modifications is to broaden the
 $\rho$-peak as well as to produce complicated structure
 around the peak.
 Since in the present work we 
restrict our calculations within the realm of MFT, {\it i.e} 
the internal nucleon loop in the $\rho$ and $\omega$ self energy 
is modified due to tadpole diagram only,
the inclusion of vertex corrections, modification of the pion 
propagator and the inclusion of baryon resonances
are not considered here.
Also, in the present work we restrict to
 zero baryon density.
%\vskip 0.2cm
%%%%%%%%%%%%%% Figure 9 %%%%%%%%%%%%%%%%%%%%%%%%%%%%%
\bef
\centerline{\psfig{figure=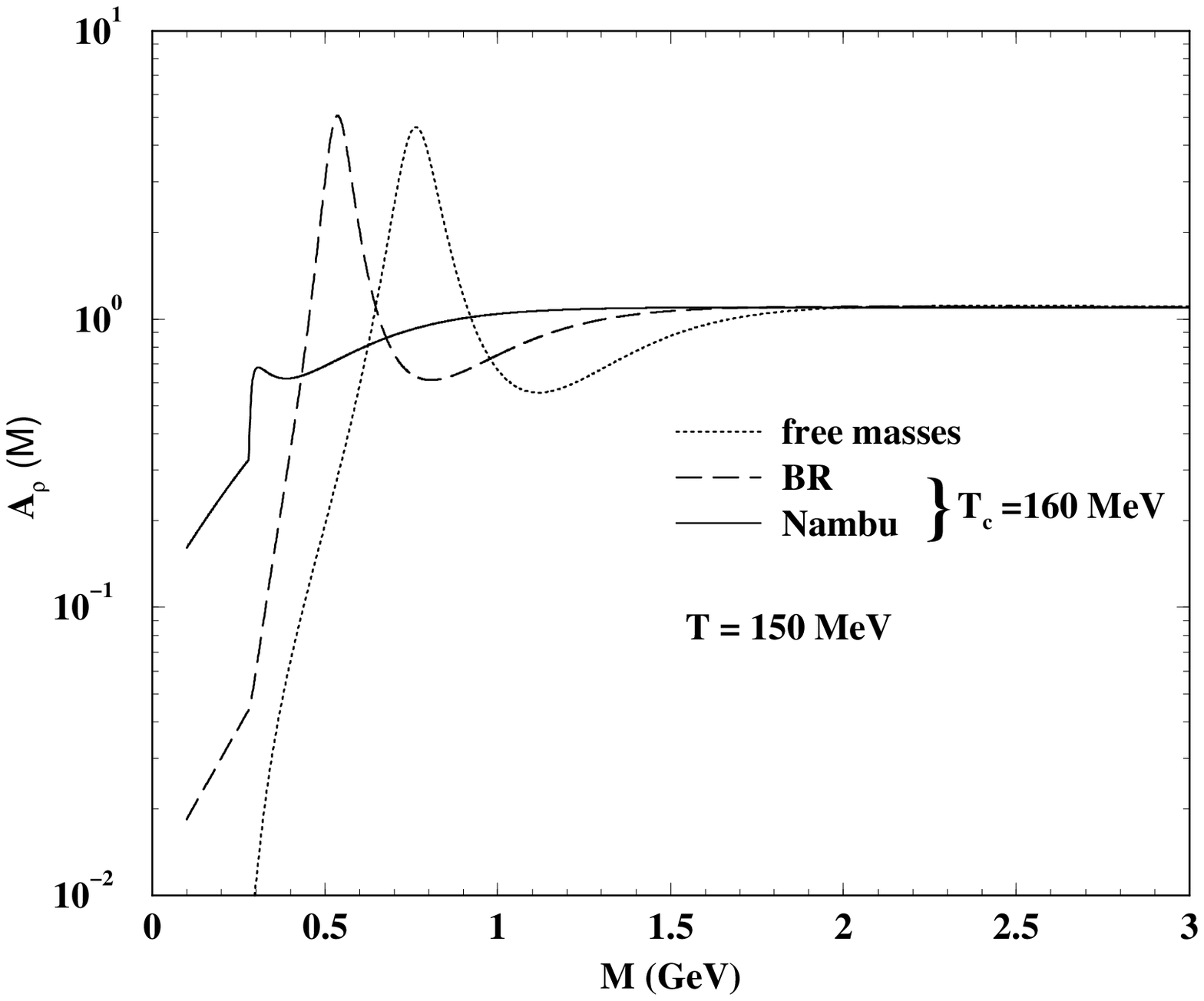,height=7cm,width=9cm}}
\caption{
Spectral function for the isovector
channel extracted from $e^+e^-$ collisions (dotted line)
as a function of invariant mass. The dashed (solid) line indicates
the spectral function when $m_\rho$ and $\omega_0$ 
vary according to BR (Nambu) scaling.
}
\label{7fig4}
\eef
%%%%%%%%%%%%%%%%%%%%%%%%%%%%%%%%%%%%%%%%%%%
%\vskip 0.2cm

In Fig.~(\ref{7fig4}) the spectral function 
($8\pi$ times Eq.~(\ref{parametrho})) for
the isovector ($\rho$) channel is plotted as a function 
of invariant mass at $T=150$ MeV and $T_c=160$ MeV. 
We find that both the peak and
the continuum threshold of the spectral function 
move towards lower invariant mass. However, in case 
of Nambu scaling the shift is more compared to BR scaling.  In the  
Nambu scaling scenario the peak of the spectral function and the 
continuum are not well separated; a merging of the two 
would take place at $T=T_c$. This could possibly indicate the 
onset of a deconfinement phase transition.
%\vskip 0.2cm
%%%%%%%%%%%%%% Figure 10 %%%%%%%%%%%%%%%%%%%%%%%%%%%%%
\bef
\centerline{\psfig{figure=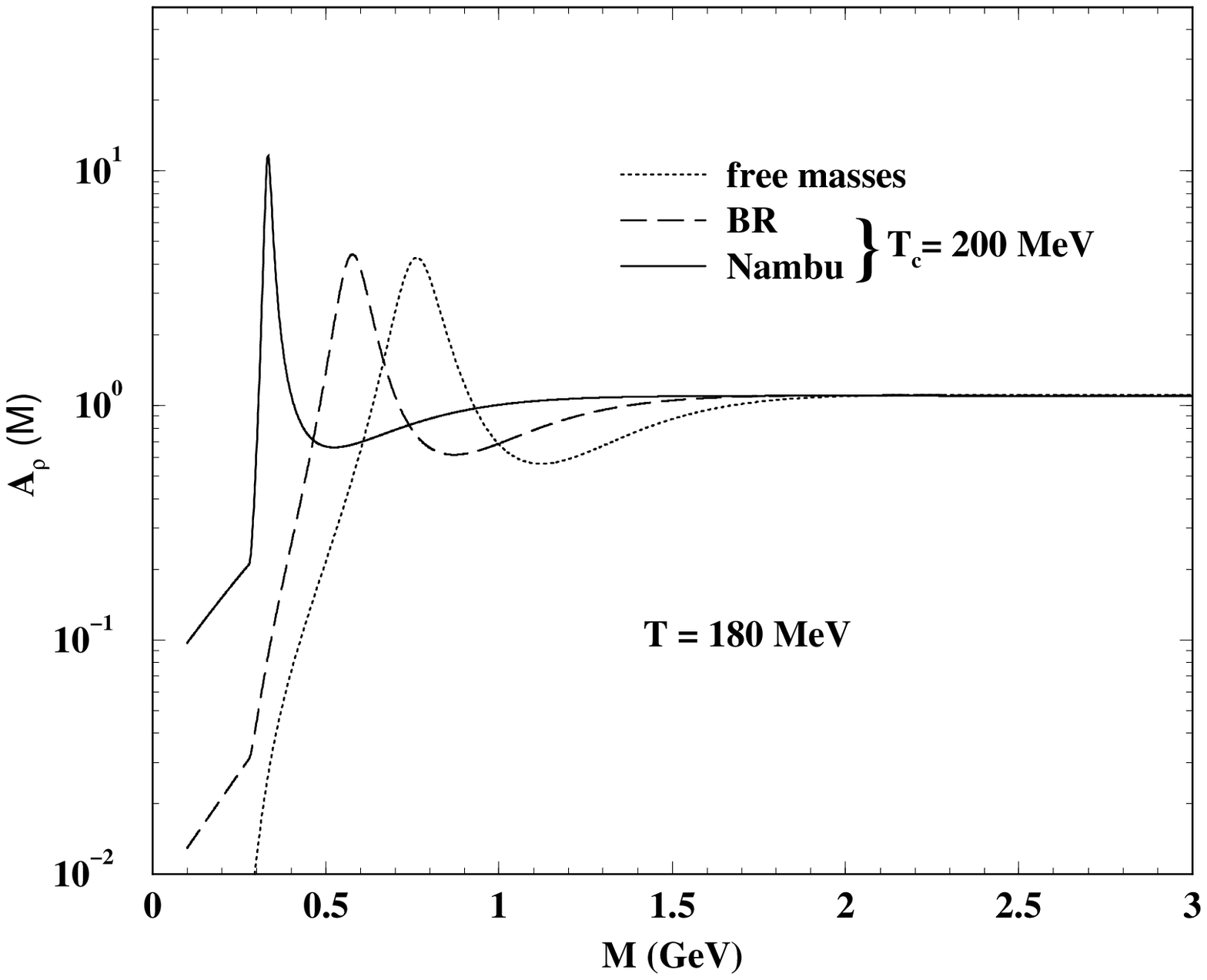,height=7cm,width=9cm}}
\caption{
Same as Fig.~\protect\ref{7fig4} at $T=180$ MeV and $T_c=200$ MeV.
}
\label{7fig5}
\eef
%%%%%%%%%%%%%%%%%%%%%%%%%%%%%%%%%%%%%%%%%%%
%\vskip 0.2cm
Fig.~(\ref{7fig5}) shows the spectral function at $T=180$ MeV
and $T_c=200$ MeV. Due to a larger separation between 
$T_c$ and $T$ compared to the 
previous case the peaks in the spectral function in all the
cases are well separated from the continuum.       

%\vskip 0.2cm
%%%%%%%%%%%%%% Figure 11 %%%%%%%%%%%%%%%%%%%%%%%%%%%%%
\bef
\centerline{\psfig{figure=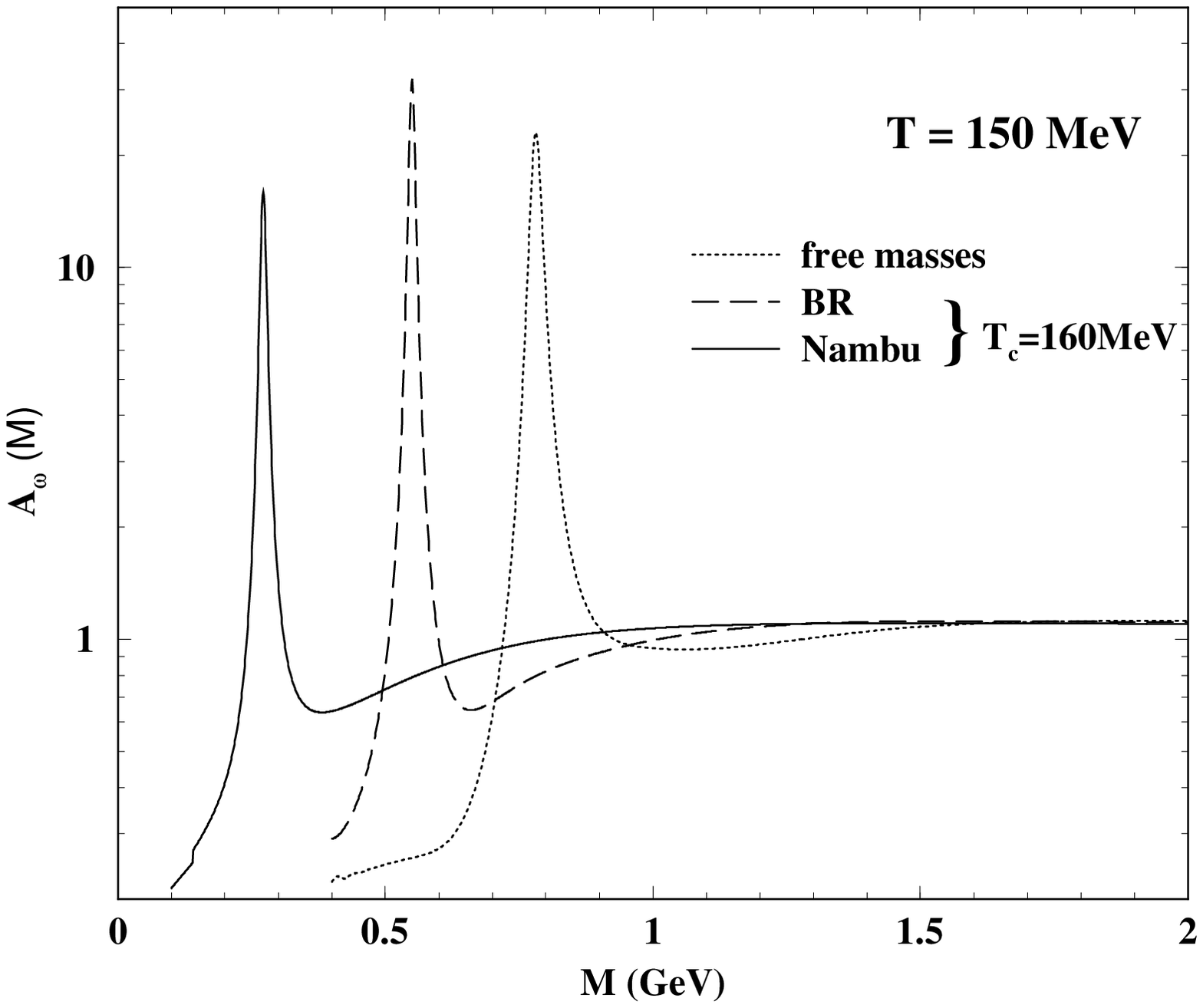,height=7cm,width=9cm}}
\caption{
Same as Fig.~\protect\ref{7fig4} for the isoscalar channel.
}
\label{7fig6}
\eef
%%%%%%%%%%%%%%%%%%%%%%%%%%%%%%%%%%%%%%%%%%%
%\vskip 0.2cm
In Figs.~(\ref{7fig6}) and (\ref{7fig7}) the spectral functions for 
the isoscalar ($\omega$) channel have been depicted. In both the cases
the peak in the spectral function is distinctly visible in all the
mass variation scenarios. The larger width in the isoscalar channel
is due to the combined processes $\omega\rightarrow\,3\pi$ and
$\omega\,\pi\rightarrow\,\pi\,\pi$ as discussed before.
%\vskip 0.2cm
%%%%%%%%%%%%%% Figure 12 %%%%%%%%%%%%%%%%%%%%%%%%%%%%%
\bef
\centerline{\psfig{figure=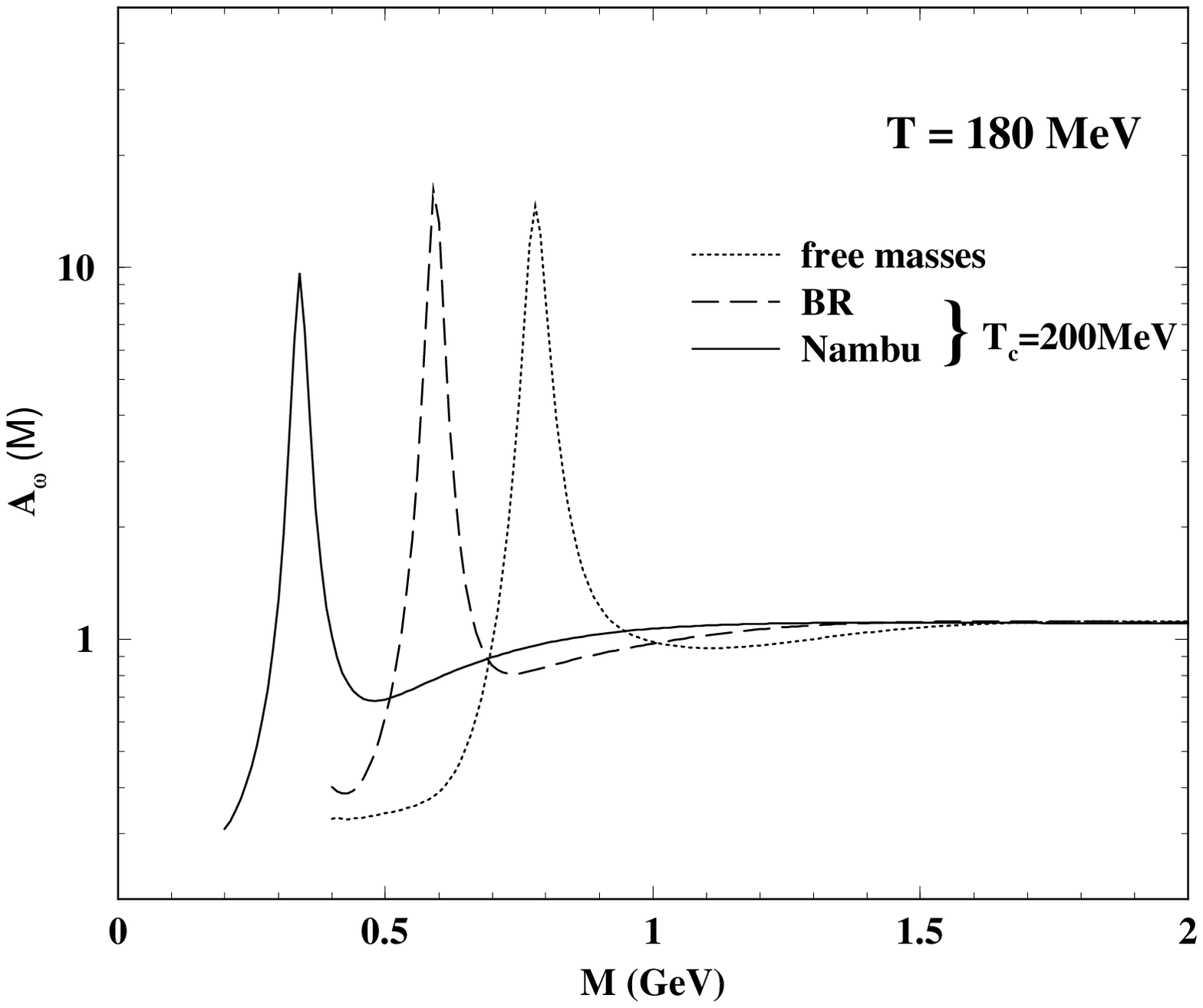,height=7cm,width=9cm}}
\caption{
Same as Fig.~\protect\ref{7fig5} for the isoscalar channel.
}
\label{7fig7}
\eef
%%%%%%%%%%%%%%%%%%%%%%%%%%%%%%%%%%%%%%%%%%%
%\vskip 0.2cm

The spectral functions for the vector mesons both in the isoscalar 
and isovector channels are plotted in Fig.~(\ref{7fig8}) 
at a temperature $T\sim T_c$. As expected from the scaling law the peak has 
vanished due to its 
overlap with the continuum. All the hadrons in the thermal bath
have melted to their fundamental constituents - the 
quarks and gluons. Such a spectral function would indicate
a transition from hot hadronic matter to QGP. 
This behaviour should,
in principle, be reflected in the dilepton spectrum
originating from these channels. Such a broad spectral
representation without any peak may be compared with that
of the huge decay width of the $\rho$ meson due to its interaction
with the baryonic medium~\cite{rcw,ek}.

%\vskip 0.2cm
%%%%%%%%%%%%%% Figure 13 %%%%%%%%%%%%%%%%%%%%%%%%%%%%%
\bef
\centerline{\psfig{figure=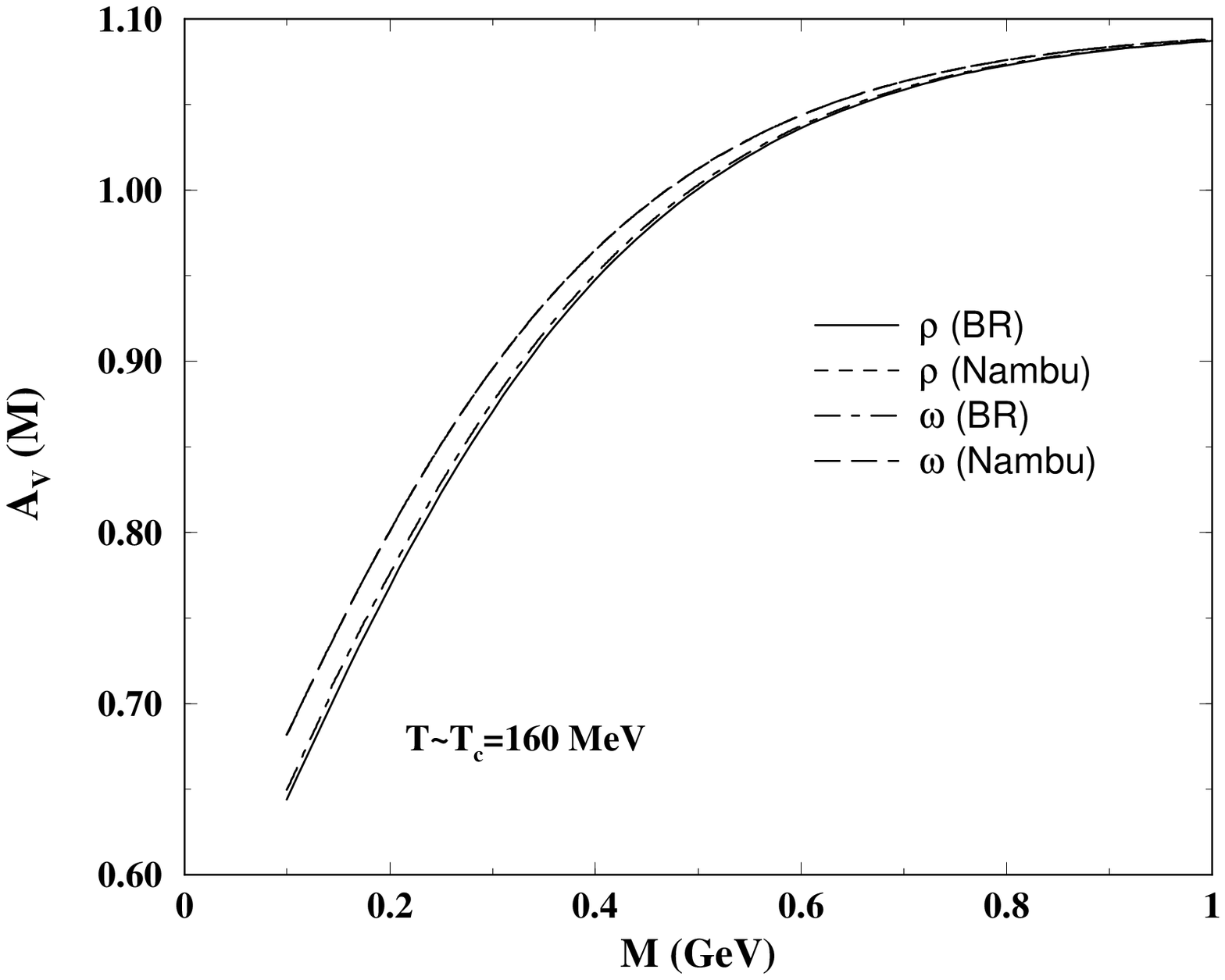,height=7cm,width=9cm}}
\caption{
Spectral functions for isovector ($\rho$) and
isoscalar ($\omega$) channels  at $T_c$ .
}
\label{7fig8}
\eef
%%%%%%%%%%%%%%%%%%%%%%%%%%%%%%%%%%%%%%%%%%%

In Fig.~(\ref{7fig9}) the shift in the pole position of 
the $\rho$-spectral function is depicted for the 
Linear Sigma Model (LSM), Non-Linear Sigma Model (NLSM), and 
Hidden Local Symmetry (HLS) approach. The width of the $\rho$
meson has been calculated for the process $\rho\,\ra\,\pi\,\pi$ at non-zero
temperature.  For the NLSM and HLS
interactions the $\rho$-mass increases by an amount 90 MeV and 10 MeV
respectively. For the enhancement of $\rho$ mass in the NLSM, a larger
phase space is available for the decay process $\rho\ra\pi\pi$ 
consequently the $\rho$ appears to be broader in this
case compared to HLS interaction.
On the other hand, for the gauged LSM the $\rho$ mass
reduces by about 45 MeV at $T=150$ MeV.
It may be noted, as mentioned in section 5.1,
that $\rho$ mass decreases in gauged
LSM for low temperatures and increases for temperatures in the vicinity
of the chiral phase transition. 

%\vskip 0.2cm
%%%%%%%%%%%%%% Figure 14 %%%%%%%%%%%%%%%%%%%%%%%%%%%%%
\bef
\centerline{\psfig{figure=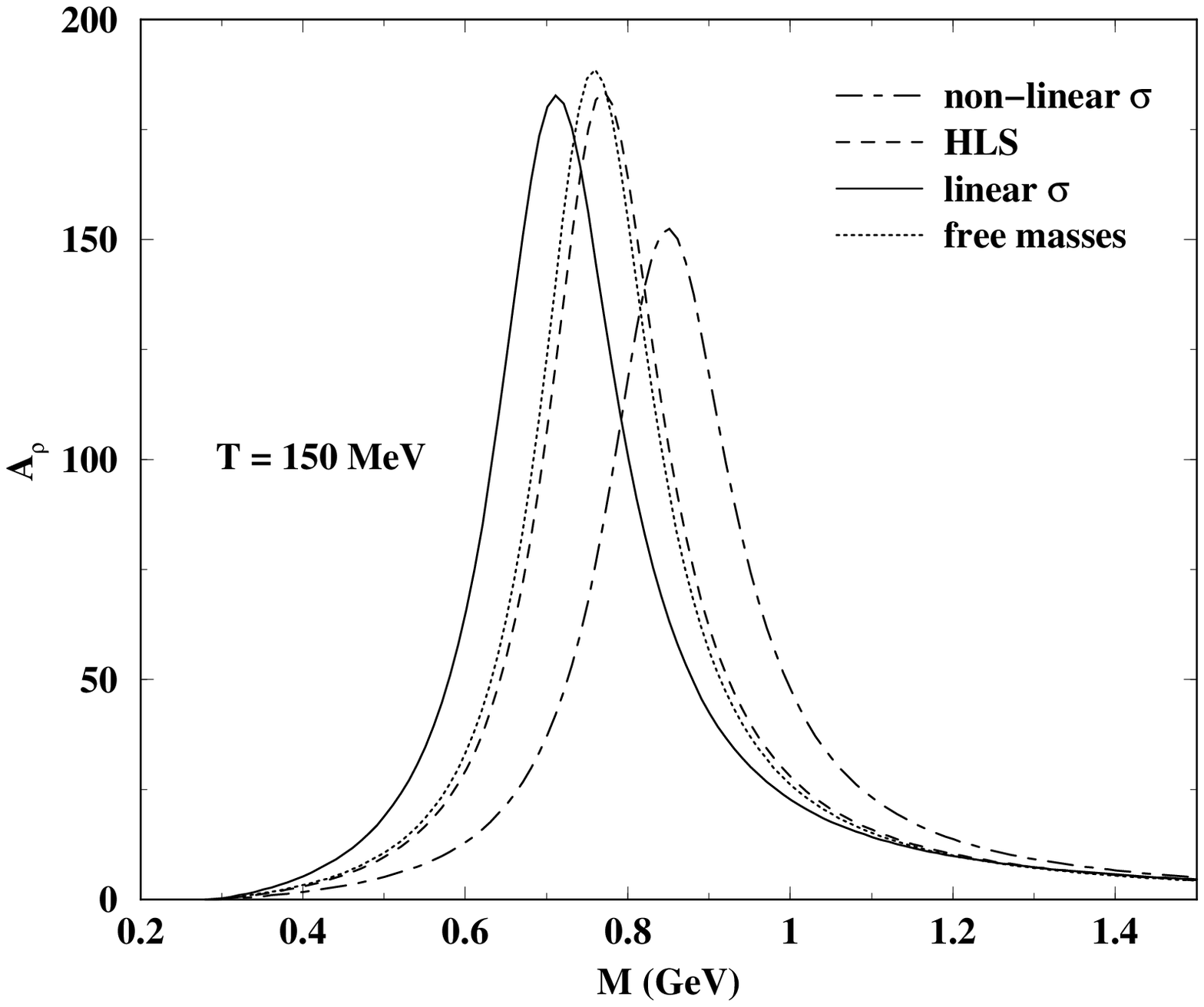,height=7cm,width=9cm}}
\caption{ Shift in the pole position of the $\rho$
spectral function for gauged Linear and Non-Linear Sigma Models and
Hidden Local Symmetry Lagrangian at $T=150$ MeV.
}
\label{7fig9}
\eef
%%%%%%%%%%%%%%%%%%%%%%%%%%%%%%%%%%%%%%%%%%%

\subsection{Static photon spectra}

In this section we consider the photon spectra from hot
hadronic matter and QGP. The medium effects enter through the 
masses and decay widths of the particles participating in the
photon producing reactions. 
It is well known~\cite{kapusta} (see also Refs.~\cite{song,halasz}) that the reactions 
$\pi\,\rho\,\ra\, \pi\,\gamma$ , 
$\pi\,\pi\,\ra\, \rho\,\gamma$ , $\pi\,\pi\,\ra\, \eta\,\gamma$ , 
$\pi\,\eta\,\ra\, \pi\,\gamma$ , and the decays $\rho\,\ra\,\pi\,\pi\,\gamma$
and $\omega\,\ra\,\pi\,\gamma$ are the most important channels 
for photon production from hadronic matter in the
energy regime of our interest. We have also included those reactions
which produce photon via intermediary axial vector meson $a_1$ as discussed
earlier. Non-zero width of vector and axial vector mesons in the
intermediate state has been taken into account.
While evaluating the photons from QGP we have considered both one loop and
two loop contributions to the photon self energy
as shown in Figs.~(\ref{loop1}) and (\ref{loop2}).
 
The total photon emission rate from QGP and hadronic matter 
at $T=160$ MeV is plotted in Fig.~(\ref{7fig10})
as a function of the energy of the emitted photon for
different values of strong charge $g_s$ in the QGP phase and for various 
mass variation scenarios in the hadronic sector. 
The photon production rate from QGP has been
evaluated in the HTL approximation, which is valid if
the hard and soft scales are well separated, i.e.
for $g_s<<1$ (which corresponds to $\alpha_s<<0.08$), 
QCD lattice calculations~\cite{karsch} however
suggest that $\alpha_s\sim 0.2-0.3$
at temperatures achievable in URHIC. This means 
that the extrapolation of the results obtained under HTL approximation 
to higher values of $g_s$ (or $\alpha_s$) corresponding to 
lattice simulation may be dubious. 
We have evaluated the photon spectra for two values of the strong coupling 
constants $g_s=0.8$ (solid square) and $2$ (solid dots)
to demonstrate the sensitivity
of the photon spectra to the value of the strong charge
and to show the uncertainties involved in the problem.
In the hadronic sector the photon
yield is seen to be  enhanced 
compared to the case when the effects of the thermal interaction
on the hadronic properties are neglected. 
This is true for almost the
entire energy range of the emitted photon under consideration. 
As a result of the similar mass shift in the Walecka model and BR scaling
the photon spectra in these two scenarios (long-dashed and dotted lines 
respectively) have a negligible difference,
whereas the enhancement in the spectrum due to hadronic mass shift
according to Nambu scaling is clearly visible (solid line).

%\vskip 0.2cm
%%%%%%%%%%%%%% Figure 15 %%%%%%%%%%%%%%%%%%%%%%%%%%%%%
\bef
\centerline{\psfig{figure=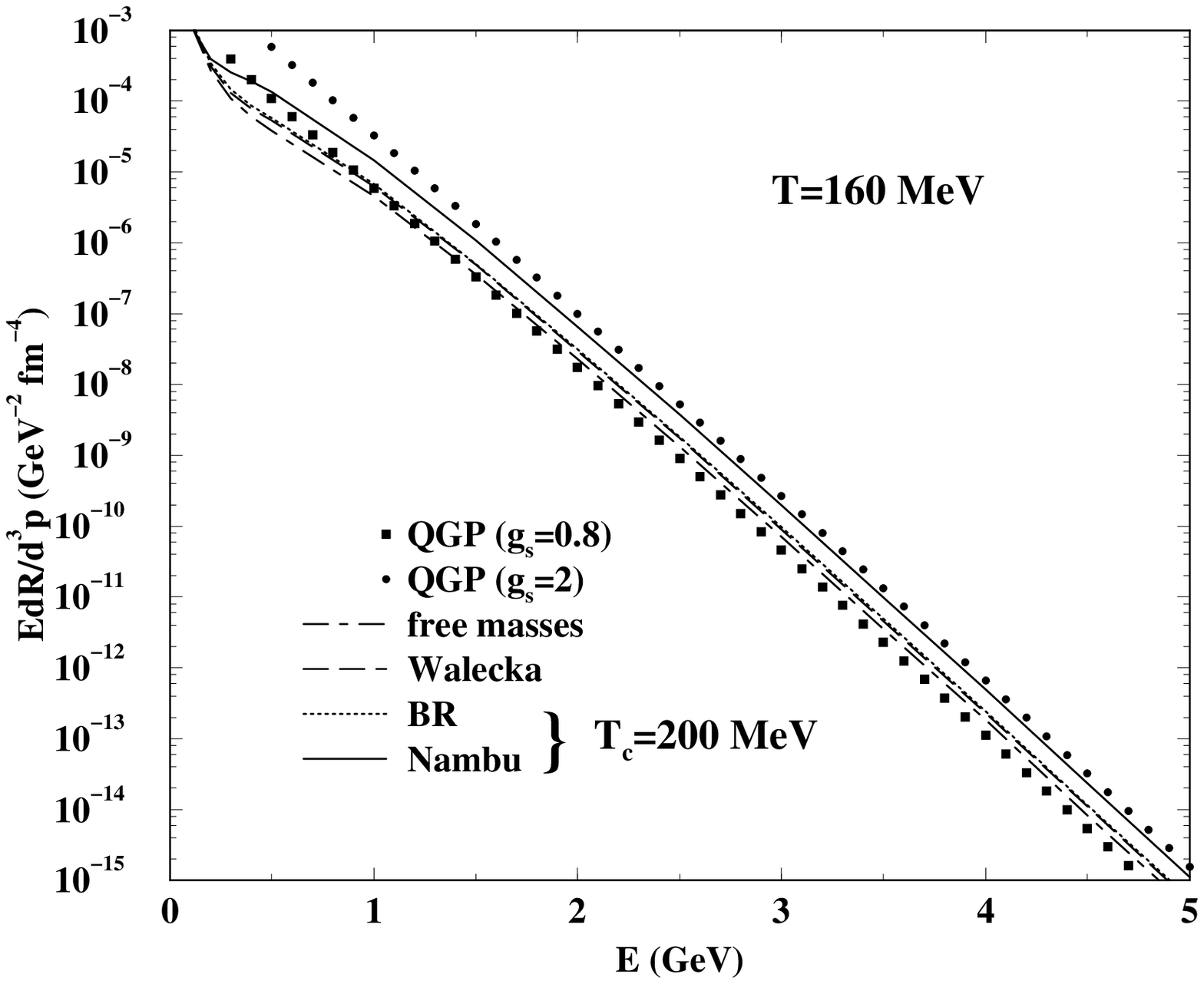,height=7cm,width=9cm}}
\caption{
Thermal photon spectra at $T=160$ MeV. Solid dots (square) indicates 
photon emission rate from QGP with both one loop and two loop 
contributions as evaluated by Kapusta et al and Aurenche {\it et al}
respectively  for $g_s=2(0.8)$. Dotdash line represents photon spectrum
from hot hadronic gas without medium effects. The result with the in-medium 
effects within the scope of the Walecka model calculations 
is shown by long dashed line. Dotted (solid) line indicates
photon spectrum with BR (Nambu) scaling mass variation scenario.
}
\label{7fig10}
\eef
%%%%%%%%%%%%%%%%%%%%%%%%%%%%%%%%%%%%%%%%%%%
%\vskip 0.2cm
In Fig.(\ref{7fig11}) we show the photon emission rate at $T=180$ MeV.
Photon spectra from hadronic matter with mass variation
according to the Nambu scaling scenario overshine the 
photons from QGP even for a larger value of $g_s$ ($\sim 2$).

At this stage one might ask: At a fixed $T$ which one is brighter --
the hot hadronic gas or QGP?. Well, within 
the scope of the present work, the answer depends on (i) the value of the strong 
coupling constant, (ii) the degree of hotness of the medium and (iii) how adversely
the hadrons are affected in the medium.  

%\vskip 0.2cm
%%%%%%%%%%%%%% Figure 16 %%%%%%%%%%%%%%%%%%%%%%%%%%%%%
\bef
\centerline{\psfig{figure=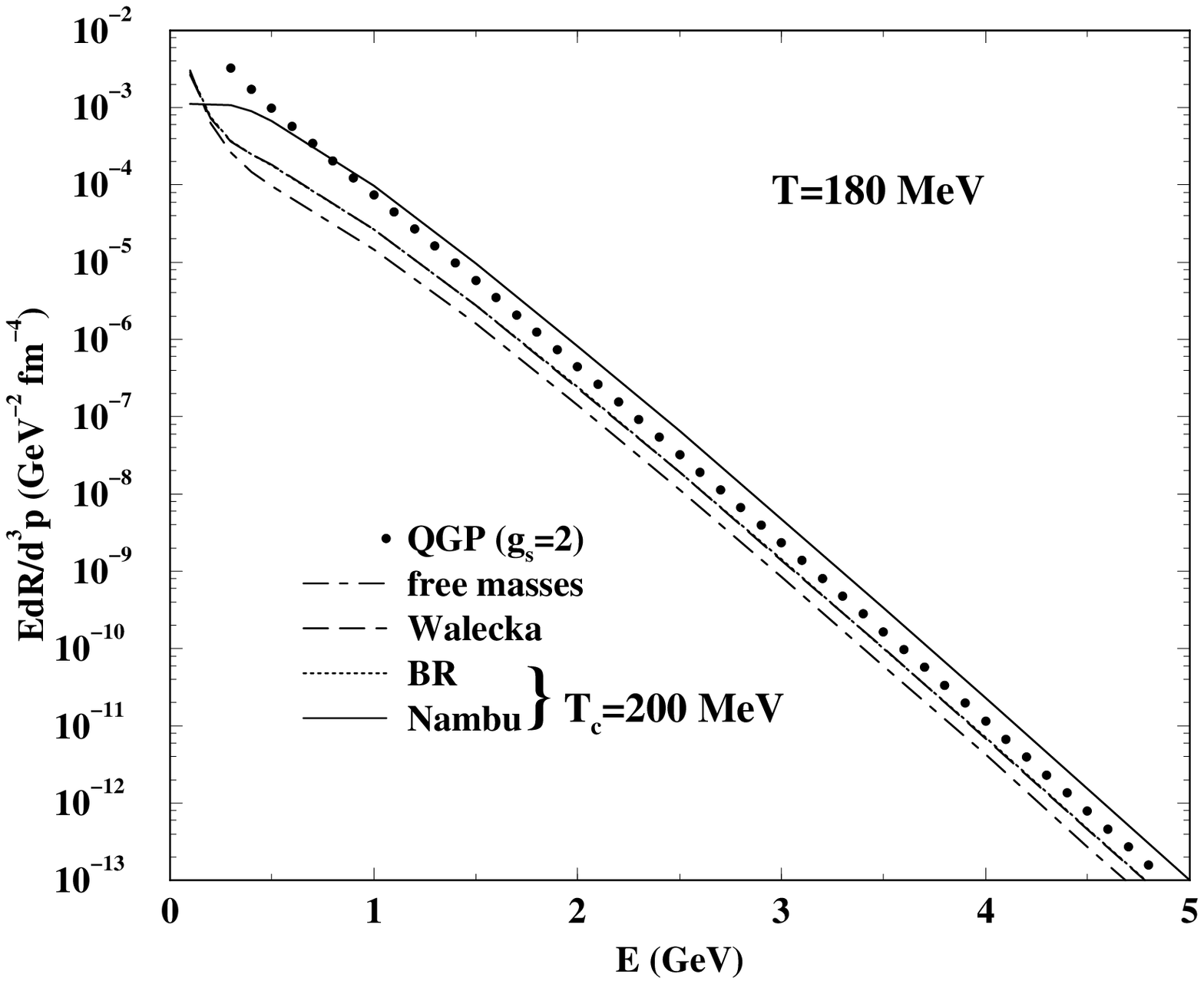,height=7cm,width=9cm}}
\caption{
Same as Fig.~(\protect\ref{7fig10}) at $T=180$ MeV and $g_s=2$.
}
\label{7fig11}
\eef
%%%%%%%%%%%%%%%%%%%%%%%%%%%%%%%%%%%%%%%%%%%
%\vskip 0.2cm
In Fig.~(\ref{7fig12}) the effect of the form factor on the
reaction $\pi\pi\rightarrow \rho\gamma$ has been demonstrated. 
We have taken same monopole form factor for both the  $\pi\pi\rho$
and $\pi\pi\gamma$ vertices~\cite{kapusta} to suppress the contribution from
very high momentum region where the quark structure of the hadrons
could be relevant. The Ward-Takahashi identity has been used to obtain   
the dressed propagator. The in-medium mass of the $\rho$ meson
has been taken from Walecka model calculations. The form factor effects 
for the above reaction reduces the photon production rate by about
10-15\%. In view of the experimental uncertainty of the photon spectra
measured in URHIC ({\it e.g.} Ref.~\cite{wa80})
such effects are not relevant at present. Therefore, we have neglected 
it in the following discussions.

%\vskip 0.2cm
%%%%%%%%%%%%%% Figure 17 %%%%%%%%%%%%%%%%%%%%%%%%%%%%%
\bef
\centerline{\psfig{figure=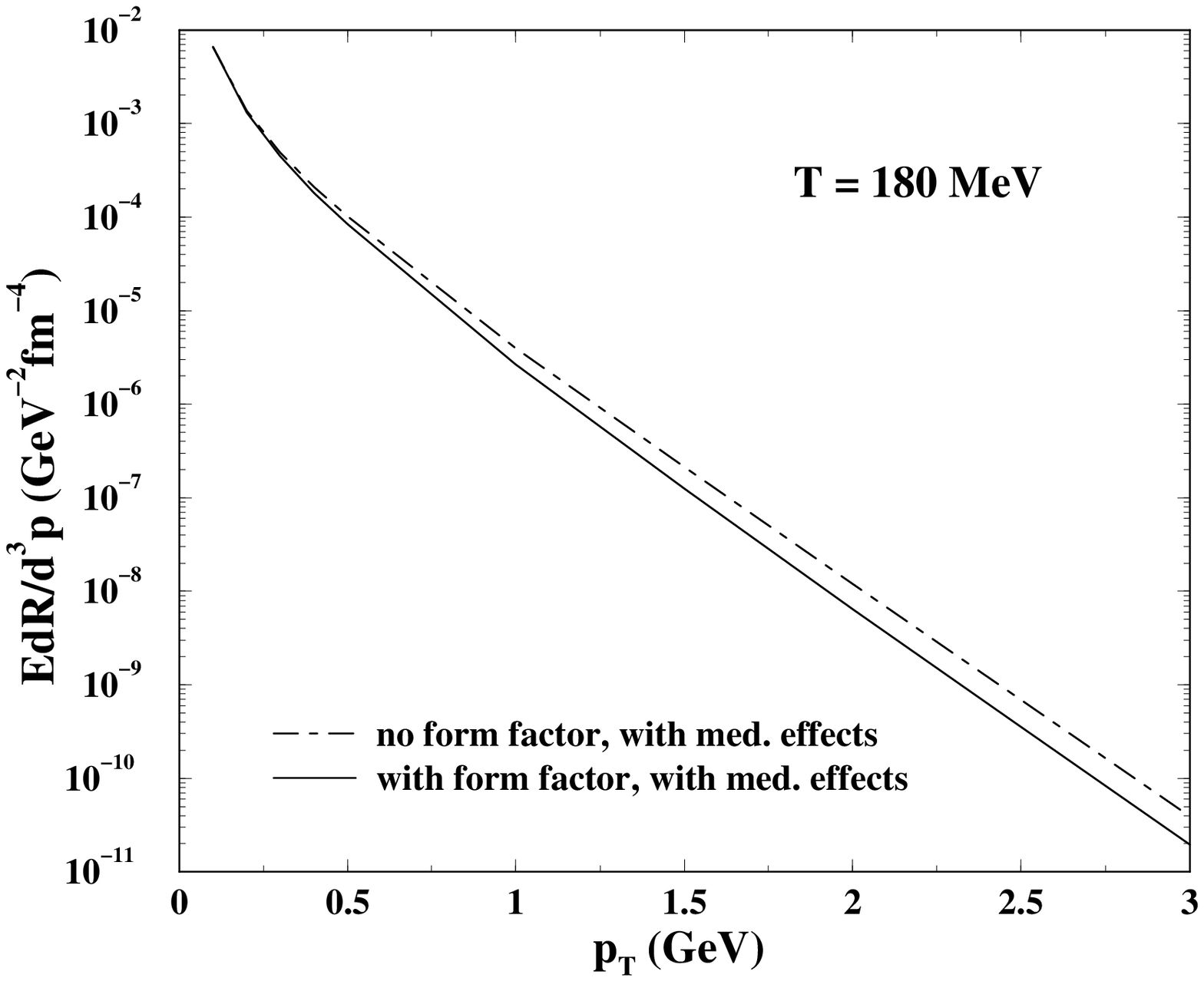,height=7cm,width=9cm}}
\caption{
The effect of the monopole form factor on the photon emission rate
from the reaction $\pi\pi\rightarrow \rho\gamma$.
}
\label{7fig12}
\eef
%%%%%%%%%%%%%%%%%%%%%%%%%%%%%%%%%%%%%%%%%%%
 \vskip 0.2cm
In Fig.~(\ref{7fig13}) the emission rate of photon is shown for
$T=150$ MeV. The change in the photon spectra with 
in-medium masses calculated in the framework of 
gauged LSM, NLSM and HLS Lagrangians are compared with that of vacuum values.
An increase of $\rho$ mass in the NLSM reduces its number due to 
Boltzmann suppression, which leads to a suppression 
in the photon emission rate
(dotted line). The production rate is enhanced due to a reduction in the
$\rho$ mass (solid line) in LSM. 
(We must remember that the $\rho$ mass decreases in gauged
LSM for low temperature and increases for temperatures close to the chiral
transition temperature $T_\chi$. Therefore, 
for $T\sim T_\chi$ it will show a reduction in
photon emission rate and the net yield would be a superposition of
all temperatures, from initial to freeze-out). The change in the 
mass of $\rho$ is so small for HLS approach (short-dashed line) 
that the production rate is 
almost indistinguishable
from the spectra with vacuum masses of the hadrons.
In Fig.~(\ref{7fig13})
we have also demonstrated how the photon spectra
is modified for a drastic change in the width of the $\rho$
meson ($\Gamma_\rho\sim 400$ MeV, such a large value of the width
is obtained in Ref.~\cite{ek} although, at non-zero baryon density) without any appreciable change 
in the pole mass ($m_\rho\sim 770$ MeV). 
Such drastic broadening of the $\rho$ meson   
have been proposed in Ref.~\cite{rcw,ek}. We observe
that the effects of such modifications in the properties
of $\rho$ on the photon spectra is rather negligible (long-dashed line). 
We will see later that such changes in the properties 
of $\rho$ modify the dilepton spectra drastically. 
%\vskip 0.2cm
%%%%%%%%%%%%%% Figure 18 %%%%%%%%%%%%%%%%%%%%%%%%%%%%%
\bef
\centerline{\psfig{figure=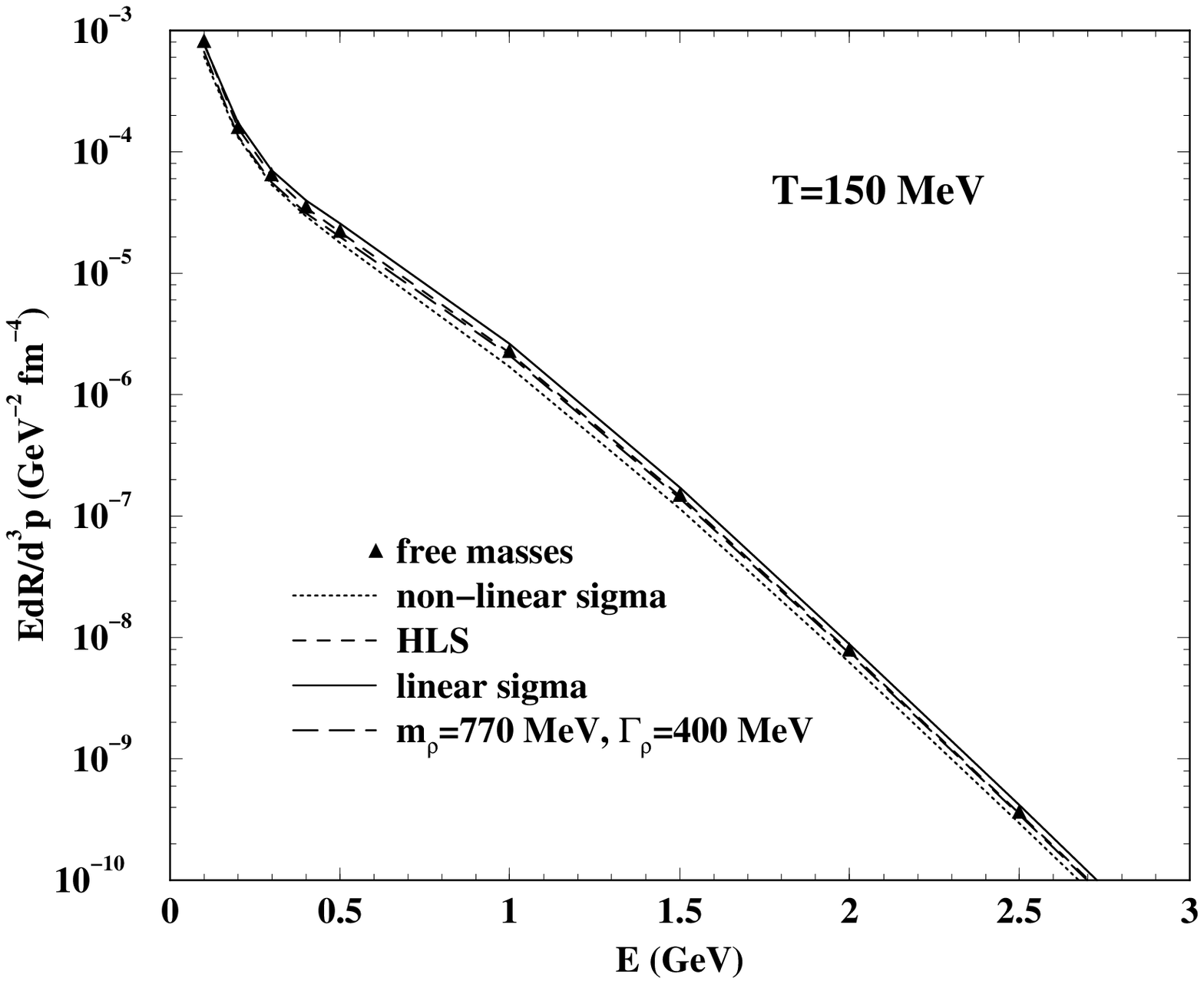,height=7cm,width=9cm}}
\caption{ The change in the photon spectra due to the finite temperature
effects on the hadronic masses in Linear, Non-Linear Sigma Model and
Hidden Local Symmetry approach at T= 150 MeV.
}
\label{7fig13}
\eef
%%%%%%%%%%%%%%%%%%%%%%%%%%%%%%%%%%%%%%%%%%%
 \vskip 0.2cm
\subsection{Static dilepton spectra}

In Fig.~(\ref{7fig14}) we display the invariant mass distribution of
$e^+e^-$ pair. The dilepton yield from $q\bar{q}$ annihilation is denoted by 
solid dots. Dotted line indicates the result obtained from the 
parametrization of the electromagnetic current-current correlation function 
in the `$\rho$' and `$\omega$' channels, when the medium effects are ignored.
A large shift towards the lower invariant mass region  of the $\rho$ peak
is seen in the Nambu scaling (solid line) as compared to the BR scaling
(dash line) consistent with the relative shift in the spectral functions in the
two cases as discussed before. In the Walecka model calculations 
the relevant reactions are $\pi\pi\rightarrow e^+e^-$, 
$\rho\rightarrow e^+e^-$ and $\omega\rightarrow e^+e^-$
(dotdash line)~\cite{ja,pr}. The two peaks corresponding to
$\rho$ and $\omega$ masses are visible in the spectra.
%\vskip 0.2cm
%%%%%%%%%%%%%% Figure 19 %%%%%%%%%%%%%%%%%%%%%%%%%%%%%
\bef
\centerline{\psfig{figure=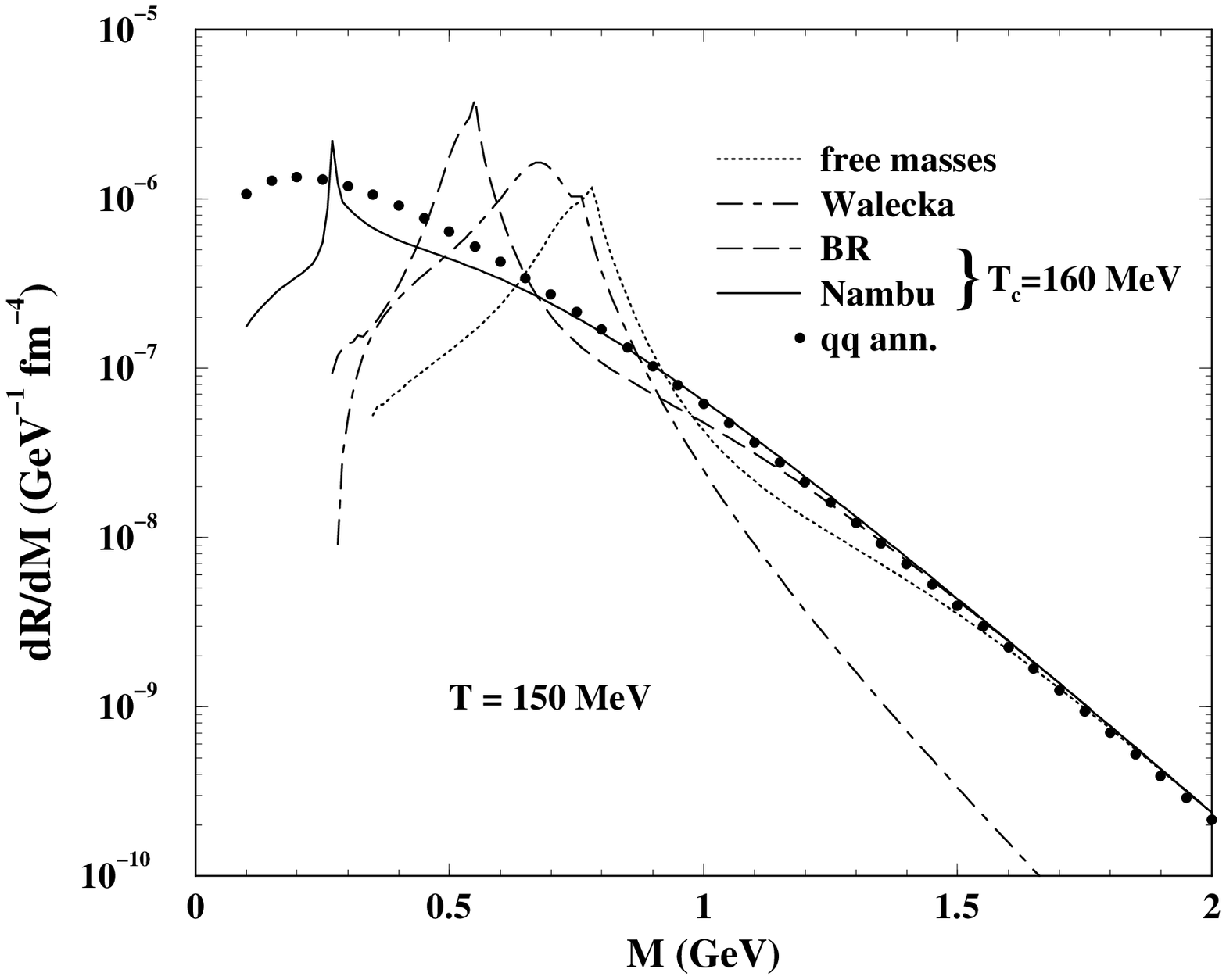,height=7cm,width=9cm}}
\caption{
Thermal dilepton spectra at $T=150$ MeV. Solid dots indicates 
dilepton emission rate from QGP.
Dotted line represents dilepton yield 
from hot hadronic gas without medium effects. 
The result with the in-medium 
effects within the scope of the Walecka model calculations 
is shown by the dot-dashed line. Long dashed (solid) line indicates
dilepton spectrum with BR (Nambu) scaling mass variation scenario.
}
\label{7fig14}
\eef
%%%%%%%%%%%%%%%%%%%%%%%%%%%%%%%%%%%%%%%%%%%
%\vskip 0.2cm
The separation between the two peaks is due to different mass shift of the $\rho$ and
$\omega$. Measurement of such separation in hadronic masses 
($\Delta m=m_{\omega}^*-m_{\rho}^*$)
would signal the in-medium effects. Validity of such results could be 
tested in URHIC by the CERES~\cite{CERES} collaboration in future. Similar shift
at zero temperature but finite baryon density could be detected 
by HADES~\cite{hades} and CEBAF~\cite{book}.  
Effects of the continuum on the dilepton
spectra is clearly visible for $M\geq 1$ GeV (please note that the value
of the continuum threshold in vacuum is 1.3 GeV). 
Due to the continuum contribution the dilepton rates from hadronic matter
and QGP shine equally brightly in the mass range $M\geq$ 1 GeV.
The lepton pair spectra at $T=180$ MeV is shown in Fig.~(\ref{7fig15}).
Since the effective mass of the $\rho$ in the Walecka and BR scaling scenario
is almost same in this case (see Fig.~(\ref{7fig1})), the corresponding rates
are very similar near the $\rho$ peak.

%\vskip 0.2cm
%%%%%%%%%%%%%% Figure 20 %%%%%%%%%%%%%%%%%%%%%%%%%%%%%
\bef
\centerline{\psfig{figure=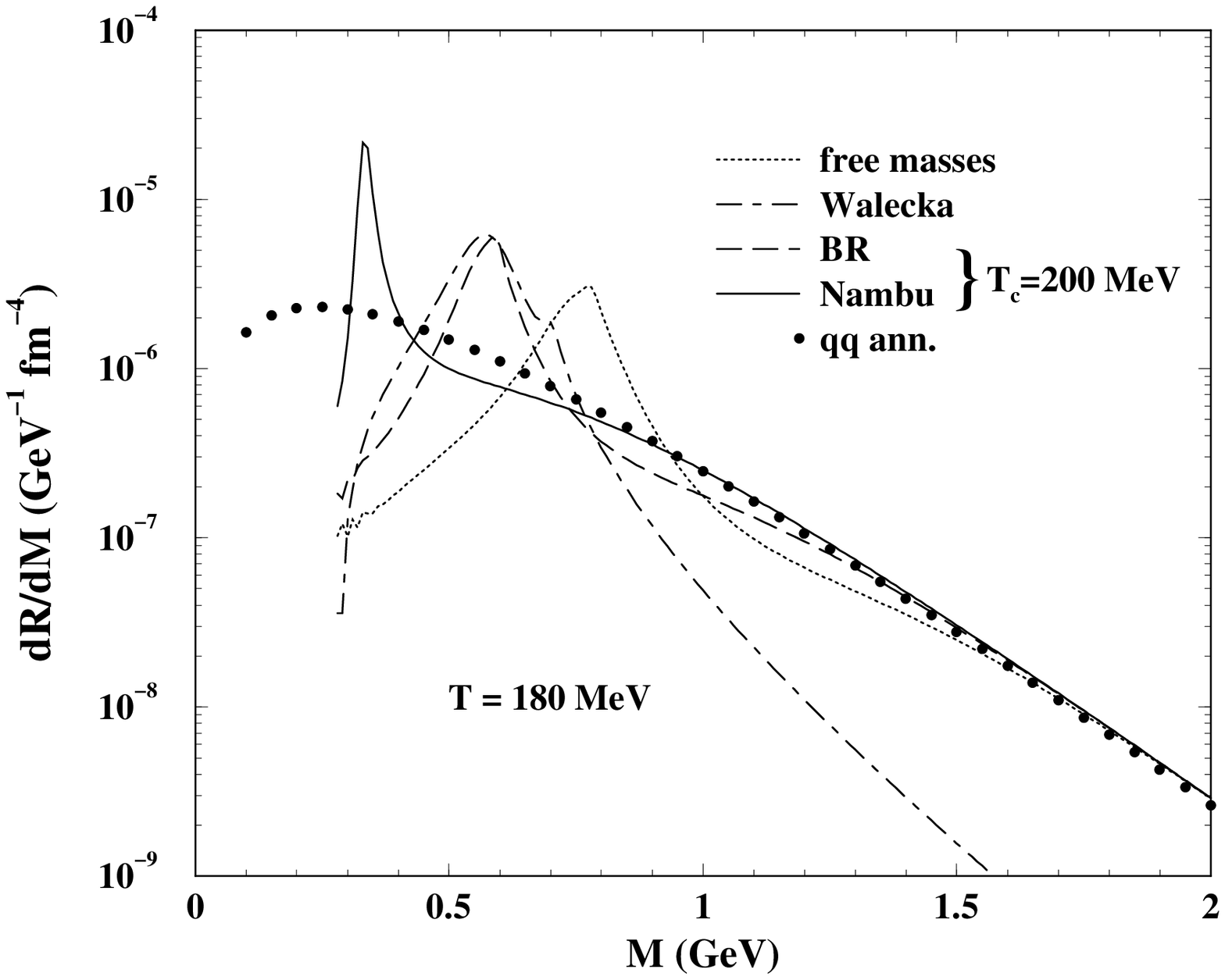,height=7cm,width=9cm}}
\caption{
Same as Fig.~(\protect\ref{7fig12}) at $T=180$ MeV. 
}
\label{7fig15}
\eef
%%%%%%%%%%%%%%%%%%%%%%%%%%%%%%%%%%%%%%%%%%%
%\vskip 0.2cm
The dilepton invariant mass distribution at $T=T_c$ is shown in 
Fig.~(\ref{7fig16}). All the peaks in the spectrum have disappeared
as expected. The rates obtained from the electromagnetic current-current
correlator is close to the rate from $q\bar{q}$ annihilation, 
indicating that the $q\bar{q}$ interaction in the vector channel
has become very weak, signaling the onset of deconfinement 
~\cite{shuryakqm99,rappqm99}.
%\vskip 0.2cm
%%%%%%%%%%%%%% Figure 21 %%%%%%%%%%%%%%%%%%%%%%%%%%%%%
\bef
\centerline{\psfig{figure=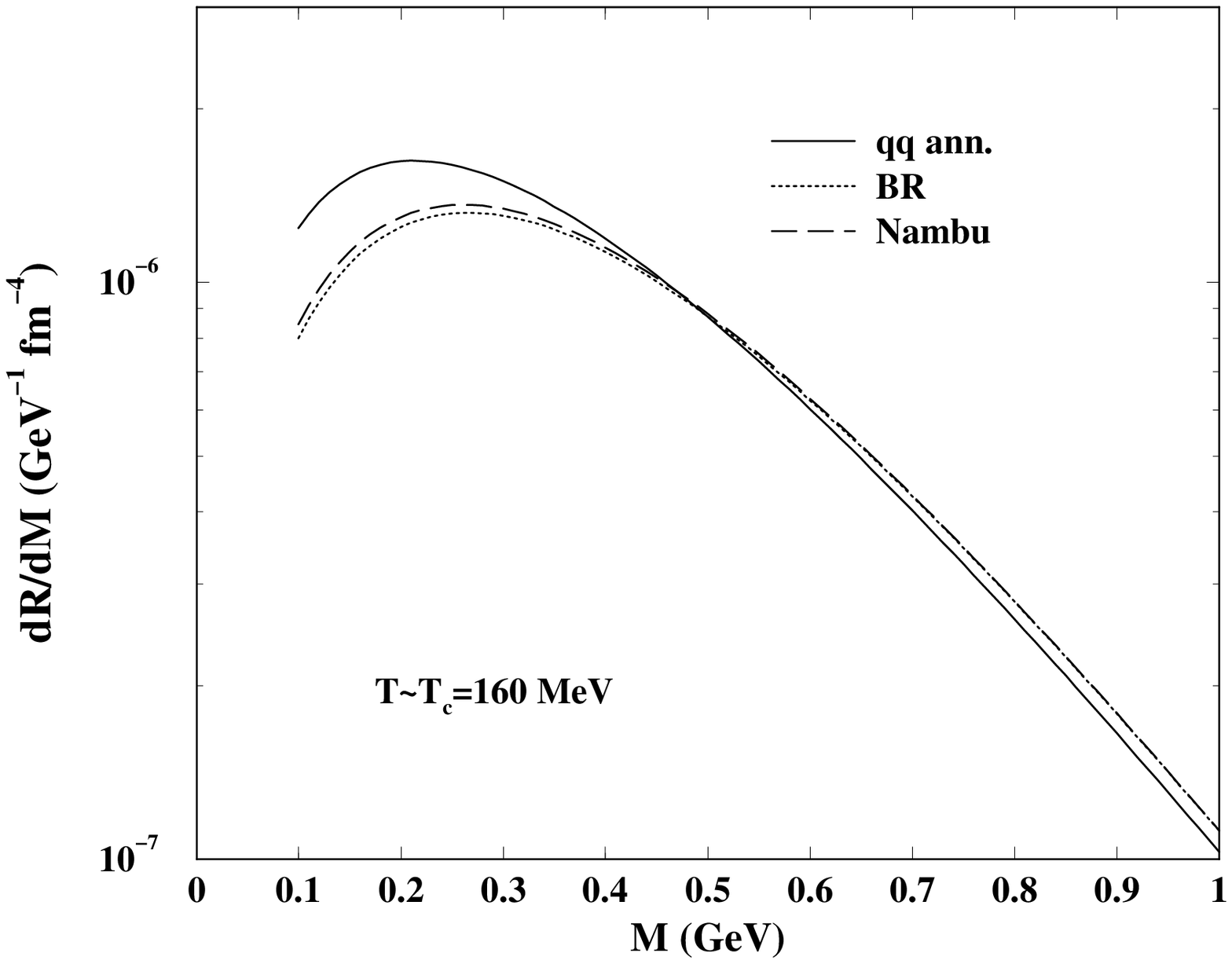,height=7cm,width=9cm}}
\caption{
Same as Fig.~(\protect\ref{7fig12}) at $T=T_c$ MeV. 
}
\label{7fig16}
\eef
%%%%%%%%%%%%%%%%%%%%%%%%%%%%%%%%%%%%%%%%%%%
%\vskip 0.2cm

In Fig.~(\ref{7fig17}) we compare the dilepton emission rate
at $T=150$ MeV for vacuum mass of $\rho$ with the rates where the effective
masses are obtained in the framework of gauged LSM, NLSM and HLS approach.
The positive shift of $\rho$ mass in NLSM is reflected in the peak position  
of the spectra towards larger value of $M$ (dotted line). A very small 
change in the $\rho$ mass in HLS approach does not cause any visible
change in the invariant mass distribution of dilepton.
The long-dashed line indicates lepton pair distribution for $m_\rho=770$ MeV
and $\Gamma_\rho=400$ MeV. The large width of $\rho$ leads to the
disappearance of the $\rho$ peak from the spectra, indicating that $\rho$
ceases to exist as a quasi-particle. However, as mentioned before, it is
interesting to note that the photon spectra is ``insensitive'' to such
drastic broadening of the $\rho$ meson.
%\vskip 0.2cm
%%%%%%%%%%%%%% Figure 22 %%%%%%%%%%%%%%%%%%%%%%%%%%%%%
\bef
\centerline{\psfig{figure=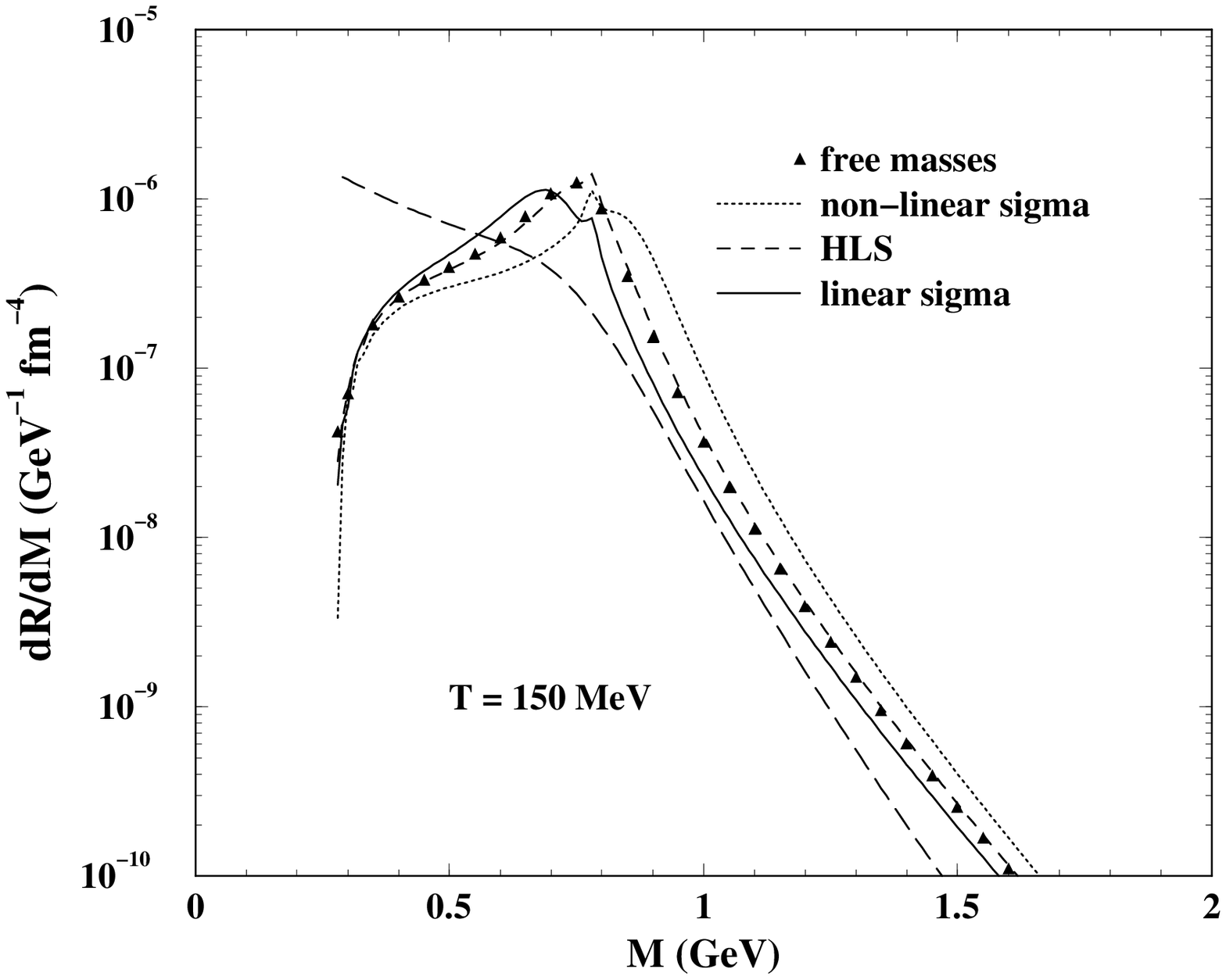,height=7cm,width=9cm}}
\caption{ The change in the invariant mass distribution 
of lepton pairs due to the finite temperature
effects on the hadronic masses in Gauged Linear, Non-Linear Sigma Model and
Hidden Local Symmetry approach at T= 150 MeV. The long-dashed line
indicates dilepton spectra for $\Gamma_\rho=400$ MeV and $m_\rho=770$
MeV (see text).
}
\label{7fig17}
\eef
%%%%%%%%%%%%%%%%%%%%%%%%%%%%%%%%%%%%%%%%%%%

\subsection{Photon and dilepton spectra with space-time evolution}

As mentioned before the basic aim of URHIC is to distinguish between the two
possibilities:

\centerline{\bf A\,+A\,$\ra$QGP$\ra$Mixed Phase$\ra$Hadronic Phase 
}
\centerline{\bf\,or\,}
\centerline{\bf A\,+\,A\,$\ra$Hadronic Phase}
The former (latter) case where the initial
state is formed in QGP (hadronic) phase will be called the `QGP scenario'
(`no phase transition scenario'). In the following we will compare
the photon and dilepton spectra originating from these two scenarios with
and without medium effects.

The observed photon and dilepton spectra originating from an expanding 
QGP or hadronic matter is obtained by convoluting the static
(fixed temperature) rate with the expansion dynamics. 
The basic ingredients required 
for a  system undergoing rapid expansion from its initial formation
stage to the final freeze-out stage with or without phase transition
have been discussed in section 6.
For the QGP sector we use a simple bag model equation of state (EOS) with
two flavour degrees of freedom. The temperature in the QGP phase evolves
according to Bjorken scaling law $T^3\,\tau=T_i^3\tau_i$.
The cooling law in the hadronic sector is quite different from that of the QGP
because of the presence of massive hadrons. These hadrons redress themselves in
the medium thereby changing their vacuum masses. This phenomenon must be taken into account
in the evolution dynamics through the equation of state. We do this by 
introducing temperature dependence in the statistical 
degeneracy which takes care of the mass varying with temperature.
  
In Fig.~(\ref{7fig18}) we depict the variation of effective degeneracy 
as a function of temperature with and without medium effects
on the hadronic masses for various scenarios. 
We observe that for $T>140$ MeV the effective
degeneracy becomes larger due to the reduction in temperature dependent 
masses compared to the free hadronic masses. 
Physically this means that the number of hadrons
in a thermal bath at a temperature $T$ is more when in-medium mass reduction
is taken into account.  Eq.~(\ref{dndy}) implies that for a given
pion multiplicity the initial temperature of the system will be 
lower (higher) when medium effects on hadronic masses are considered
(ignored). This is clearly demonstrated in Fig.~(\ref{7fig19})
where we show the variation of temperature 
with proper time for different initial conditions. 
%\vskip 0.2cm
%%%%%%%%%%%%%% Figure 23 %%%%%%%%%%%%%%%%%%%%%%%%%%%%%
\bef
\centerline{\psfig{figure=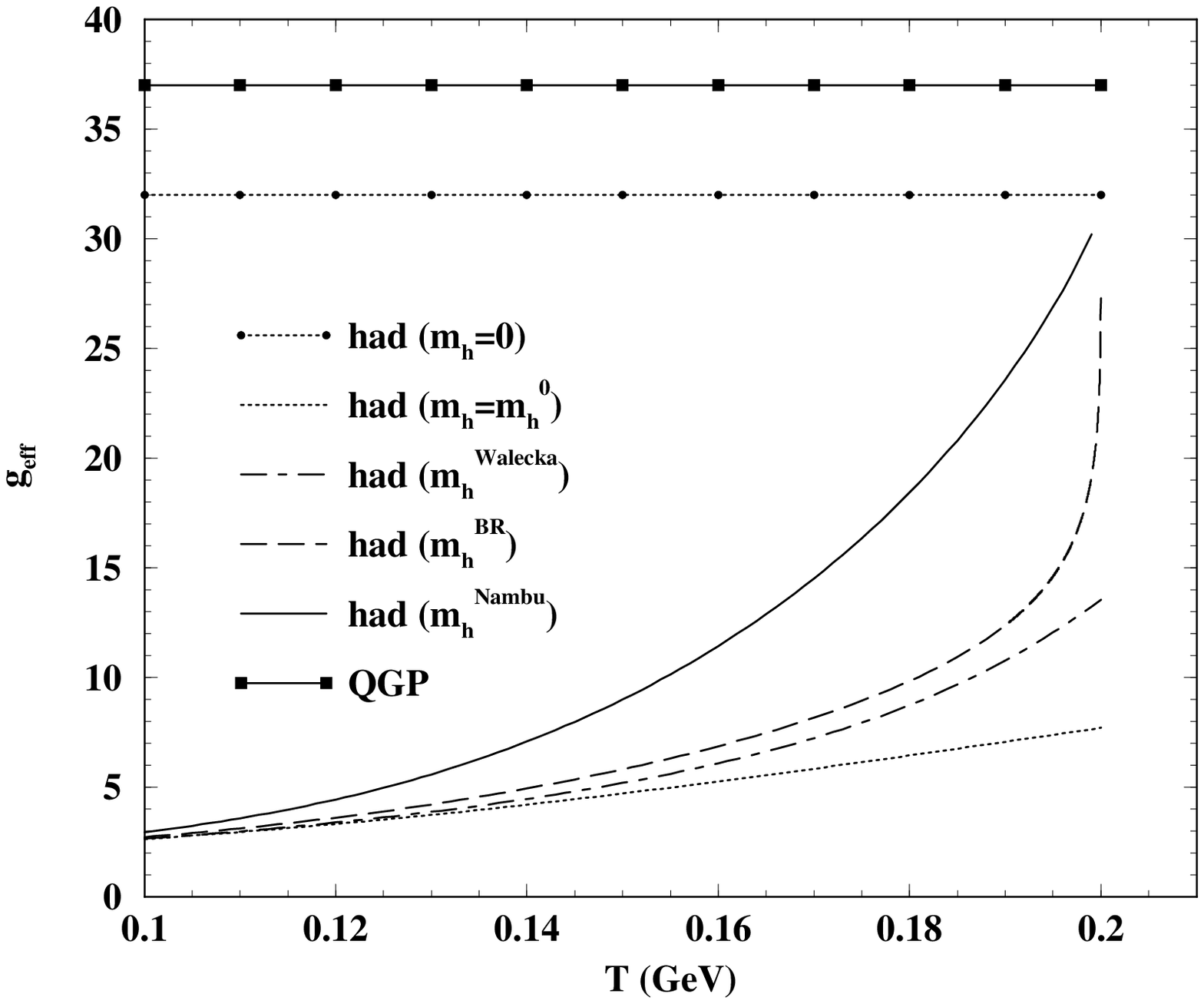,height=7cm,width=9cm}}
\caption{
Variation of effective degeneracy as a function of temperature.
}
\label{7fig18}
\eef
%%%%%%%%%%%%%%%%%%%%%%%%%%%%%%%%%%%%%%%%%%%
%\vskip 0.2cm
The solid dots indicate the scenario where QGP is formed initially 
at $T_i=185$ MeV and cools down according to 
Bjorken law upto a temperature $T_c$ at proper time $\tau_Q$,
at which a phase transition
takes place; it remains constant at $T_c$ up to a time
$\tau_H=9.4$ fm/c after which the temperature decreases as 
$T=0.247/\tau^{0.194}$ (when medium effects  
are taken from Walecka model) to a temperature $T_f$($=130$ MeV). 
If the system is considered to be formed in the hadronic phase 
then the initial temperature is obtained as $T_i=220$ MeV (270 MeV)
when in-medium effects on the hadronic masses from Walecka model is 
taken into account
(ignored). The corresponding cooling laws are
displayed in Fig.~(\ref{7fig19}). 
The above parametrizations of the cooling law in the hadronic phase
have been obtained by solving Eq.~(\ref{entro1}) self consistently.
An initial state with the vanishing meson masses at 
$T_i=195$ MeV ($\tau_i=1$ fm/c) could be realised in 
the case of BR and Nambu scaling scenarios
for the value of pion multiplicity, $dN/dy=600$. 
%\vskip 0.2cm
%%%%%%%%%%%%%% Figure 24 %%%%%%%%%%%%%%%%%%%%%%%%%%%%%
\bef
\centerline{\psfig{figure=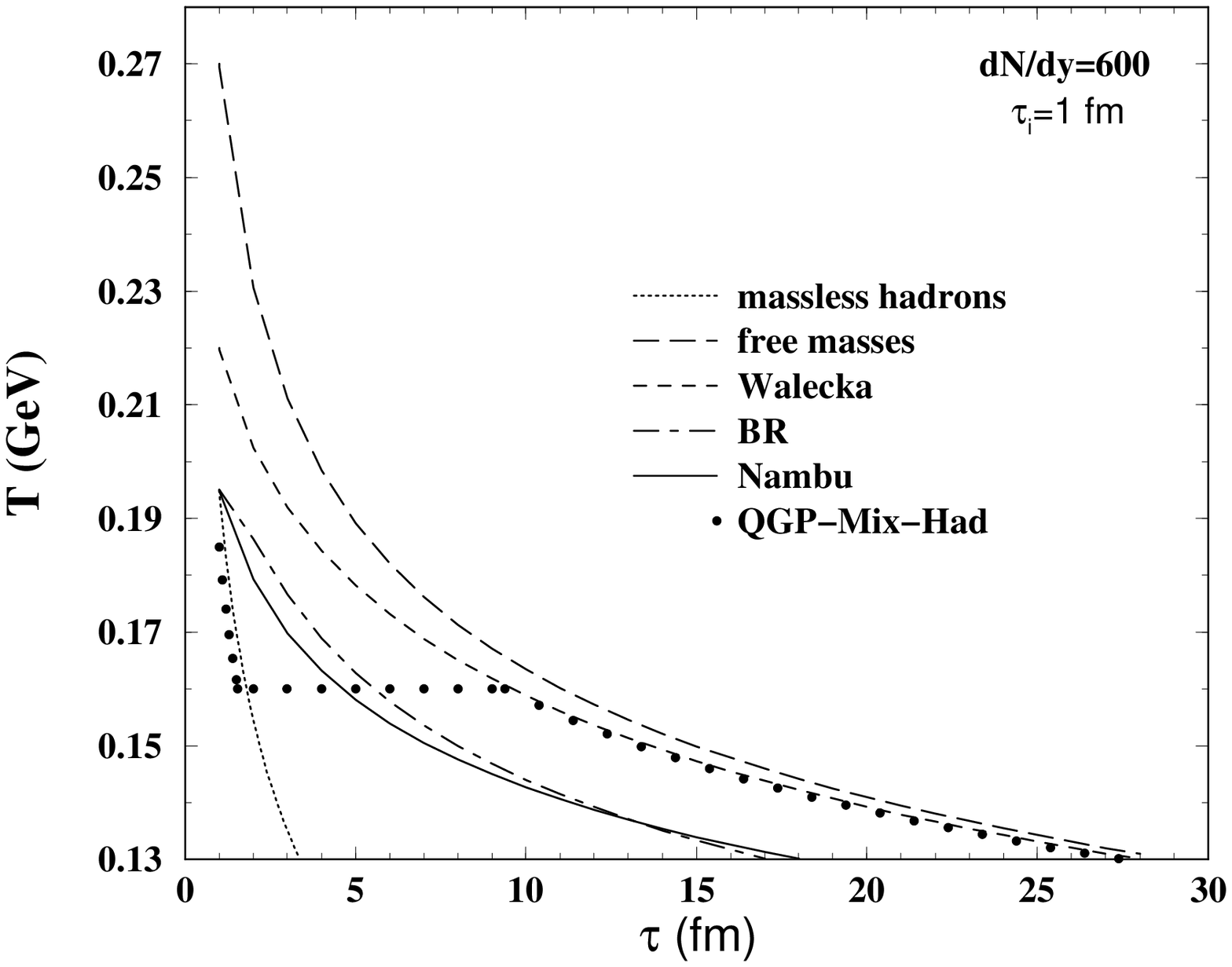,height=7cm,width=9cm}}
\caption{
Variation of temperature as a function of proper time.
}
\label{7fig19}
\eef
%%%%%%%%%%%%%%%%%%%%%%%%%%%%%%%%%%%%%%%%%%%
%\vskip 0.2cm

In Table~1 we quote the values of the initial temperatures
obtained by assuming various mass variation scenarios. The value of initial
thermalization time has been assumed as 1 fm/c both for SPS ($dN/dy=600$)
and RHIC ($dN/dy=1735$) energies. $\tau_Q$ ($\tau_H$) indicates the starting
(end) point of the mixed phase. $\tau_H-\tau_Q$ is  the life time of the mixed
phase in a first order phase transition scenario. $\vartheta$ and $\delta$ 
dictate the variation of temperature with proper time for the 
hadronic matter according to the cooling law  $T=\vartheta/\tau^{\delta}$. 
The values of $\delta$ indicate a slower cooling in the hadronic phase as 
compared to that in the QGP phase ($T\sim\,1/\tau^{0.33}$). 
\vskip 0.2in
%%%%%%%%%%%%%%%%%%%%% Table 1 %%%%%%%%%%%%%%%% 
\begin{center}
\begin{tabular}{|l|c|c|c|c|c|c|c|}
\hline
&
\multicolumn{4}{c|}{$dN/dy$=600 $\tau_i$=1 fm}&
\multicolumn{3}{c|}{$dN/dy$=1735 $\tau_i$=1 fm} \\
\cline{2-8}
& hadronic gas &
\multicolumn{3}{c|}{QGP + Mix + Had} &
\multicolumn{3}{c|}{QGP + Mix + Had}  \\
& initial state & \multicolumn{3}{c|}{$T_i$=185 MeV $\tau_Q$=1.6 fm} &
\multicolumn{3}{c|}{$T_i$=265 MeV $\tau_Q$=4.6 fm}  \\
\cline{2-8}
& $T_i$ (MeV) & $\tau_H$ (fm) & $\vartheta$ & $\delta$ & 
$\tau_H$ (fm) & $\vartheta$ & $\delta$  \\
\hline
free mass & 270 & 10.8  & 0.267 &  0.215  &  31.9  &  0.337  &  0.215  \\
Walecka &  220 & 9.4  &  0.247  &  0.194  & 27.6  &  0.305  &  0.194  \\
BR  & 195 &  8.2  &  0.236  &  0.184  &  23.9  &  0.288  &  0.185  \\
Nambu & 195 & 4.7  &  0.203  &  0.151  &  13.9  &  0.239  &  0.152  \\
\hline
\end{tabular}
\end{center}
\begin{center}
Table 1 : Values of initial temperatures and various time scales for
SPS and RHIC energies.
\end{center}
%%%%%%%%%%%%%%%%%%%%% End of Table 1 %%%%%%%%%%%%%%%% 
Having obtained the finite temperature effects on hadronic properties
and the cooling laws we now
integrate the rates obtained in the previous sections over the space-time
evolution of the collision. We must account for the fact that the thermal rates
are evaluated in the rest frame of the emitting matter and hence the momenta of the
emitted photons or dileptons are expressed in that frame. Accordingly,
the integral over the expanding matter is of the form
\be
\frac{dN}{d\Gamma}=\begin{array}{c}
{\small {freeze-out}}\\{\displaystyle\int}\\{\small {formation}}\end{array}\,
d^4x\frac{dR(E^\ast,T(x))}{d\Gamma}
\ee 
where $d\Gamma$ stands for invariant phase space elements:
$d^3p/E$ for photons and $d^4q$ for dileptons.
$E^\ast$ is the energy of the photon or lepton pair 
in the rest frame of the emitting matter
and $T(x)$ is the local temperature.
In a fixed frame like the laboratory or the centre of mass frame, where
the 4-momentum of the photon or lepton pair is $q_\mu=(E,\vec q)$ 
and the emitting matter element $d^3x$ moves with a velocity
$u_\mu=\gamma(1,\vec v)$, the energy in the rest frame of the fluid 
element is given by $E^\ast=u_\mu q^\mu$.

In a first order
phase transition scenario 
the photon and dilepton spectra
from a $(1+1)$ dimensionally expanding system is obtained as
\begin{eqnarray}
\frac{dN}{d\Gamma}&=&\pi\,R_A^2\,\int\left[
\left(\frac{dR}{d\Gamma}\right)_{QGP}
\Theta(s-s_Q^c)\right.\nonumber\\
& &+\left[\left(\frac{dR}{d\Gamma}\right)_{QGP}\frac{s-s_H^c}
{s_Q^c-s_H^c}\right.\nonumber\\
& &+\left.\left(\frac{dR}{d\Gamma}\right)_{H}\frac{s_Q^c-s}
{s_Q^c-s_H^c}\right]
\Theta(s_Q^c-s)
\Theta(s-s_H^c)\nonumber\\
& &+ 
\left.\left(\frac{dR}{d\Gamma}\right)_{H}\Theta(s_H^c-s)\right]
\tau\,d\tau\,d\eta 
\end{eqnarray}
where $R_A$ is the radius of the nuclei 
and $\Theta$ functions are 
introduced to get the contribution from individual phases.
$s_H^c$ and $s_Q^c$ are defined in section 7.2.

As discussed earlier, $g_{\s eff}$ is obtained as a function of $T$  
by solving Eq.~(\ref{entro}). A smaller (larger) value of $g_{\s eff}$ 
is obtained in the free (effective) mass scenario. 
As a result we get a larger (smaller) initial temperature
by solving Eq.~(\ref{dndy}) in the free (dropping) mass scenario 
for a given multiplicity.
Naively we expect that at a given temperature if a meson mass drops
its Boltzmann factor will be enhanced and more of those mesons will
be produced leading to more photons~\cite{sourav,csong}.
However, a larger drop in the hadronic masses results in smaller initial
temperature, implying that the
space time integrated spectra crucially depends on these two
competitive factors.
Therefore, with (without) medium effects one integrates an enhanced (depleted) 
static rate over smaller (larger) temperature range for a fixed
freeze-out temperature ($T_f=130$ MeV in the present case).
In the present calculation the enhancement in the photon emission due to 
the higher initial temperature in the free mass scenario (where static rate is
smaller) overwhelms the enhancement of the rate due to negative shift
in the vector meson masses (where the initial temperature is smaller).
%%%%%%%%%%%%%% Figure.25 %%%%%%%%%%%%%%%%%%%%%%%%%%%%%
\begin{figure}
\centerline{\psfig{figure=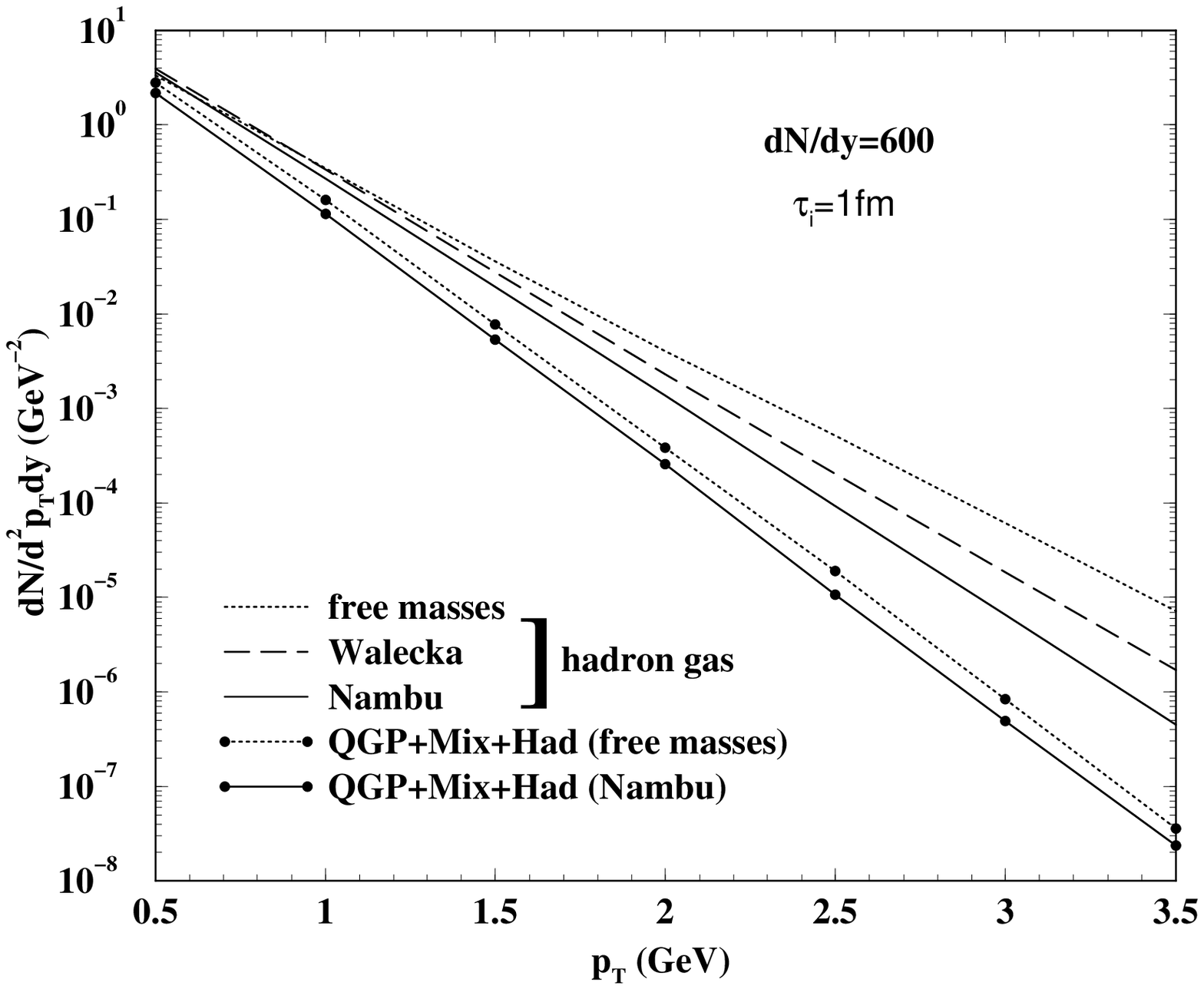,height=7cm,width=9cm}}
\caption{Total thermal photon yield corresponding to $dN/dy=600$ and
$\tau_i=1$ fm/c. 
The solid (long-dash) line indicates photon spectra  
when hadronic matter formed in the initial state 
at  $T_i=195$ MeV ($T_i=220$ MeV) 
and the medium effects are taken from Nambu scaling (Walecka model).
The dotted line  represents the photon spectra without 
medium effects
with $T_i=270$ MeV. The solid (dotted) line with solid dots
represent the yield for the `QGP scenario' when the hadronic
mass variations are taken from Nambu scaling (free mass).
}
\label{7fig20}
\end{figure}
%%%%%%%%%%%%%%%%%%%%%%%%%%%%%%%%%%%%%%%%%%%
Accordingly, in the case of free mass ( Nambu scaling) scenario the 
photon yield is the highest (lowest). In case of the Walecka model,
the photon yield lies between the above two limits.
This is demonstrated in Fig.~(\ref{7fig20}).
%%%%%%%%%%%%%% Figure.26 %%%%%%%%%%%%%%%%%%%%%%%%%%%%%
\begin{figure}
\centerline{\psfig{figure=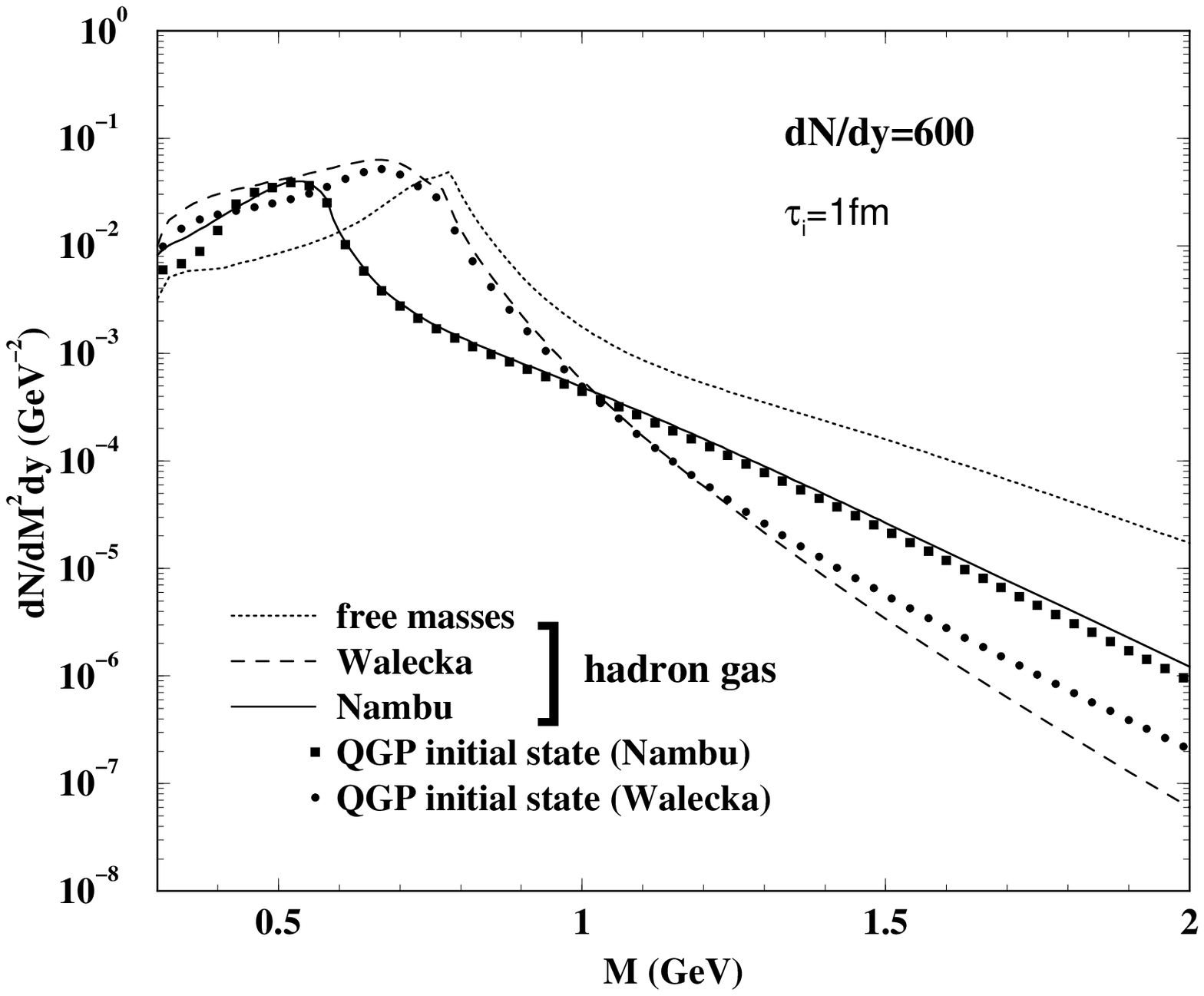,height=7cm,width=9cm}}
\caption{Total thermal dilepton yield corresponding to $dN/dy=600$ and
$\tau_i=1$ fm/c. 
The solid (long-dash) line indicates dilepton spectra  
when hadronic matter formed in the initial state 
at  $T_i=195$ MeV ($T_i=220$ MeV) 
and the medium effects are taken from Nambu scaling (Walecka model).
The dotted line  represents the spectra without medium effects with
$T_i=270$ MeV.
The square (solid dots) represent the yield for the `QGP scenario' 
when the hadronic
mass variations are taken from Nambu scaling (Walecka model).
}
\label{7fig21}
\end{figure}
%%%%%%%%%%%%%% %%%%%%%%%%%%%%%%%%%%%%%%%%%%%

In the `QGP scenario' the  photon
yield with in-medium mass is lower than the case where free masses of hadrons are
considered. However, the difference is considerably less than the
`no phase transition scenario'.   
This is because, in this case
the initial temperature is determined by the quark and gluon 
degrees of freedom and the main difference between the two
is due to the different lifetimes of the mixed phase.  
In Fig.~(\ref{7fig20}), the photon spectra
from `QGP scenario' is compared with that from `no phase
transition scenario'; the latter overshines the former.

The space time integrated dilepton spectra  for
the `QGP scenario' and `no phase transition scenario' 
with different mass variation are shown in Fig.(\ref{7fig21}).
The shifts in the invariant mass distribution of the spectra
due to the reduction in the hadronic masses according
to different models are distinctly visible. Similar to the
photon spectra, the dilepton spectra from `no phase
transition scenario' dominates over the `QGP' scenario
for invariant mass beyond $\rho$ peak.

The space time integrated photon yield  corresponding to 
the static (fixed temperature) rate shown in Fig.~(\ref{7fig13})
(when in-medium effects are taken from gauged Linear and Non-Linear
Sigma models and Hidden Local Symmetry)
are not displayed in Fig.(\ref{7fig20}) because the resulting
spectra does not differ appreciably from those already shown 
and it makes the Fig.(\ref{7fig20}) clumsy. 
For the same reason the corresponding dilepton spectra are also not shown.

\begin{figure}
\centerline{\psfig{figure=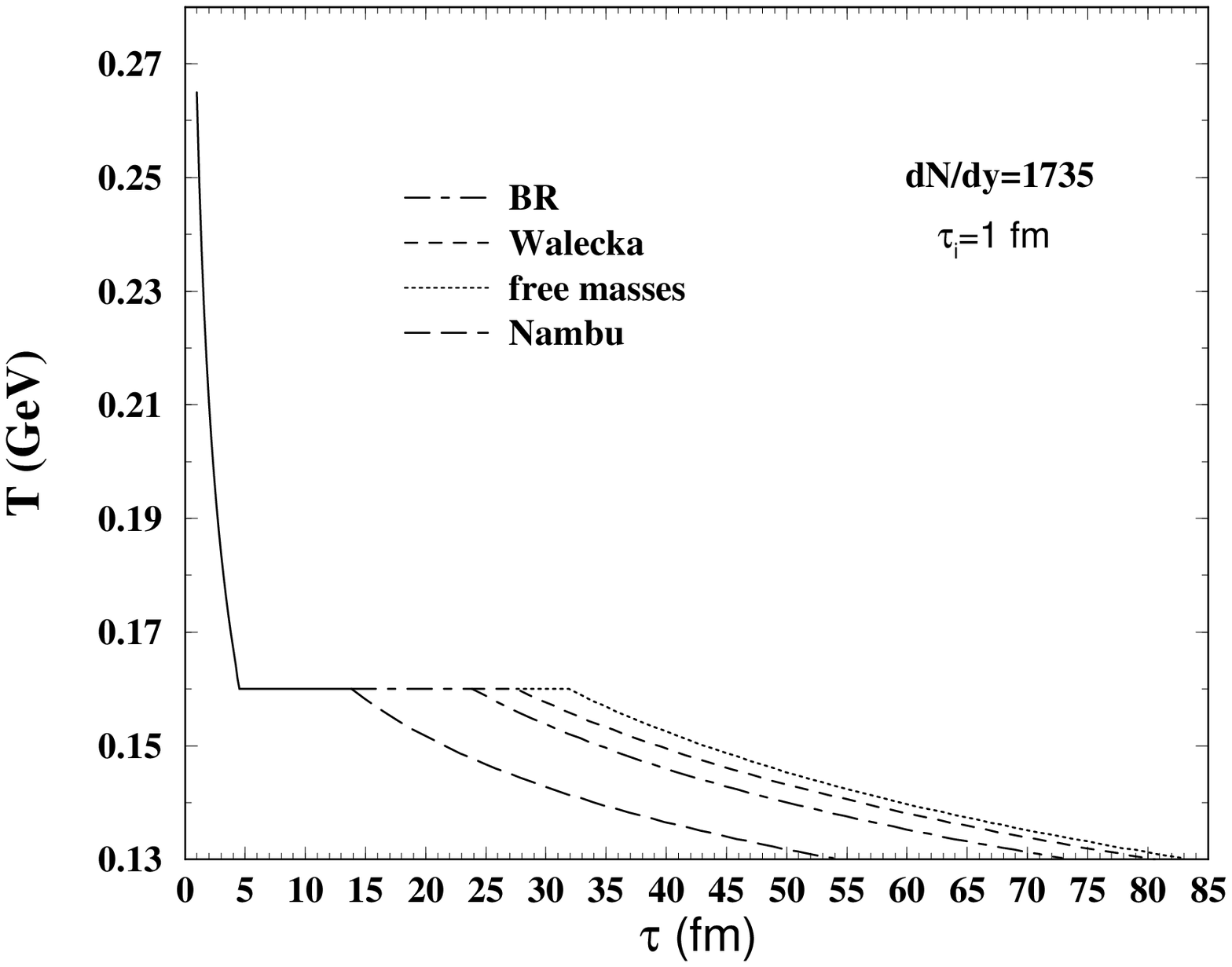,height=7cm,width=9cm}}
\caption{Variation of temperature as a function of proper time.
The initial temperature has been determined by assuming `QGP
scenario'. The initial temperature $T_i=265$ MeV for $\tau_i=1$ fm/c 
and $dN/dy=1735$.
}
\label{7fig22}
\end{figure}

Finally we study the electromagnetic probes for RHIC energies.
At RHIC a scenario of a pure hot hadronic system
within the format of the model used here, appears to be
unrealistic. The initial temperature considering free hadronic
masses turns out to be $\sim$ 340 MeV whereas for the other extreme
case of massless hadrons it is $\sim$ 290 MeV. With temperature
dependent masses the initial temperature will lie somewhere between these
two values. For such high temperatures, clearly a hot dense hadronic
system cannot be a reality, the hadrons would have melted away even for 
lower temperatures. Thus, for RHIC we have treated the case of a QGP
initial state only. The temperature profile for RHIC is depicted 
in Fig.~(\ref{7fig22}) where we observe that 
the length of the plateau, 
which indicates the life time of the mixed 
phase $\tau_{mix}^{life}=\tau_H-\tau_Q$,
depends on the masses of the hadrons in the hadronic phase. 
The effective degeneracy plays an important role here.
At the transition point there is a large decrease in the entropy density.
This decrease has to be compensated by the
expansion (increasing the volume) to keep the total entropy constant. 
Since we are considering
(1+1) dimensional expansion this change in the entropy density will
be compensated by increasing $\tau$ ($s\tau=$ const.). We have
seen earlier (Fig.~(\ref{7fig18})) that the effective degeneracy in the 
hadronic phase is the largest for the Nambu scaling and smallest for 
the free mass scenario, resulting in smallest (largest) discontinuity in the
entropy density for the former (latter) case. Consequently the time
taken for the system to compensate the decrease of the entropy density 
in the Nambu scaling scenario is smaller as compared to free mass case.
Hence the life time of the mixed phase for the Nambu scaling case is 
smaller than all other cases.

%%%%%%%%%%%%%% Figure. 27 %%%%%%%%%%%%%%%%%%%%%%%%%%%%%
%%%%%%%%%% %%%%%%%%%%%%%%%%%%%%%%%%%%%%%%%%

\begin{figure}
\centerline{\psfig{figure=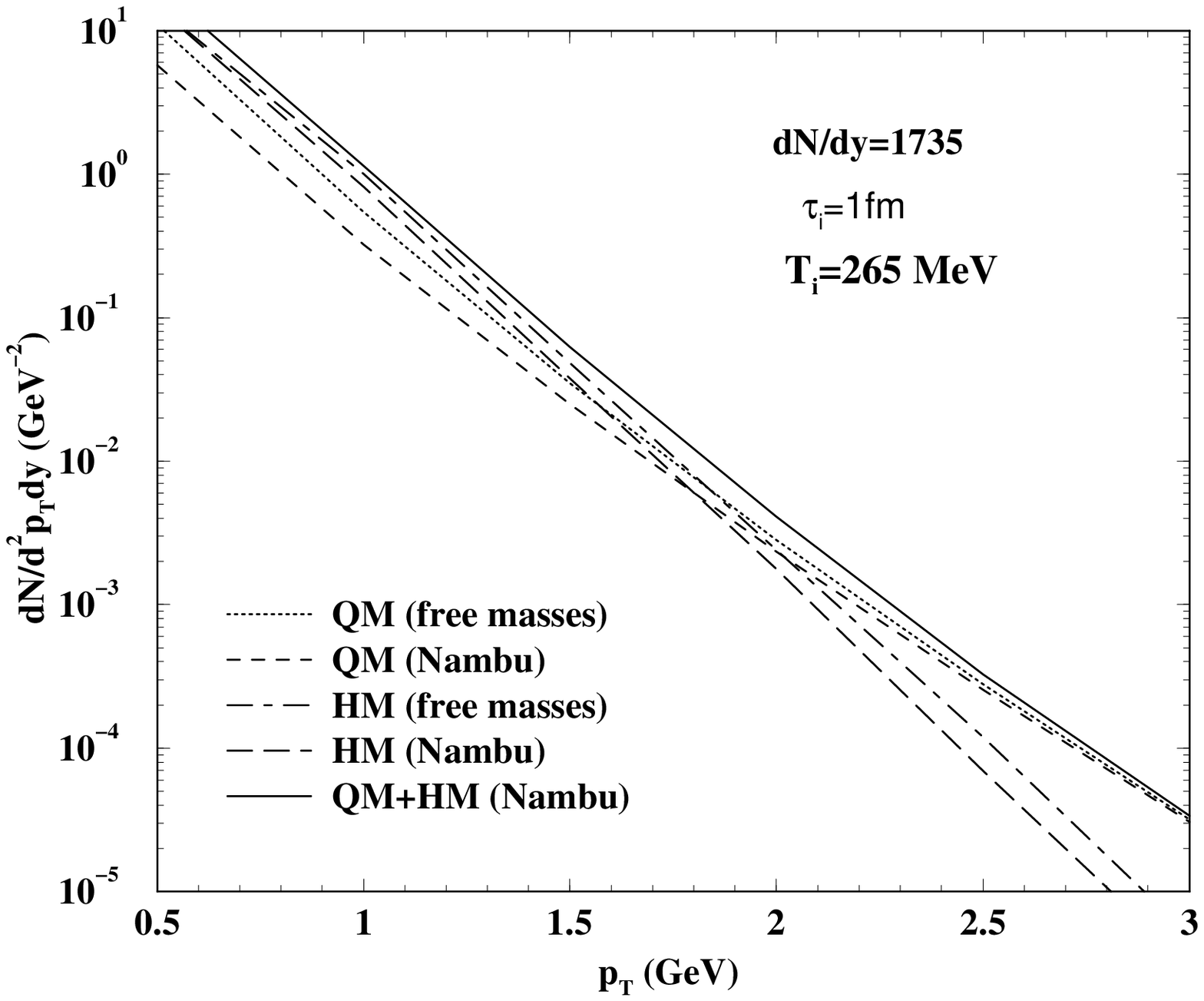,height=7cm,width=9cm}}
\caption{Thermal photon spectra at RHIC energies.
}
\label{7fig23}
\end{figure}

The thermal photon spectra for RHIC is displayed in Fig.~(\ref{7fig23}).
The solid line represents the total thermal photon yield 
originating from initial QGP state, mixed phase and the pure hadronic
phase. The short dash line indicates photons from quark 
matter (QM) (= pure QGP phase + QGP part of the mixed phase)
and the long dash line represents photons from hadronic matter (HM)
(= hadronic part of the mixed phase + pure hadronic phase).
In all these cases the effective masses of the hadrons have been taken 
from Nambu scaling. For $p_T>2$ GeV photons from QM overshines
those from HM since most of these high $p_T$ photons originate
from the high temperature QGP phase. We arrive at the similar
conclusions when
in-medium effects in the Walecka model is considered 
(not shown in Fig.~(\ref{7fig23}) to avoid clumsiness).
The dotted and the dotdash lines indicate photon yields from QM
and HM respectively with free masses in the hadronic sector. 
The HM contribution for the free mass is larger than the
effective mass (Nambu) scenario because of the larger value of the life
time of the mixed phase in the earlier case (see Table~1).
We note here that for $p_T>2$ GeV, photons from QM 
overshine those from HM irrespective of the models used
for calculating the in-medium modifications of the hadrons.

%%%%%%%%%%%%%% Figure. 28 %%%%%%%%%%%%%%%%%%%%%%%%%%%%%
%%%%%%%%%%%%%%%%%%%%%%%%%%%%%%%%%%%%%%%%%%

\begin{figure}
\centerline{\psfig{figure=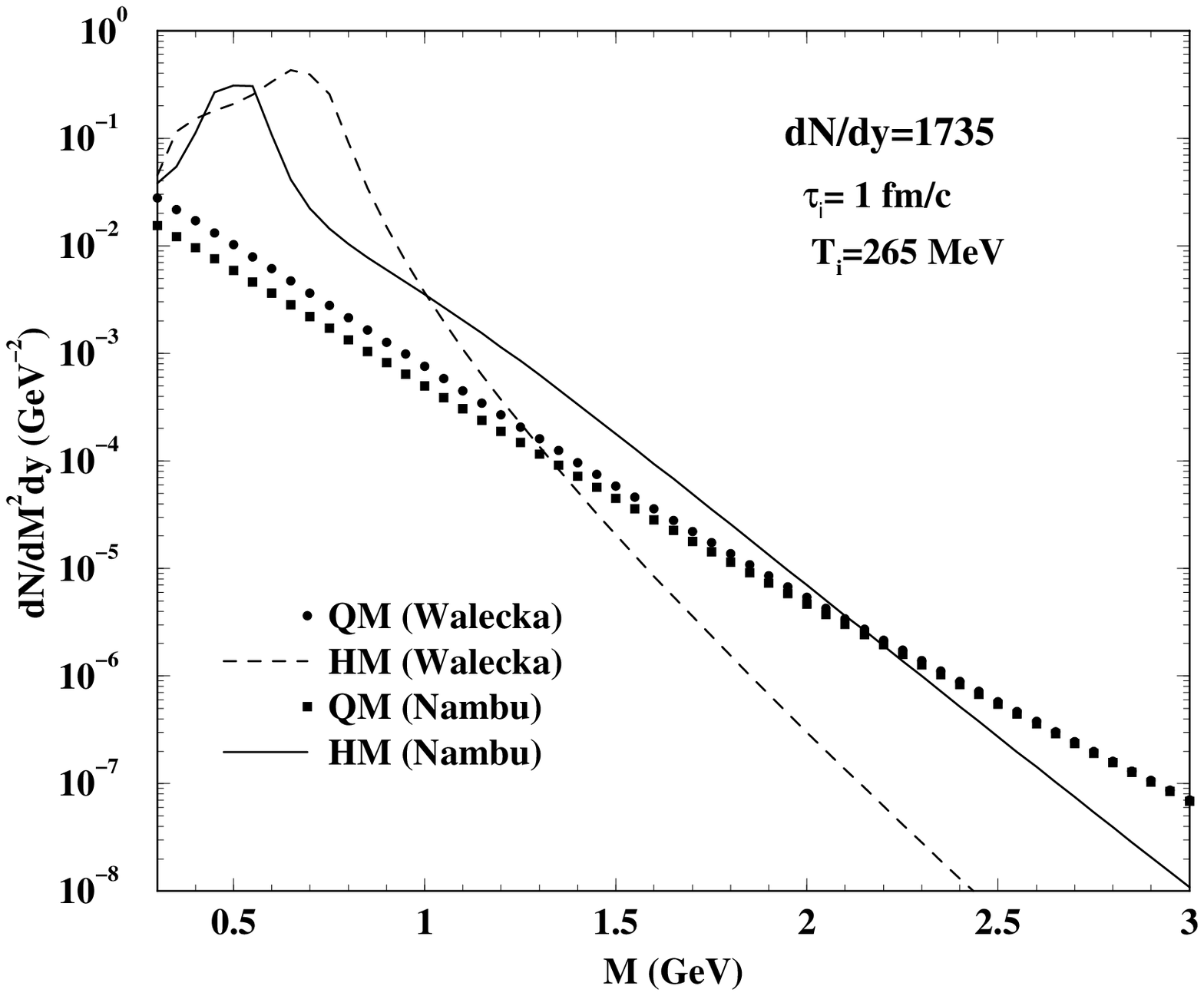,height=7cm,width=9cm}}
\caption{Thermal dilepton spectra at RHIC energies.
}
\label{7fig24}
\end{figure}

Thermal dilepton yield at RHIC energies for QGP initial state 
and for different mass variation scenarios are shown in 
Fig.~(\ref{7fig24}). The shape of the peak in the dilepton
spectra for Walecka model is slightly different(broader) from that of 
Nambu scenario because of the larger mass separation 
between $\rho$ and $\omega$
mesons in the former case (see Fig.~(\ref{7fig1})). 
The shift in the peak of the dilepton spectra
towards lower invariant mass 
for both the Walecka model and 
Nambu scaling scenario is clearly visible.
However, the dilepton yield from QGP dominates 
over the contribution from the hadronic phase
for $M\geq 1.4$ ($M\geq 2.3$) GeV 
for Walecka model (Nambu scaling).

%%%%%%%%%%%%%% Figure. 29 %%%%%%%%%%%%%%%%%%%%%%%%%%%%%
%%%%%%%%%%%%%% %%%%%%%%%%%%%%%%%%%%%%%%%%%%%

%%%%%%%%%%%%%%%%%%%%%%%%%%%%%%%%summary.tex %%%%%%%%%%%%%%%%%%%%%%%

\section{Summary and Outlook}
In the present work we have reviewed the formulation of the
production of photon and lepton pair from QGP and hot hadronic
gas based on finite temperature field theory. The changes in the
spectral functions of the hadrons appearing in the internal
loop of the photon self energy diagram have been considered in the
Walecka model, gauged linear sigma model, non-linear sigma model,
hidden local symmetry approach
and QCD sum rule approach. The hadronic spectral
functions (in vacuum) for the isovector and isoscalar channel have been
constrained from the experimental data of $e^+e^-\rightarrow hadrons$.
Due to the lack of our
understanding of the critical behaviour of scalar and tensor condensates 
we have parametrized the vector meson masses and continuum threshold
as a function of temperature according to BR and Nambu scaling.

We observe that the in-medium effects on the hadronic
properties within the frame work of 
the Gauged Linear and Non-Linear Sigma Model, 
Hidden Local Symmetry approach are too small
to affect the electromagnetic spectra substantially.
However, the shift in the hadronic properties of
different magnitude  within the frame work of the
Walecka model, Brown-Rho scaling and Nambu scaling
scenarios  are prominently visible through the low invariant mass 
distribution of dilepton and transverse momentum spectra 
of photon at SPS energies. The photons from `no phase transition
scenario' outshine those originating from the `QGP scenario' for the
entire range of $p_T$ at SPS.
At RHIC energies a scenario of a pure hot hadronic system
appears to be unrealistic because of the very high initial temperature
obtained within the format of the model used in the present work.
We observe that at RHIC energies the thermal photon (dilepton)
spectra originating from Quark Gluon Plasma overshines those from
hadronic matter for high transverse momentum (invariant mass) irrespective
of the models used for evaluating the finite temperature effects
on the hadronic properties.

We note that Walecka model calculation 
gives different mass shift for $\rho$ and
$\omega$ mesons (because $\rho$ and $\omega$ couples to 
nucleons with different strength). The disentanglement of the
$\rho$ and $\omega$ peaks in the dilepton spectrum resulting
from URHIC would be an excellent evidence of in-medium mass
shift of vector mesons ~\cite{CERES} and/or validity of 
such model calculations for the situation under consideration.
It is also interesting to note that the dilepton
spectra is affected both by the changes in the decay width as
well as in the mass of the vector mesons.
However, the photon spectra is affected
only by the change in the mass of the vector mesons but rather 
insensitive to the change in its width. The effects
of the continuum on the dilepton spectra are seen to be substantial.

In the following we would like to reiterate some of the 
assumptions made in the present calculations. Further
work is needed for the justifications of these assumptions.

The exact value of the critical temperature ($T_c$) for 
the deconfinement phase transition is still uncertain. 
However, recent lattice simulation~\cite{ukawa}
for two flavour QCD indicates a value of $T_c$ for chiral transition 
$\sim 130-160$ MeV.  We have taken $T_c=160$ MeV, although till 
now it is not known whether the values of $T_c$ for the chiral 
and deconfinement transition are the same or not. 
The value of the initial thermalization time $\tau_i$ is 
unfortunately also an unknown quantity (though considerable
progress has been made for the understanding of the initial
condition~\cite{lmrv}, a lot is yet to be done).
We take $\tau_i=1$ fm/c as a canonical value following Bjorken
\cite{bjorken}. A similar value of $\tau_i$ has been 
considered in the literature, e.g. see 
Refs.~\cite{janepr,dumitru,csong,solf}.

The photon production from QGP has been evaluated using 
HTL resummation based on the assumption $g_s<<1$, which is
unlikely to meet in URHIC even at the highest energy to
be available at the CERN LHC in future. 
The strong coupling constant is likely to attain a value $g_s\sim$ 2 at
RHIC/LHC. Evaluation of the photon spectra at such high values is
a formidable task.
In this respect the development of methods suitable for
addressing non-perturbative effects near and 
above the QCD phase
transition point is of paramount importance. 
Extension of the 
self-consistent resummation
scheme developed in $\phi^4$ theory~\cite{chiku} to 
non-abelian gauge theory~\cite{abs,bir} would be a 
very important step towards the understanding of the phenomena 
near the QCD phase transition. 

In this work we have neglected transverse expansion of the evolving matter.
This is because our main emphasis has been to study
the shift in the photon and dilepton spectra
due to the change in the hadronic  spectral function 
at non-zero temperature. Inclusion of the transverse 
expansion will shift the momentum distribution of
the photon both for the free mass and in-medium mass
scenarios but the relative shift will remain approximately
unchanged and it is this relative change in which
we are interested.
The effects of the non-zero baryonic chemical potential are ignored
in the present work, calculation addressing these issues is in
progress~\cite{spjhb}.

Throughout this work we have
assumed thermal equilibrium, which may not be realized 
practically~\cite{geiger,xnwang,janeprl,pkr,rqmd}. 
Unfortunately, although considerable progress has been made~\cite{an,mt,mb},
the general techniques for solving non-equilibrium 
quantum field theoretical problems is
still in the early stages of development~\cite{kajantie}. 

In spite of some of the remaining uncertainties,
the progress in this 
field is remarkable, especially when one considers that not a 
tremendous number of experiments to seek out QGP have been performed.
Many of the ambiguities pertaining to 
the ``pre-data'' theory have been removed by the experimental
data from AGS and SPS. We have all the reasons to look forward
to RHIC and LHC where the initial energy density will be so 
high that the formation of QGP is almost inevitable.

\vspace{1cm}
{\bf Acknowledgement:} J. A. is grateful to Japan Society
for Promotion of Science (JSPS) for financial support.
T. H. was partly supported by Grant-in-Aid for Scientific
Research No. 10874042 of the Japanese Ministry of Education,
Science, and Culture. J. A. and T. H. were also supported 
 by Grant-in-Aid for Scientific Research No. 98360 of
 JSPS.
\newpage
%%%%%%%%%%%%%%%%%%%%%%%%%%%%%%% appendix %%%%%%%%%%%%%%%%%%%%%%%
\addcontentsline{toc}{section}{Appendix : Thermal Propagators}
\setcounter{equation}{0}
\def\theequation{I.\arabic{equation}}
\section*{Appendix - Thermal Propagators}

In this appendix we will briefly discuss
the in-medium (thermal)
propagators in Quantum Field Theory~\cite{bellac,adas,kapustaft,landsman}
 which have been used extensively in
this work. We will begin by first defining the
propagators in vacuum.

The free propagator of a complex scalar field $\phi$ propagating with
a momentum $p$ in vacuum is defined as (see {\it e.g.}~\cite{peskin})
\bea
i\bar\D^0(p)&\equiv&\int\,d^4x\,e^{ip\cdot x}\langle T\{\phi(x)\phi^*(0)\} \rangle_0
\nonumber\\
          &=&\frac{i}{p^2-m^2+i\eps}
\label{sprop0}
\eea
The operator $T$ appearing within angular brackets ensures that the field
operators are time-ordered and the subscript `0' indicates that there are
no interactions.
The vacuum propagator for fermions is defined as
\bea
i\bar G^0_{\ab}(p)&\equiv&\int\,d^4x\,e^{ip\cdot x}\langle T\{\psi_\alpha(x)\bar\psi_\beta(0)\} \rangle_0
\nonumber\\
          &=&\frac{i(p\!\!\!/ + m)_{\ab}}{p^2-m^2+i\eps}
\label{fprop0}
\eea
where $\alpha$ and $\beta$ denote the spinor indices of the fermion field $\psi$.
For massive vector particles we define the free propagator in vacuum as
\bea
i\bar D^0_{\mn}(p)&\equiv&\int\,d^4x\,e^{ip\cdot x}\langle T\{A_\mu(x)A_\nu(0)\} \rangle_0
\nonumber\\
          &=&\frac{i(-g^{\mn}+p^\mu p^\nu/m^2)}{p^2-m^2+i\eps}
\label{vprop0}
\eea
where $A_\mu$ is the massive vector field. For charged vector particles
the fields $A_\mu$ will also have isospin indices.
It must be noted that the fields $\phi$, $\psi$ and $A_\mu$ appearing above
are free fields and the expectation values are
calculated between noninteracting vacuum states. The propagators defined 
through Eqs.~(\ref{sprop0}), (\ref{fprop0}) and (\ref{vprop0}) are
referred to as Feynman propagators.

In the presence of interactions these propagators have to be redefined
with interacting Heisenberg fields in place of the free fields
and interacting vacua instead of the free vacua.
The interacting propagator can be expressed in terms of the free(non-interacting) 
propagator through perturbation theory. 
In the scalar case, for example, the exact propagator in the presence
of interactions, $\bar \D$, is 
obtained as
\be
\bar\D=\bar\D_0 + \bar\D_0 \Pi \bar\D
\ee
Where, $\Pi$ is the self energy of the particle due to interactions.
This equation is known as the Dyson-Schwinger equation for propagators.

Let us now study the situation in a medium at finite temperature (and density).
We will be interested in a system in thermal equilibrium. Hence we will 
assume that the interaction slowly switches off as we go into the
remote past and the fields become noninteracting fields
satisfying the free equations of motion.
These fields appear in the definition of the free propagators in the
medium. The thermal propagator has more
structure than the vacuum one arising due to different combinations
of time-ordering on the real time contour~\cite{bellac,adas,kobes}.
In the real time formalism there are four non-trivial propagator structures
possible which are collected in a 2$\times$2 matrix~\cite{nieves,pvl}. 
For scalars the free thermal propagator is defined as
\bea
i\bf \Delta_0 &\equiv& \left[ \begin{array}{ll}
i\Delta^{11}_0(p)& 
i\Delta^{12}_0(p)\\ 
i\Delta^{21}_0(p)& 
i\Delta^{22}_0(p)\\
\end{array}\right] \nonumber\\
&=&\left[ \begin{array}{ll}
\int\,d^4x\,e^{ip\cdot x}\langle T\{\phi(x)\phi^*(0)\}\rangle_T^0 &
\int\,d^4x\,e^{ip\cdot x}\langle\{\phi^*(0)\phi(x)\}\rangle_T^0\\
\int\,d^4x\,e^{ip\cdot x}\langle\{\phi(x)\phi^*(0)\}\rangle_T^0 &
\int\,d^4x\,e^{ip\cdot x}\langle\bar T\{\phi(x)\phi^*(0)\}\rangle_T^0 \\
\end{array}\right]
\label{s0propft}
\eea
where the operator $\bar T$ denotes anti-time-ordered product. The subscript `$T$'
indicates that a thermal average is being performed.

In order to obtain the thermal propagators in momentum space one
follows the usual procedure of expanding the field operators in terms
of the creation and annihilation operators and making use of the 
commutation relations between them.
The four components are then obtained as
\bea
\Delta^{11}_0(p)&=&\frac{1}{p^2-m^2+i\eps}
-2\pi i\delta(p^2-m^2)\eta(p.u)\nonumber\\ 
\Delta^{12}_0(p)&=&-2\pi i\delta(p^2-m^2)[\eta(p.u)+\theta(-p.u)]\nonumber\\
\Delta^{21}_0(p)&=&-2\pi i\delta(p^2-m^2)[\eta(p.u)+\theta(p.u)]\nonumber\\
\Delta^{22}_0(p)&=&\frac{-1}{p^2-m^2-i\eps}
-2\pi i\delta(p^2-m^2)\eta(p.u)
\label{s0propft1}
\eea
where $\eta(p.u)=\theta(p.u)f_{BE}(z)+\theta(-p.u)f_{BE}(-z)$.  
$f_{BE}=[e^z-1]^{-1}$ is the Bose distribution 
with $z=(p\cdot u-\mu)/T$ and $u^{\mu}$ is
the four velocity of the thermal bath.
We observe that the elements of the matrix propagator $\bf \D_0$ are not independent. From
their definitions one can see that $\D_0^{11}$ and $\D_0^{22}$ can be expressed in terms of
$\D_0^{12}$ and $\D_0^{21}$. Also, the Kubo-Martin-Schwinger periodicity
condition yields $\D_0^{12}(p)=e^{-z}\D_0^{21}(p)$.

Similarly, for fermions the corresponding thermal propagators are,
\bea
i\bf G^0_{\ab} &\equiv& \left[ \begin{array}{ll}
iG_{\ab}^{0(11)}(p)& 
iG_{\ab}^{0(12)}(p)\\ 
iG_{\ab}^{0(21)}(p)& 
iG_{\ab}^{0(22)}(p)\\
\end{array}\right] \nonumber\\
&=&\left[ \begin{array}{ll}
\int\,d^4x\,e^{ip\cdot x}\langle T\{\psi_\alpha(x)\bar\psi_\beta(0)\}\rangle_T^0 &
-\int\,d^4x\,e^{ip\cdot x}\langle\{\bar\psi_\beta(0)\psi_\alpha(x)\}\rangle_T^0\\
\int\,d^4x\,e^{ip\cdot x}\langle\{\psi_\alpha(x)\bar\psi_\beta(0)\}\rangle_T^0 &
\int\,d^4x\,e^{ip\cdot x}\langle\bar T\{\psi_\alpha(x)\bar\psi_\beta(0)\}\rangle_T^0 \\
\end{array}\right] 
\label{f0propft}
\eea
Explicitly, the four components are
\bea
G_{\ab}^{0(11)}(p)&=&(p\sls+m)_{\ab}\left[\frac{1}{p^2-m^2+i\eps}
+2\pi i\delta(p^2-m^2)\eta(p.u)\right]\nonumber\\ 
G_{\ab}^{0(12)}(p)&=&2\pi i(p\sls +m)_{\ab}\delta(p^2-m^2)[\eta(p.u)-\theta(-p.u)]\nonumber\\
G_{\ab}^{0(21)}(p)&=&2\pi i(p\sls +m)_{\ab}\delta(p^2-m^2)[\eta(p.u)-\theta(p.u)]\nonumber\\
G_{\ab}^{0(22)}(p)&=&(p\sls+m)_{\ab}\left[\frac{-1}{p^2-m^2-i\eps}
+2\pi i\delta(p^2-m^2)\eta(p.u)\right] 
\label{f0propft1}
\eea
where,
$\eta(p.u)=\theta(p.u)f_{FD}(z)+\theta(-p.u)f_{FD}(-z)$,
$f_{FD}=[e^z+1]^{-1}$, the Fermi-Dirac distribution.
For the fermions the KMS anti-periodicity condition leads to
$G^{12}_0=-e^{-z}G^{21}_0$. 

Lastly, we define the finite temperature propagators for massive vector particles:
\bea
i\bf D^0_{\mn} &\equiv& \left[ \begin{array}{ll}
iD_{\mn}^{0(11)}(p)& 
iD_{\mn}^{0(12)}(p)\\ 
iD_{\mn}^{0(21)}(p)& 
iD_{\mn}^{0(22)}(p)\\
\end{array}\right] \nonumber\\
&=&\left[ \begin{array}{ll}
\int\,d^4x\,e^{ip\cdot x}\langle T\{A_\mu(x)A_\nu(0)\}\rangle_T^0 &
\int\,d^4x\,e^{ip\cdot x}\langle\{A_\nu(0)A_\mu(x)\}\rangle_T^0\\
\int\,d^4x\,e^{ip\cdot x}\langle\{A_\mu(x)A_\nu(0)\}\rangle_T^0 &
\int\,d^4x\,e^{ip\cdot x}\langle\bar T\{A_\mu(x)A_\nu(0)\}\rangle_T^0 \\
\end{array}\right]
\label{v0propft}
\eea
where the isospin indices relevant for charged vector particles like the
$\rho$ meson have been suppressed.

It is important to note that the real time propagators as given
by Eqs.~(\ref{s0propft1}) and (\ref{f0propft1}) consist of two
parts  - one corresponding to the vacuum, describing the exchange 
of virtual particle and the other, the temperature dependent part,
describing the participation of real (on-shell) particles present in the
thermal bath in the emission and absorption processes. The temperature
dependent part does not change the ultra-violet behaviour 
of the theory as it contains on-shell contributions
which has a natural cut-off due to the Boltzmann
factor. Therefore, the zero temperature counter term 
is adequate for the renormalization of the theory. 
However, the infra-red problem becomes more severe at finite
temperature~\cite{bellac,thoma}.

Thermal field theory in the real time approach can be reformulated by 
diagonalising the 2$\times$2 matrix propagators described above.
A well-known possibility is to diagonalise to a matrix constructed from
the Feynman propagators. The non-interacting propagator defined by 
Eq.~(\ref{s0propft}) can be written as
\be
\bf \Delta_0 ={\bf U} \left[ \begin{array}{cc}
\bar\Delta_0 & 0 \\ 
0 & -\bar\Delta^*_0 \\
\end{array}\right]{\bf U} 
\label{diago}
\ee
where 
\[
{\bf U} = \left[ \begin{array}{cc}
\sqrt{1+\eta} & {\displaystyle \frac{\eta+\theta(-p\cdot u)}{\sqrt{1+\eta}}} \\ 
 & \\
{\displaystyle\frac{\eta+\theta(p\cdot u)}{\sqrt{1+\eta}}} & \sqrt{1+\eta} \\ 
\end{array}\right]. 
\]
The exact propagators in the medium
can be defined analogously as Eqs.~(\ref{s0propft}), (\ref{f0propft})
and (\ref{v0propft}) with interacting Heisenberg fields instead of the
free fields. In this case we write
\be
\bf \Delta ={\bf U} \left[ \begin{array}{cc}
\Delta & 0 \\ 
0 & -\Delta^* \\
\end{array}\right]{\bf U} 
\ee
where $\bf \D$ is the matrix of interacting thermal propagators.
Using thermal perturbation theory $\bf\D$ can be expanded in terms of $\bf\D_0$.
One obtains
\be
{\bf \D} = {\bf \D_0} + {\bf\D_0}{\bf \Pi}{\bf \D}
\label{d-sft}
\ee
where ${\bf \Pi}$ now is a matrix of self energies;
\be
\bf \Pi ={\bf U}^{-1} \left[ \begin{array}{cc}
\Pi & 0 \\ 
0 & -\Pi^* \\
\end{array}\right]{\bf U}^{-1} 
\ee

The diagonal component of Eq.~(\ref{d-sft}) is given by
\be
\D =  \bar\D_0 + \bar\D_0 {\s \Pi} \D
\ee

In the following we will discuss
the vector (spin 1) propagator in some detail.
It is very similar to the scalar case except now we have to take
into account the Lorentz structure of the propagator and the self energy.
The exact propagator(matrix) $\bf D$ can be diagonalised as above 
and the diagonal element satisfies Dyson equation
\be
D_{\mn}=\bar D^0_{\mn}+\bar D^0_{\mu\rho}{\s \Pi}^{\rho\sigma}D_{\sigma\nu}
\ee
which gives
\be
D_{\mu \nu}^{-1} = (\bar D_{\mu \nu}^{0})^{-1} - \Pi_{\mu \nu},
\label{dyson}
\ee
where 
$\bar D_{\mu \nu}^{0}$ is the vacuum Feynman propagator for massive vector
particles as defined above and
$\Pi_{\mu \nu}$ is the self energy which can be written as a sum of two
contributions:
\be
\Pi^{\mu \nu}=\Pi_{\rm {vac}}^{\mu \nu}+\Pi_{\rm {med}}^{\mu \nu},
\label{pitot}
\ee
where
\be
\Pi_{\rm {vac}}^{\mu \nu}=(g^{\mu \nu} - \frac{p^{\mu}p^{\nu}}{p^2})\,
\Pi_{\rm {vac}}(p^2), 
\label{pivac}
\ee
is the vacuum contribution to the self energy. 
$p^{\mu} = (\omega, {\vec p})$
is the four-momentum of the propagating particle.
In a thermal 
bath moving with four-velocity $u^{\mu}$, $\Pi_{\rm {med}}^{\mu \nu}$
has transverse and longitudinal components~\cite{adas}:
\be
\Pi_{{\rm med}}^{\mn}(\omega,{\vec p}) = A^{\mn}\Pi_{T,{{\rm med}}} +
 B^{\mn}\Pi_{L,{\rm {med}}}.
\label{pimed}
\ee
$A^{\mu \nu}$ and $ B^{\mu \nu}$ are the transverse and longitudinal
projection tensors given by
\be
A^{\mn}=\frac{1}{p^2-\omega^2}\left[(p^2-\omega^2)(g^{\mn}-u^{\mu}u^{\nu})
\frac{}{}-p^{\mu}p^{\nu}-\omega^2u^{\mu}u^{\nu}+\omega(u^{\mu}p^{\nu}+
p^{\mu}u^{\nu})\right],
\ee
and
\be
B^{\mn}=\frac{1}{p^2(p^2-\omega^2)}\left[\frac{}{}\omega^2p^{\mu}p^{\nu}+
p^4u^{\mu}u^{\nu}-\omega p^2(u^{\mu}p^{\nu}+p^{\mu}u^{\nu})\right],
\ee
which satisfy the following algebra:
\bea
A_{\mu \rho}A^{\rho \nu} &=& A^\nu_\mu\nonumber\\
B_{\mu \rho}B^{\rho \nu} &=& B^\nu_\mu\nonumber\\
A_{\mu \rho}B^{\rho \nu} &=& 0\nonumber\\
A^{\mu \nu} + B^{\mu \nu} &=& g^{\mu \nu} - \frac{p^{\mu}p^{\nu}}{p^2}.
\label{ab}
\eea
Using Eqs.~(\ref{dyson}-\ref{ab}) the effective propagator becomes
\be
D_{\mu \nu} = -\,\frac{A_{\mu \nu}}{p^2-m^2+\Pi_{T}}
-\,\frac{B_{\mu \nu}}{p^2-m^2+\Pi_{L}} + \frac{p_{\mu}p_{\nu}}{p^4},
\label{deff}
\ee
where
\be
\Pi_{T(L)}=\Pi_{T(L),{\rm {med}}}+\Pi_{\rm {vac}}
\ee
and $m$, we recall, is the bare mass of the particle.
The real part of $\Pi_{T(L)}$
affects the dispersion relation of the particle in the medium.
The displaced pole position of the effective propagator
in the rest frame of the propagating particle (i.e. where the
three momentum of the particle is zero) gives the effective
mass of the particle in the medium. The imaginary part of ${\s \Pi}_{T(L)}$
is connected to the decay width.

A different scheme in the formulation of finite temperature field
theory known as the `R/A' formalism~\cite{bech} (see also~\cite{eijck}), 
is to diagonalise to a matrix composed of retarded and
advanced propagators which are known to have better analyticity properties 
than the Feynman ones. 
In this case the analogue of Eq.~(\ref{diago}) is
\be
\bf \Delta_0 ={\bf V} \left[ \begin{array}{cc}
\Delta^R_0 & 0 \\ 
0 & \Delta^A_0 \\
\end{array}\right]{\bf W}. 
\ee
The matrices $\bf V$ and $\bf W$ depend on the momentum as well as
the thermal factor containing the distribution functions. Their exact forms
are given in Ref.~\cite{bech}.
The retarded and advanced propagators are
defined as
\bea
i\D^R_0&\equiv&\int\,d^4x\,e^{ip\cdot x}\theta(x_0)\lgl[\phi(x),\phi(0)]\rangle_0 \nonumber\\
i\D^A_0&\equiv&\int\,d^4x\,e^{ip\cdot x}\theta(-x_0)\lgl[\phi(x),\phi(0)]\rangle_0 .
\eea
For the case of vectors one arrives at the following
expression for the effective retarded propagator at finite temperature:
\be
D^R_{\mu \nu} = -\,\frac{A_{\mu \nu}}{p^2-m^2+\Pi^R_{T}}
-\,\frac{B_{\mu \nu}}{p^2-m^2+\Pi^R_{L}} + \frac{p_{\mu}p_{\nu}}{p^4},
\label{deffr}
\ee
where ${\s \Pi}^R_T$ and ${\s \Pi}^R_L$ are respectively the retarded
transverse and longitudinal components of the self energy.

Before we end our discussion on finite temperature propagators let us briefly mention 
about the imaginary time formalism or Matsubara formalism
which has been used extensively in the literature.
In the imaginary
time formalism, the form of the propagator at finite temperature 
remains same as that of vacuum but the time component of the four-momentum
takes discrete values, i.e. $p_0=2n\pi\,iT (=(2n+1)\pi\,iT)$ for bosons
(fermions) with $n=-\infty$ to $+\infty$, the vertices
are the  same as the zero temperature theory and the loop integral 
$\int d^4p/(2\pi)^4$ is replaced by the sum $iT\sum_{n}\int d^3p/(2\pi)^3$.
There are standard tricks to evaluate the sum over the 
frequencies~\cite{bellac}. 
Another method (known as SACLAY method) 
which uses the mixed representation
of the propagator i.e. it depends on the three momentum and Euclidean time
has also been used extensively in the literature~\cite{braaten}, 
a detailed discussion of which can be found in Ref.~\cite{npb309}.
The propagators in the imaginary time formalism
can also be obtained by proper analytic continuation of the
real time propagators~\cite{agd,fradkin}.

%%%%%%%%%%%%%%%%%%%%%%%%%%%%%%% ref.tex %%%%%%%%%%%%%%%%%%%%%%%%%%

\end{document}